\documentclass[pdflatex,sn-mathphys-num]{sn-jnl}%

\usepackage{graphicx,color}
\usepackage{multirow}%
\usepackage{amsmath,amssymb,amsfonts}%
\usepackage{amsthm}%
\usepackage{mathrsfs}%
\usepackage[title]{appendix}%
\usepackage{xcolor}%
\usepackage{textcomp}%
\usepackage{manyfoot}%
\usepackage{booktabs}%
\usepackage{algorithm}%
\usepackage{algorithmic}
\usepackage{listings}%

\usepackage{latexsym}
\usepackage{indentfirst}

\usepackage{placeins}
\usepackage{booktabs}
\usepackage{multirow}
\usepackage[caption=false]{subfig}
\usepackage{diagbox}
\usepackage{braket}
\usepackage[normalem]{ulem}
\usepackage{hyperref}
\usepackage{cleveref}
\usepackage{newfloat}
\usepackage{bbm}

\usepackage{amsmath,amsfonts,bm}

\def\eqref#1{(\ref{#1})}

\def\ceil#1{\lceil #1 \rceil}

\def\va{{\bm{a}}}

\def\vr{{\bm{r}}}

\def\mI{{\bm{I}}}

\DeclareMathAlphabet{\mathsfit}{\encodingdefault}{\sfdefault}{m}{sl}
\SetMathAlphabet{\mathsfit}{bold}{\encodingdefault}{\sfdefault}{bx}{n}

\def\gA{{\mathcal{A}}}

\def\gC{{\mathcal{C}}}

\def\gM{{\mathcal{M}}}

\def\gO{{\mathcal{O}}}

\def\gU{{\mathcal{U}}}

\def\sC{{\mathbb{C}}}

\def\sN{{\mathbb{N}}}

\def\sR{{\mathbb{R}}}

\newcommand{\R}{\mathbb{R}}

\newcommand{\KL}{D_{\mathrm{KL}}}

\DeclareMathOperator*{\argmin}{arg\,min}

\usepackage{amsmath,tikz}
\usetikzlibrary{positioning,calc}

\tikzset{
  tensor/.style={draw,rectangle,fill=blue!10,minimum width=1.8cm,
                 minimum height=1.1cm,thick,align=center},
  tensorv2/.style={draw,rounded corners=2pt,fill=blue!10,minimum width=1.8cm,
                 minimum height=1.1cm,thick,align=center},
  leg/.style={thick},
  tn/vertical leg/.store in=\TNVerticalLeg,
  tn/vertical leg=0.3cm,
  tn/horizontal leg/.store in=\TNHorizontalLeg,
  tn/horizontal leg=0.5cm,
}

\newcommand{\tninline}[2][1]{%
  \vcenter{\hbox{%
    \tikz[baseline=-0.6ex,scale=#1,transform shape,
          every node/.style={font=\Large}]{#2}%
  }}%
}

\newcommand{\PsiBlock}[1][1]{%
  \tninline[#1]{%
    \node[tensor,minimum width=4cm] (psi) {$C$};
    \foreach \x/\lab in {0.5/{i_1}, 1.5/{i_2}, 3.5/{i_n}}{
      \draw[leg] ($(psi.south west)+(\x,0)$) -- ++(0,-\TNVerticalLeg);
      \node[below] at ($(psi.south west)+(\x,-\TNVerticalLeg)$) {$\lab$};
    }
    \foreach \x/\lab in {2.5/{\cdots}}{
      \node[below] at ($(psi.south west)+(\x,-\TNVerticalLeg)$) {$\lab$};
    }
  }%
}

\newcommand{\BkBlock}[1][1]{%
  \tninline[#1]{%
    \node[tensor,minimum width=3cm] (psi) {$B_k$};
    \foreach \x/\lab in {0.4/{\zeta}, 1.5/{i_k}, 2.6/{\mu}}{
      \draw[leg] ($(psi.south west)+(\x,0)$) -- ++(0,-0.6cm);
      \node[below] at ($(psi.south west)+(\x,-0.6cm)$) {$\lab$};
    }
  }%
}

\newcommand{\PsiLeftBlock}[1][1]{%
  \tninline[#1]{%
    \node[tensor,minimum width=3cm] (psi) {$C_{<k}$};
    \node[right=\TNHorizontalLeg of psi] (empty) {};
    \draw[leg] (empty.west)--(psi.east);

    \foreach \x/\lab in {0.4/{i_1}, 2.6/{i_{k-1}}}{
      \draw[leg] ($(psi.south west)+(\x,0)$) -- ++(0,-\TNVerticalLeg);
      \node[below] at ($(psi.south west)+(\x,-\TNVerticalLeg)$) {$\lab$};
    }
    \foreach \x/\lab in {1.5/{\cdots}}{
      \node[below] at ($(psi.south west)+(\x,-\TNVerticalLeg)$) {$\lab$};
    }
  }%
}

\newcommand{\PsiLeftBlockGeneral}[2][1]{%
  \tninline[#1]{%
    \node[tensor,minimum width=3cm] (psi) {$#2_{<k}$};
    \node[right=\TNHorizontalLeg of psi] (empty) {};
    \draw[leg] (empty.west)--(psi.east);

    \foreach \x/\lab in {0.4/{i_1}, 2.6/{i_{k-1}}}{
      \draw[leg] ($(psi.south west)+(\x,0)$) -- ++(0,-\TNVerticalLeg);
      \node[below] at ($(psi.south west)+(\x,-\TNVerticalLeg)$) {$\lab$};
    }
    \foreach \x/\lab in {1.5/{\cdots}}{
      \node[below] at ($(psi.south west)+(\x,-\TNVerticalLeg)$) {$\lab$};
    }
  }%
}

\newcommand{\PsiRightBlock}[1][1]{%
  \tninline[#1]{%
    \node[tensor,minimum width=3cm] (psi) {$C_{>k}$};
    \node[left=\TNHorizontalLeg of psi] (empty) {};
    \draw[leg] (empty.east)--(psi.west);

    \foreach \x/\lab in {0.4/{i_{k+1}}, 2.6/{i_{n}}}{
      \draw[leg] ($(psi.south west)+(\x,0)$) -- ++(0,-\TNVerticalLeg);
      \node[below] at ($(psi.south west)+(\x,-\TNVerticalLeg)$) {$\lab$};
    }
    \foreach \x/\lab in {1.5/{\cdots}}{
      \node[below] at ($(psi.south west)+(\x,-\TNVerticalLeg)$) {$\lab$};
    }
  }%
}

\newcommand{\PsiRightBlockGeneral}[3][1]{%
  \tninline[#1]{%
    \node[tensor,minimum width=3cm] (psi) {$#2_{>k}$};
    \node[left=\TNHorizontalLeg of psi] (empty) {};
    \draw[leg] (empty.east)--(psi.west);

    \foreach \x/\lab in {0.4/{i_{k+1}}, 2.6/{i_{#3}}}{
      \draw[leg] ($(psi.south west)+(\x,0)$) -- ++(0,-\TNVerticalLeg);
      \node[below] at ($(psi.south west)+(\x,-\TNVerticalLeg)$) {$\lab$};
    }
    \foreach \x/\lab in {1.5/{\cdots}}{
      \node[below] at ($(psi.south west)+(\x,-\TNVerticalLeg)$) {$\lab$};
    }
  }%
}

\newcommand{\SketchLeftBlock}[2][1]{%
  \tninline[#1]{%
    \node[tensorv2,minimum width=3cm, minimum height= 0.9cm] (psi) {$S_{<k}$};
    \node[below=\TNVerticalLeg of psi] (label1) {$\zeta$};
    \draw[leg] (label1.north)--(psi.south);

    \foreach \x/\lab in {0.4/{i_1}, 2.6/{i_{k-1}}}{
      \draw[leg] ($(psi.north west)+(\x,0)$) -- ++(0,\TNVerticalLeg);
      \node[above] at ($(psi.north west)+(\x,\TNVerticalLeg)$) {$\lab$};
    }
    \foreach \x/\lab in {1.5/{\cdots}}{
      \node[above] at ($(psi.north west)+(\x,\TNVerticalLeg)$) {$\lab$};
    }
  }%
}

\newcommand{\SketchRightBlock}[1][1]{%
  \tninline[#1]{%
    \node[tensorv2,minimum width=3cm, minimum height=0.9cm] (psi) {$S_{>k}$};
    \node[below=\TNVerticalLeg of psi] (label2) {$\mu$};
    \draw[leg] (label2.north)--(psi.south);

    \foreach \x/\lab in {0.4/{i_{k+1}}, 2.6/{i_{n}}}{
      \draw[leg] ($(psi.north west)+(\x,0)$) -- ++(0,\TNVerticalLeg);
      \node[above] at ($(psi.north west)+(\x,\TNVerticalLeg)$) {$\lab$};
    }
    \foreach \x/\lab in {1.5/{\cdots}}{
      \node[above] at ($(psi.north west)+(\x,\TNVerticalLeg)$) {$\lab$};
    }
  }%
}

\newcommand{\MPSChain}[1][1]{%
  \tninline[#1]{%
    \node[tensor] (A1) {$G_1$};
    \node[tensor,right=\TNHorizontalLeg of A1] (A2) {$G_2$};
    \node[right=\TNHorizontalLeg of A2] (dots) {\ldots};
    \node[tensor,right=\TNHorizontalLeg of dots] (AN) {$G_n$};

    \draw[leg] (A1.east)--(A2.west);
    \draw[leg] (A2.east)--(dots.west);
    \draw[leg] (dots.east)--(AN.west);

    \foreach \T/\lab in {A1/{i_1}, A2/{i_2}, AN/{i_n}}{
      \draw[leg] (\T.south)--++(0,-\TNVerticalLeg);
      \node[below] at ($(\T.south)+(0,-\TNVerticalLeg)$) {$\lab$};

    }
  }%
}

\newcommand{\MPSChainGeneral}[3][1]{%
  \tninline[#1]{%
    \node[tensor] (A1) {$#2_1$};
    \node[tensor,right=\TNHorizontalLeg of A1] (A2) {$#2_2$};
    \node[right=\TNHorizontalLeg of A2] (dots) {\ldots};
    \node[tensor,right=\TNHorizontalLeg of dots] (AN) {$#2_{#3}$};

    \draw[leg] (A1.east)--(A2.west);
    \draw[leg] (A2.east)--(dots.west);
    \draw[leg] (dots.east)--(AN.west);

    \foreach \T/\lab in {A1/{i_1}, A2/{i_2}, AN/{i_{#3}}}{
      \draw[leg] (\T.south)--++(0,-\TNVerticalLeg);
      \node[below] at ($(\T.south)+(0,-\TNVerticalLeg)$) {$\lab$};

    }
  }%
}

\newcommand{\MPSChainPerturbed}[1][1]{%
  \tninline[#1]{%
    \node[tensor] (A1) {$\hat{F}_1$};
    \node[right=\TNHorizontalLeg of A1] (dots) {\ldots};
    \node[tensor,right=\TNHorizontalLeg of dots] (Ak) {$\hat{F}_k$};
    \node[tensor,right=\TNHorizontalLeg of Ak] (Ak1) {$F_{k+1}$};

    \node[right=\TNHorizontalLeg of Ak1] (dots2) {\ldots};
    \node[tensor,right=\TNHorizontalLeg of dots2] (AN) {$F_d$};

    \draw[leg] (A1.east)--(dots.west);
    \draw[leg] (dots.east)--(Ak.west);
    \draw[leg] (Ak.east)--(Ak1.west);
    \draw[leg] (Ak1.east)--(dots2.west);
    \draw[leg] (dots2.east)--(AN.west);

    \foreach \T/\lab in {A1/{i_1}, Ak/{i_k}, Ak1/{i_{k+1}} , AN/{i_d}}{
      \draw[leg] (\T.south)--++(0,-\TNVerticalLeg);
      \node[below] at ($(\T.south)+(0,-\TNVerticalLeg)$) {$\lab$};

    }
  }%
}

\newcommand{\MPSLeftChain}[1][1]{%
  \tninline[#1]{%
    \node[tensor] (A1) {$G_1$};
    \node[right=\TNHorizontalLeg of A1] (dots) {\ldots};
    \node[tensor,right=\TNHorizontalLeg of dots] (AN) {$G_{k-1}$};
    \node[right=\TNHorizontalLeg of AN] (empty) {};

    \draw[leg] (A1.east)--(dots.west);
    \draw[leg] (dots.east)--(AN.west);
    \draw[leg] (AN.east)--(empty.west);

    \foreach \T/\lab in {A1/{i_1}, AN/{i_{k-1}}}{
      \draw[leg] (\T.south)--++(0,-\TNVerticalLeg);
      \node[below] at ($(\T.south)+(0,-\TNVerticalLeg)$) {$\lab$};
    }
  }%
}

\newcommand{\MPSLeftChainGeneral}[2][1]{%
  \tninline[#1]{%
    \node[tensor] (A1) {$#2_1$};
    \node[right=\TNHorizontalLeg of A1] (dots) {\ldots};
    \node[tensor,right=\TNHorizontalLeg of dots] (AN) {$#2_{k-1}$};
    \node[right=\TNHorizontalLeg of AN] (empty) {};

    \draw[leg] (A1.east)--(dots.west);
    \draw[leg] (dots.east)--(AN.west);
    \draw[leg] (AN.east)--(empty.west);

    \foreach \T/\lab in {A1/{i_1}, AN/{i_{k-1}}}{
      \draw[leg] (\T.south)--++(0,-\TNVerticalLeg);
      \node[below] at ($(\T.south)+(0,-\TNVerticalLeg)$) {$\lab$};
    }
  }%
}

\newcommand{\MPSRightChain}[1][1]{%
  \tninline[#1]{%
    \node[tensor] (A1) {$G_{k+1}$};
    \node[left=\TNHorizontalLeg of A1] (empty) {};
    \node[right=\TNHorizontalLeg of A1] (dots) {\ldots};
    \node[tensor,right=\TNHorizontalLeg of dots] (AN) {$G_{n}$};

    \draw[leg] (empty.east)--(A1.west);
    \draw[leg] (A1.east)--(dots.west);
    \draw[leg] (dots.east)--(AN.west);

    \foreach \T/\lab in {A1/{i_{k+1}}, AN/{i_{n}}}{
      \draw[leg] (\T.south)--++(0,-\TNVerticalLeg);
      \node[below] at ($(\T.south)+(0,-\TNVerticalLeg)$) {$\lab$};
    }
  }%
}

\newcommand{\MPSRightChainGeneral}[3][1]{%
  \tninline[#1]{%
    \node[tensor] (A1) {$#2_{k+1}$};
    \node[left=\TNHorizontalLeg of A1] (empty) {};
    \node[right=\TNHorizontalLeg of A1] (dots) {\ldots};
    \node[tensor,right=\TNHorizontalLeg of dots] (AN) {$#2_{#3}$};

    \draw[leg] (empty.east)--(A1.west);
    \draw[leg] (A1.east)--(dots.west);
    \draw[leg] (dots.east)--(AN.west);

    \foreach \T/\lab in {A1/{i_{k+1}}, AN/{i_{#3}}}{
      \draw[leg] (\T.south)--++(0,-\TNVerticalLeg);
      \node[below] at ($(\T.south)+(0,-\TNVerticalLeg)$) {$\lab$};
    }
  }%
}

\newcommand{\MPSChainGk}[1][1]{%
  \tninline[#1]{%
    \node[tensor,minimum width=3cm] (psi) {$C_{<k}$};
    \node[tensor, right=\TNHorizontalLeg of psi] (A1) {$G_k$};
    \node[tensor,minimum width=3cm, right=\TNHorizontalLeg of A1] (psi2) {$C_{>k}$};

    \draw[leg] (psi.east)--(A1.west);
    \draw[leg] (A1.east)--(psi2.west);

    \draw[leg] (A1.south)--++(0,-\TNVerticalLeg);
    \node[below] at ($(A1.south)+(0,-\TNVerticalLeg)$) {$i_k$};

    \foreach \x/\lab in {0.4/{i_{1}}, 2.6/{i_{k-1}}}{
      \draw[leg] ($(psi.south west)+(\x,0)$) -- ++(0,-\TNVerticalLeg);
      \node[below] at ($(psi.south west)+(\x,-\TNVerticalLeg)$) {$\lab$};
    }
    \foreach \x/\lab in {1.5/{\cdots}}{
      \node[below] at ($(psi.south west)+(\x,-\TNVerticalLeg)$) {$\lab$};
    }

    \foreach \x/\lab in {0.4/{i_{k+1}}, 2.6/{i_{n}}}{
      \draw[leg] ($(psi2.south west)+(\x,0)$) -- ++(0,-\TNVerticalLeg);
      \node[below] at ($(psi2.south west)+(\x,-\TNVerticalLeg)$) {$\lab$};
    }
    \foreach \x/\lab in {1.5/{\cdots}}{
      \node[below] at ($(psi2.south west)+(\x,-\TNVerticalLeg)$) {$\lab$};
    }
  }%
}

\newcommand{\MPSChainPerturbation}[1][1]{%
  \tninline[#1]{%
    \node[tensor,minimum width=3cm] (psi) {$\tilde{D}_{<k}$};
    \node[tensor, right=\TNHorizontalLeg of psi] (A1) {$\Delta F_k$};
    \node[tensor,minimum width=3cm, right=\TNHorizontalLeg of A1] (psi2) {$D_{>k}$};

    \draw[leg] (psi.east)--(A1.west);
    \draw[leg] (A1.east)--(psi2.west);

    \draw[leg] (A1.south)--++(0,-\TNVerticalLeg);
    \node[below] at ($(A1.south)+(0,-\TNVerticalLeg)$) {$i_k$};

    \foreach \x/\lab in {0.4/{i_{1}}, 2.6/{i_{k-1}}}{
      \draw[leg] ($(psi.south west)+(\x,0)$) -- ++(0,-\TNVerticalLeg);
      \node[below] at ($(psi.south west)+(\x,-\TNVerticalLeg)$) {$\lab$};
    }
    \foreach \x/\lab in {1.5/{\cdots}}{
      \node[below] at ($(psi.south west)+(\x,-\TNVerticalLeg)$) {$\lab$};
    }

    \foreach \x/\lab in {0.4/{i_{k+1}}, 2.6/{i_{d}}}{
      \draw[leg] ($(psi2.south west)+(\x,0)$) -- ++(0,-\TNVerticalLeg);
      \node[below] at ($(psi2.south west)+(\x,-\TNVerticalLeg)$) {$\lab$};
    }
    \foreach \x/\lab in {1.5/{\cdots}}{
      \node[below] at ($(psi2.south west)+(\x,-\TNVerticalLeg)$) {$\lab$};
    }
  }%
}

\newcommand{\MPSChainPerturbationLeftCase}[1][1]{%
  \tninline[#1]{%
    \node[tensor] (A1) {$\Delta F_1$};
    \node[tensor,minimum width=3cm, right=\TNHorizontalLeg of A1] (psi2) {$D_{>1}$};

    \draw[leg] (A1.east)--(psi2.west);

    \draw[leg] (A1.south)--++(0,-\TNVerticalLeg);
    \node[below] at ($(A1.south)+(0,-\TNVerticalLeg)$) {$i_1$};

    \foreach \x/\lab in {0.4/{i_{2}}, 2.6/{i_{d}}}{
      \draw[leg] ($(psi2.south west)+(\x,0)$) -- ++(0,-\TNVerticalLeg);
      \node[below] at ($(psi2.south west)+(\x,-\TNVerticalLeg)$) {$\lab$};
    }
    \foreach \x/\lab in {1.5/{\cdots}}{
      \node[below] at ($(psi2.south west)+(\x,-\TNVerticalLeg)$) {$\lab$};
    }
  }%
}

\newcommand{\MPSChainIllustration}[1][1]{%
  \tninline[#1]{%
    \node[tensor] (A1) {$G_{k}$};
    \node[left=\TNHorizontalLeg of A1] (empty) {};
    \node[right=\TNHorizontalLeg of A1] (dots) {\ldots};
    \node[tensor,right=\TNHorizontalLeg of dots] (AN) {$G_{k+L-1}$};

    \draw[leg] (empty.east)--(A1.west);
    \draw[leg] (A1.east)--(dots.west);
    \draw[leg] (dots.east)--(AN.west);

    \foreach \T/\lab in {A1/{i_{k}}, AN/{i_{k+L-1}}}{
      \draw[leg] (\T.south)--++(0,-\TNVerticalLeg);
      \node[below] at ($(\T.south)+(0,-\TNVerticalLeg)$) {$\lab$};
    }
    \draw[leg] (AN.east)--++(\TNHorizontalLeg,0);

  }%
}

\newcommand{\MPSChainIllustrationTwo}[1][1]{%
  \tninline[#1]{%
    \node[tensor,minimum width=3cm] (psi) {$F$};
    \node[right=\TNHorizontalLeg of psi] (empty) {};
    \draw[leg] (empty.west)--(psi.east);

    \foreach \x/\lab in {0.4/{i_{k}}, 2.6/{i_{k+L-1}}}{
      \draw[leg] ($(psi.south west)+(\x,0)$) -- ++(0,-\TNVerticalLeg);
      \node[below] at ($(psi.south west)+(\x,-\TNVerticalLeg)$) {$\lab$};
    }
    \foreach \x/\lab in {1.5/{\cdots}}{
      \node[below] at ($(psi.south west)+(\x,-\TNVerticalLeg)$) {$\lab$};
    }
    \draw[leg] (psi.west)--++(-\TNHorizontalLeg,0);

  }%
}

\newcommand{\SketchEquationLHS}[1][1]{%
  \tninline[#1]{%
    \node[tensor,minimum width=2.5cm] (psi) {$A_{<k}$};
    \node[tensor, right=\TNHorizontalLeg of psi] (A1) {$G_k$};
    \node[tensor,minimum width=2.5cm, right=\TNHorizontalLeg of A1] (psi2) {$A_{>k}$};
    \node[below=0.6cm of psi] (label1) {$\zeta$};
    \draw[leg] (label1.north)--(psi.south);
    \node[below=0.6cm of psi2] (label2) {$\mu$};
    \draw[leg] (label2.north)--(psi2.south);
    
    \draw[leg] (psi.east)--(A1.west);
    \draw[leg] (A1.east)--(psi2.west);

    \draw[leg] (A1.south)--++(0,-0.6cm);
    \node[below] at ($(A1.south)+(0,-0.6cm)$) {$i_k$};

  }%
}

\newcommand{\SketchBigLeftBlock}[1][1]{%
  \tninline[#1]{%
    \node[tensor,minimum width=3cm] (psi) {$C_{<k}$};
    \node[tensor, right=\TNHorizontalLeg of psi] (A1) {$G_k$};
    \node[tensor,minimum width=3cm, right=\TNHorizontalLeg of A1] (psi2) {$C_{>k}$};
    \draw[leg] (psi.east)--(A1.west);
    \draw[leg] (A1.east)--(psi2.west);
    \node[tensorv2,minimum width=3cm, minimum height= 1.1cm
    ] at ($(psi.south)+(0,-1.2cm)$) (S1) {$S_{<k}$};
    \node[tensorv2,minimum width=3cm, minimum height= 1.1cm
    ] at ($(psi2.south)+(0,-1.2cm)$) (S2) {$S_{>k}$};
    
    \foreach \lab in {S1, S2}{\foreach \x in {0.3, 0.6, 0.9, 2.7, 2.4, 2.1}{
      \draw[leg] ($(\lab.north west)+(\x,0)$) -- ++(0,0.65cm);
    }
    \node[above] at ($(\lab.north)$) {$\cdots$};
    }
    \node[below=0.6cm of S2] (label2) {$\mu$};
    \draw[leg] (label2.north)--(S2.south);
    \node[below=0.6cm of S1] (label1) {$\zeta$};
    \draw[leg] (label1.north)--(S1.south);

    \node[below=0.6cm of A1] (label3) {$i_k$};
    \draw[leg] (label3.north)--(A1.south);
    
  }%
}

\newcommand{\SketchBigRightBlock}[1][1]{%
  \tninline[#1]{%

    \node[tensor,minimum width=8cm] (psi) {$C$};
    \node[tensorv2,minimum width=3cm, minimum height= 1.1cm
    ] at ($(psi.south)+(-2.4cm,-1.2cm)$) (S1) {$S_{<k}$};
    \node[tensorv2,minimum width=3cm, minimum height= 1.1cm
    ] at ($(psi.south)+(2.4cm,-1.2cm)$) (S2) {$S_{>k}$};
    
    \foreach \lab in {S1, S2}{\foreach \x in {0.3, 0.6, 0.9, 2.7, 2.4, 2.1}{
      \draw[leg] ($(\lab.north west)+(\x,0)$) -- ++(0,0.65cm);
    }
    \node[above] at ($(\lab.north)$) {$\cdots$};
    }
    \node[below=0.6cm of S2] (label2) {$\mu$};
    \draw[leg] (label2.north)--(S2.south);
    \node[below=0.6cm of S1] (label1) {$\zeta$};
    \draw[leg] (label1.north)--(S1.south);

    \node[below=0.6cm of psi] (label3) {$i_k$};
    \draw[leg] (label3.north)--(psi.south);
    
  }%
}

\newcommand{\ZBigLeftBlock}[1][1]{%
  \tninline[#1]{%
    \node[tensor,minimum width=2.5cm] (psi) {$A_{<(k+1)}$};
    \node[tensor,minimum width=2.5cm, right=\TNHorizontalLeg of psi] (psi2) {$A_{>k}$};
    \draw[leg] (psi.east)--(psi2.west);
    \node[below=0.6cm of psi2] (label2) {$$};
    \draw[leg] (label2.north)--(psi2.south);
    \node[below=0.6cm of psi] (label1) {$$};
    \draw[leg] (label1.north)--(psi.south);

  }%
}

\newcommand{\ZBigRightBlock}[1][1]{%
  \tninline[#1]{%

    \node[tensor,minimum width=7cm] (psi) {$C$};
    \node[tensorv2,minimum width=3cm, minimum height= 1.1cm
    ] at ($(psi.south)+(-2cm,-1.2cm)$) (S1) {$S_{<(k+1)}$};
    \node[tensorv2,minimum width=3cm, minimum height= 1.1cm
    ] at ($(psi.south)+(2cm,-1.2cm)$) (S2) {$S_{>k}$};
    
    \foreach \lab in {S1, S2}{\foreach \x in {0.3, 0.6, 0.9, 2.7, 2.4, 2.1}{
      \draw[leg] ($(\lab.north west)+(\x,0)$) -- ++(0,0.65cm);
    }
    \node[above] at ($(\lab.north)$) {$\cdots$};
    }
    \node[below=0.6cm of S2] (label2) {$$};
    \draw[leg] (label2.north)--(S2.south);
    \node[below=0.6cm of S1] (label1) {$$};
    \draw[leg] (label1.north)--(S1.south);

  }%
}

\newcommand{\KetPsiBlock}[1][1]{%
  \tninline[#1]{%
    \node[tensor,minimum width=4cm] (psi) {$\Ket{\psi}$};
    \foreach \x/\lab in {0.5/{j_1}, 1.5/{j_2}, 3.5/{j_n}}{
      \draw[leg] ($(psi.south west)+(\x,0)$) -- ++(0,-\TNVerticalLeg);
      \node[below] at ($(psi.south west)+(\x,-\TNVerticalLeg)$) {$\lab$};
    }
    \foreach \x/\lab in {2.5/{\cdots}}{
      \node[below] at ($(psi.south west)+(\x,-\TNVerticalLeg)$) {$\lab$};
    }
  }%
}

\newcommand{\PsiBlockGeneral}[3][1]{%
  \tninline[#1]{%
    \node[tensor,minimum width=4cm] (psi) {#2};
    \foreach \x/\lab in {0.5/{i_1}, 1.5/{i_2}, 3.5/{i_{#3}}}{
      \draw[leg] ($(psi.south west)+(\x,0)$) -- ++(0,-\TNVerticalLeg);
      \node[below] at ($(psi.south west)+(\x,-\TNVerticalLeg)$) {$\lab$};
    }
    \foreach \x/\lab in {2.5/{\cdots}}{
      \node[below] at ($(psi.south west)+(\x,-\TNVerticalLeg)$) {$\lab$};
    }
  }%
}

\newcommand{\KetPsiMPSChain}[1][1]{%
  \tninline[#1]{%
    \node[tensor] (A1) {$F_1$};
    \node[tensor,right=\TNHorizontalLeg of A1] (A2) {$F_2$};
    \node[right=\TNHorizontalLeg of A2] (dots) {\ldots};
    \node[tensor,right=\TNHorizontalLeg of dots] (AN) {$F_n$};

    \draw[leg] (A1.east)--(A2.west);
    \draw[leg] (A2.east)--(dots.west);
    \draw[leg] (dots.east)--(AN.west);

    \foreach \T/\lab in {A1/{j_1}, A2/{j_2}, AN/{j_n}}{
      \draw[leg] (\T.south)--++(0,-\TNVerticalLeg);
      \node[below] at ($(\T.south)+(0,-\TNVerticalLeg)$) {$\lab$};
    }
  }%
}

\newcommand{\RhoBlock}[1][1]{%
  \tninline[#1]{%
    \node[tensor,minimum width=4cm] (psi) {$\rho$};
    \foreach \x/\lab in {0.5/{j_1}, 1.5/{j_2}, 3.5/{j_n}}{
      \draw[leg] ($(psi.south west)+(\x,0)$) -- ++(0,-\TNVerticalLeg);
      \node[below] at ($(psi.south west)+(\x,-\TNVerticalLeg)$) {$\lab$};
      \draw[leg] ($(psi.north west)+(\x,0)$) -- ++(0,\TNVerticalLeg);
      \node[above] at ($(psi.north west)+(\x,\TNVerticalLeg)$) {$\lab'$};
    }
    \foreach \x/\lab in {2.5/{\cdots}}{
      \node[below] at ($(psi.south west)+(\x,-\TNVerticalLeg)$) {$\lab$};
      \node[above] at ($(psi.north west)+(\x,\TNVerticalLeg)$) {$\lab$};
    }
  }%
}

\newcommand{\RhoMPSChain}[1][1]{%
  \tninline[#1]{%
    \node[tensor] (A1) {$F_1$};
    \node[tensor, above = \TNVerticalLeg of A1] (B1) {$\Bar{F_1}$};

    \node[tensor,right=\TNHorizontalLeg of A1] (A2) {$F_2$};
    \node[tensor, above = \TNVerticalLeg of A2] (B2) {$\Bar{F_2}$};

    \node[right=\TNHorizontalLeg of A2] (Adots) {\ldots};
    \node[right=\TNHorizontalLeg of B2] (Bdots) {\ldots};
    \node[tensor,right=\TNHorizontalLeg of Adots] (AN) {$F_n$};
    \node[tensor, above = \TNVerticalLeg of AN] (BN) {$\Bar{F_n}$};

    \foreach \a/\b in {A1/A2, A2/Adots, Adots/AN,
    B1/B2, B2/Bdots, Bdots/BN}{
        \draw[leg] (\a.east)--(\b.west);
    }

    \foreach \T/\lab in {A1/{j_1}, A2/{j_2}, AN/{j_n}}{
      \draw[leg] (\T.south)--++(0,-\TNVerticalLeg);
      \node[below] at ($(\T.south)+(0,-\TNVerticalLeg)$) {$\lab$};
    }
    \foreach \T/\lab in {B1/{j_1'}, B2/{j_2'}, BN/{j_n'}}{
      \draw[leg] (\T.north)--++(0,\TNVerticalLeg);
      \node[above] at ($(\T.north)+(0,\TNVerticalLeg)$) {$\lab$};
    }
  }%
}

\newcommand{\GkdefLHS}[1][1]{%
  \tninline[#1]{%

    \node[tensor] (A1) {$G_k$};
    \node[left=\TNHorizontalLeg of A1] (A0) {};
    \node[right=\TNHorizontalLeg of A1] (A2) {};
  
    \foreach \x/\lab in {-0.4/{i_1}, -0.6/{i_{k-1}}}{
      \draw[leg] ($(A1.north east)+(0,\x)$) -- ++(\TNHorizontalLeg,0);
      \draw[leg] ($(A1.north west)+(0,\x)$) -- ++(-\TNHorizontalLeg,0);
    }
      \draw[leg] (A1.north) -- ++(0,\TNVerticalLeg);
      \node[above] at ($(A1.north)+(0,\TNVerticalLeg)$) {$i$};
      \node[below] at ($(A1.south)+(0,-\TNVerticalLeg)$) {};

  }%
}

\newcommand{\GkdefRHS}[1][1]{%
  \tninline[#1]{%
  
    \node[tensor] (A1) {$F_k$};
    \node[tensor, above = \TNVerticalLeg of A1] (B1) {$\sigma_{k}^{i}$};
    \node[tensor, above = \TNVerticalLeg of B1] (C1) {$\Bar{F_k}$};
    \draw[leg] (A1.north)--(B1.south);
    \draw[leg] (B1.north)--(C1.south);

    \node[left=\TNHorizontalLeg of A1] (A0) {};
    \node[left=\TNVerticalLeg of C1] (C0) {};

    \node[right=\TNHorizontalLeg of A1] (A2) {};
    \node[right=\TNVerticalLeg of C1] (C2) {};

    \foreach \a/\b in {A0/A1, C0/C1, A1/A2, C1/C2}{
        \draw[leg] (\a.east)--(\b.west);
    }

  }%
}

\newcommand{\GOnedefLHS}[1][1]{%
  \tninline[#1]{%

    \node[tensor] (A1) {$G_1$};
    \node[right=\TNHorizontalLeg of A1] (A2) {};
  
    \foreach \x/\lab in {-0.4/{i_1}, -0.6/{i_{k-1}}}{
      \draw[leg] ($(A1.north east)+(0,\x)$) -- ++(\TNHorizontalLeg,0);
    }
      \draw[leg] (A1.north) -- ++(0,\TNVerticalLeg);
      \node[above] at ($(A1.north)+(0,\TNVerticalLeg)$) {$i$};
      \node[below] at ($(A1.south)+(0,-\TNVerticalLeg)$) {};

  }%
}

\newcommand{\GOnedefRHS}[1][1]{%
  \tninline[#1]{%
  
    \node[tensor] (A1) {$F_1$};
    \node[tensor, above = \TNVerticalLeg of A1] (B1) {$\sigma_{1}^{i}$};
    \node[tensor, above = \TNVerticalLeg of B1] (C1) {$\Bar{F_1}$};
    \draw[leg] (A1.north)--(B1.south);
    \draw[leg] (B1.north)--(C1.south);

    \node[right=\TNHorizontalLeg of A1] (A2) {};
    \node[right=\TNVerticalLeg of C1] (C2) {};

    \foreach \a/\b in {A1/A2, C1/C2}{
        \draw[leg] (\a.east)--(\b.west);
    }

  }%
}

\newcommand{\GNdefLHS}[1][1]{%
  \tninline[#1]{%

    \node[tensor] (A1) {$G_n$};
    \node[left=\TNHorizontalLeg of A1] (A0) {};
  
    \foreach \x/\lab in {-0.4/{i_1}, -0.6/{i_{k-1}}}{
      \draw[leg] ($(A1.north west)+(0,\x)$) -- ++(-\TNHorizontalLeg,0);
    }
      \draw[leg] (A1.north) -- ++(0,\TNVerticalLeg);
      \node[above] at ($(A1.north)+(0,\TNVerticalLeg)$) {$i$};
      \node[below] at ($(A1.south)+(0,-\TNVerticalLeg)$) {};

  }%
}

\newcommand{\GNdefRHS}[1][1]{%
  \tninline[#1]{%
  
    \node[tensor] (A1) {$F_n$};
    \node[tensor, above = \TNVerticalLeg of A1] (B1) {$\sigma_{n}^{i}$};
    \node[tensor, above = \TNVerticalLeg of B1] (C1) {$\Bar{F_n}$};
    \draw[leg] (A1.north)--(B1.south);
    \draw[leg] (B1.north)--(C1.south);

    \node[left=\TNHorizontalLeg of A1] (A0) {};
    \node[left=\TNVerticalLeg of C1] (C0) {};

    \foreach \a/\b in {A0/A1, C0/C1}{
        \draw[leg] (\a.east)--(\b.west);
    }

  }%
}

\newcommand{\PsiBlockCustom}[1][1]{%
  \tninline[#1]{%
    \node[tensor,minimum width=4cm] (psi) {$C$};
    \foreach \x/\lab in {0.5/{i_1}, 1.5/{i_2}, 3.5/{i_n}}{
      \draw[leg] ($(psi.south west)+(\x,0)$) -- ++(0,-\TNVerticalLeg);
      \node[below] at ($(psi.south west)+(\x,-\TNVerticalLeg)$) {$\lab$};
    }
    \foreach \x/\lab in {2.5/{\cdots}}{
      \node[below] at ($(psi.south west)+(\x,-\TNVerticalLeg)$) {$\lab$};
      \node[above] at ($(psi.north west)+(\x,2*\TNVerticalLeg)$) {};
    }
  }%
}

\newcommand{\MPSChainCustom}[1][1]{%
  \tninline[#1]{%
    \node[tensor] (A1) {$G_1$};
    \node[tensor,right=\TNHorizontalLeg of A1] (A2) {$G_2$};
    \node[right=\TNHorizontalLeg of A2] (dots) {\ldots};
    \node[tensor,right=\TNHorizontalLeg of dots] (AN) {$G_n$};

    \draw[leg] (A1.east)--(A2.west);
    \draw[leg] (A2.east)--(dots.west);
    \draw[leg] (dots.east)--(AN.west);

    \foreach \T/\lab in {A1/{i_1}, A2/{i_2}, AN/{i_n}}{
      \draw[leg] (\T.south)--++(0,-\TNVerticalLeg);
      \node[below] at ($(\T.south)+(0,-\TNVerticalLeg)$) {$\lab$};
      \node[above] at ($(\T.north)+(0,2*\TNVerticalLeg)$) {};

    }
  }%
}

\newcommand{\RhoPauliMPSChain}[1][1]{%
  \tninline[#1]{%
    \node[tensor] (A1) {$F_1$};
    \node[tensor, above = \TNVerticalLeg of A1] (B1) {$\sigma_{1}^{i_1}$};
    \node[tensor, above = \TNVerticalLeg of B1] (C1) {$\Bar{F_1}$};

    \node[tensor, right=\TNHorizontalLeg of A1] (A2) {$F_2$};
    \node[tensor, above = \TNVerticalLeg of A2] (B2) {$\sigma_{2}^{i_2}$};
    \node[tensor, above = \TNVerticalLeg of B2] (C2) {$\Bar{F_2}$};

    \node[right=\TNHorizontalLeg of A2] (Adots) {\ldots};
    \node[right=\TNHorizontalLeg of C2] (Cdots) {\ldots};

    \node[tensor, right=\TNHorizontalLeg of Adots] (AN) {$F_n$};
    \node[tensor, above = \TNVerticalLeg of AN] (BN) {$\sigma_{n}^{i_n}$};
    \node[tensor, above = \TNVerticalLeg of BN] (CN) {$\Bar{F_n}$};

    \foreach \a/\b in {A1/A2, A2/Adots, Adots/AN,
    C1/C2, C2/Cdots, Cdots/CN}{
        \draw[leg] (\a.east)--(\b.west);
    }

    \foreach \T in {A1,A2,AN}{
      \draw[leg] (\T.north)--++(0,\TNVerticalLeg);
    }
    \foreach \T in {C1,C2,CN}{
      \draw[leg] (\T.south)--++(0,-\TNVerticalLeg);
    }
  }%
}

\newcommand{\KetPhiBlock}[1][1]{%
  \tninline[#1]{%
    \node[tensor,minimum width=4cm] (psi) {$\Ket{\phi}$};
    \foreach \x/\lab in {0.5/{j_1}, 1.5/{j_2}, 3.5/{j_n}}{
      \draw[leg] ($(psi.south west)+(\x,0)$) -- ++(0,-\TNVerticalLeg);
      \node[below] at ($(psi.south west)+(\x,-\TNVerticalLeg)$) {$\lab$};
    }
    \foreach \x/\lab in {2.5/{\cdots}}{
      \node[below] at ($(psi.south west)+(\x,-\TNVerticalLeg)$) {$\lab$};
    }
  }%
}

\newcommand{\KetPhiMPSChain}[1][1]{%
  \tninline[#1]{%
    \node[tensor] (A1) {$F_1$};
    \node[tensor,right=\TNHorizontalLeg of A1] (A2) {$F_2$};
    \node[right=\TNHorizontalLeg of A2] (dots) {\ldots};
    \node[tensor,right=\TNHorizontalLeg of dots] (AN) {$F_n$};

    \draw[leg] (A1.east)--(A2.west);
    \draw[leg] (A2.east)--(dots.west);
    \draw[leg] (dots.east)--(AN.west);

    \foreach \T/\lab in {A1/{j_1}, A2/{j_2}, AN/{j_n}}{
      \draw[leg] (\T.south)--++(0,-\TNVerticalLeg);
      \node[below] at ($(\T.south)+(0,-\TNVerticalLeg)$) {$\lab$};
    }
  }%
}

\newcommand{\KetPhiZDef}[1][1]{%
  \tninline[#1]{%
    \node[tensor] (A1) {$F_1$};
    \node[tensor, above = \TNVerticalLeg of A1] (B1) {$\Bar{F_1}$};

    \node[tensor,right=\TNHorizontalLeg of A1] (A2) {$F_2$};
    \node[tensor, above = \TNVerticalLeg of A2] (B2) {$\Bar{F_2}$};

    \node[right=\TNHorizontalLeg of A2] (Adots) {\ldots};
    \node[right=\TNHorizontalLeg of B2] (Bdots) {\ldots};
    \node[tensor,right=\TNHorizontalLeg of Adots] (AN) {$F_n$};
    \node[tensor, above = \TNVerticalLeg of AN] (BN) {$\Bar{F_n}$};

    \foreach \a/\b in {A1/A2, A2/Adots, Adots/AN,
    B1/B2, B2/Bdots, Bdots/BN}{
        \draw[leg] (\a.east)--(\b.west);
    }

    \foreach \T/\lab in {A1/B1, A2/B2, AN/BN}{
      \draw[leg] (\T.north)--++(0,\TNVerticalLeg);
    }
  }%
}

\newcommand{\KetPhiEvalDef}[1][1]{%
  \tninline[#1]{%
    \node[tensor] (A1) {$F_1$};
    \node[tensor, above = \TNVerticalLeg of A1] (B1) {$U_1^{(j)}$};

    \node[tensor,right=\TNHorizontalLeg of A1] (A2) {$F_2$};
    \node[tensor, above = \TNVerticalLeg of A2] (B2) {$U_2^{(j)}$};

    \node[right=\TNHorizontalLeg of A2] (Adots) {\ldots};
    \node[right=\TNHorizontalLeg of B2] (Bdots) {\ldots};
    \node[tensor,right=\TNHorizontalLeg of Adots] (AN) {$F_n$};
    \node[tensor, above = \TNVerticalLeg of AN] (BN) {$U_n^{(j)}$};

    \foreach \a/\b in {A1/A2, A2/Adots, Adots/AN,
    B1/B2, B2/Bdots, Bdots/BN}{
        \draw[leg] (\a.east)--(\b.west);
    }

    \foreach \T/\lab in {A1/B1, A2/B2, AN/BN}{
      \draw[leg] (\T.north)--++(0,\TNVerticalLeg);
    }
    \foreach \T/\lab in {B1/{b_1^{(j)}}, B2/{b_2^{(j)}}, BN/{b_n^{(j)}}}{
      \draw[leg] (\T.north)--++(0,\TNVerticalLeg);
      \node[above] at ($(\T.north)+(0,\TNVerticalLeg)$) {$\lab$};
    }
  }%
}

\newcommand{\singleton}[1]{\left\{ #1 \right\}}
\newcommand{\NLL}{\mathcal{L}}

\newcommand{\tr}[1]{\mathrm{tr}\left[#1\right]}

\theoremstyle{thmstyleone}%
\newtheorem{theorem}{Theorem}%
\newtheorem{proposition}[theorem]{Proposition}%

\theoremstyle{thmstyletwo}%
\newtheorem{remark}{Remark}%

\theoremstyle{thmstylethree}%
\newtheorem{definition}{Definition}%
\newtheorem{lemma}{Lemma}

\newtheorem{assumption}{Assumption}

\begin{document}

\title[Sketch Tomography: Hybridizing Classical Shadow and Matrix Product State]{Sketch Tomography: Hybridizing Classical Shadow and Matrix Product State}

\author*[1]{\fnm{Xun} \sur{Tang}}\email{xuntang@stanford.edu}

\author[2]{\fnm{Haoxuan} \sur{Chen}}\email{haoxuanc@stanford.edu}

\author[3,4]{\fnm{Yuehaw} \sur{Khoo}}\email{ykhoo@uchicago.edu}

\author[1,2]{\fnm{Lexing} \sur{Ying}}\email{lexing@stanford.edu}

\affil[1]{\orgdiv{Department of Mathematics}, \orgname{Stanford University}, \orgaddress{\city{Stanford}, \state{CA}, \postcode{94305}, \country{USA}}}

\affil[2]{\orgdiv{Institute for Computational and Mathematical Engineering (ICME)}, \orgname{Stanford University}, \orgaddress{\city{Stanford}, \state{CA}, \postcode{94305}, \country{USA}}}

\affil[3]{\orgdiv{Department of Statistics}, \orgname{University of Chicago}, \orgaddress{\city{Chicago}, \state{IL}, \postcode{60637}, \country{USA}}}

\affil[4]{\orgdiv{Committee on Computational and Applied Mathematics (CCAM)}, \orgname{University of Chicago}, \orgaddress{\city{Chicago}, \state{IL}, \postcode{60637}, \country{USA}}}

\abstract{
We introduce Sketch Tomography, an efficient procedure for quantum state tomography based on the classical shadow protocol used for quantum observable estimations. The procedure applies to the case where the ground truth quantum state is a matrix product state (MPS). The density matrix of the ground truth state admits a tensor train ansatz as a result of the MPS assumption, and we estimate the tensor components of the ansatz through a series of observable estimations, thus outputting an approximation of the density matrix. The procedure is provably convergent with a sample complexity that scales quadratically in the system size. We conduct extensive numerical experiments to show that the procedure outputs an accurate approximation to the quantum state. For observable estimation tasks involving moderately large subsystems, we show that our procedure gives rise to a more accurate estimation than the classical shadow protocol. We also show that sketch tomography is more accurate in observable estimation than quantum states trained from the maximum likelihood estimation formulation.
}

\keywords{Quantum state tomography, Classical shadow, Low-rank methods}

\maketitle

\section{Introduction}\label{sec1}

Quantum state tomography (QST) is a crucial part of quantum computing for verifying the output of quantum algorithms and quantum devices \cite{vogel1989determination, leonhardt1995quantum, cramer2010efficient, torlai2018neural}. This work considers a broad class of QST tasks where the ground truth quantum state \(\ket{\psi}\) admits a matrix product state (MPS) representation \cite{white1992density}. The MPS ansatz is a special 1D tensor network characterized by limited entanglements between sites. For example, the MPS ansatz can represent qubits entangled by shallow local quantum circuits \cite{glasser2019expressive}.

We introduce Sketch Tomography, a sketching-based tomography procedure that recovers the density matrix \(\rho\) of \(\ket{\psi}\) through classical shadow. The MPS assumption on \(\ket{\psi}\) implies \(\rho\) is representable by a tensor train (TT). The classical shadow protocol \cite{huang2020predicting} provides a density matrix approximation \(\hat{\rho} \approx \rho\). Our sketching procedure uses \(\hat{\rho}\) to output a TT approximation \(\tilde{\rho} \approx \rho\). For global observables, using \(\tilde{\rho}\) for observable estimation is more accurate than using \(\hat{\rho}\).

Essentially, sketch tomography is a procedure that conducts observable estimation to approximate each tensor component in the TT representation of \(\rho\). Those tensor components can be obtained by solving linear equations formed from observable estimations on \(\rho\). Thus, we use \(\hat{\rho}\) from classical shadow to formulate approximate linear equations, and solving these linear equations leads to a TT approximation \(\tilde{\rho} \approx \rho\). Our procedure only involves postprocessing of \(\hat{\rho}\) and can be carried out classically.

After the procedure is carried out, the output \(\tilde{\rho}\) can be used for observable estimation. For global observables, we show empirically that using our tensor train approximation for observable estimation leads to a higher accuracy when compared with the estimate obtained by classical shadow under random Pauli measurements. The underlying cause of the improved accuracy is that \(\tilde{\rho}\) provably approximates \(\rho\) in the Frobenius norm. As a result, \(\tilde{\rho}\) can accurately perform observable estimation for generic observables \(O\), whereas the classical shadow estimation has a large variance for global observables. In addition, the output \(\tilde{\rho}\) also allows for other prediction tasks such as entanglement entropy prediction.

Several prior works on QST have considered the MPS ansatz. We include a detailed discussion on related work in Appendix~\ref{sec: related work}. Our work bears the most resemblance to \cite{landon2010efficient,cramer2010efficient}, where a direct tomography procedure measures the local density matrices to reconstruct the tensor network structure of the full density matrix. Our work essentially improves on this approach by obtaining the tensor components from general sketches that are not necessarily the local density matrix. As a result, our proposal allows direct tomography procedures to handle systems with non-local interactions. Moreover, as we form the sketches from classical shadows, our proposal does not require direct access to copies of \(\rho\).

\section{Procedure}\label{sec: procedure}
Throughout this text, we focus on the setting of \(n\)-qubit systems.
We assume that \(\ket{\psi} \in \sC^{2^n}\) is a fixed but unknown target quantum state, and \(\rho  \in  \sC^{2^n \times 2^n}\) is its corresponding density matrix. The goal is to approximate \(\rho\) accurately. We go through the main idea of the procedure for sketch tomography, and equations are primarily illustrated with tensor diagrams. The derivation and implementation details can be found in Appendix~\ref{sec: sketching}. For an integer \(n > 0\), we write \([n]:= \{1, \ldots, n\}\). For \(k < n\), we write \([n] - [k] = \{k+1, \ldots, n\}\), where the \(-\) symbol denotes set difference.

The procedure requires access to the output of the classical shadow protocol \cite{huang2020predicting}. Through repeated random Pauli measurements, the protocol outputs a collection of \(W\) approximations \(\hat{\rho}_{1}, \ldots, \hat{\rho}_{W}\) and uses a median-of-means estimator \cite{nemirovskij1983problem} for observable estimation. For simplicity, we illustrate the procedure with \(W = 1\) so that there is only one approximation \(\hat{\rho} \approx \rho\).\footnote{Given \(\hat{\rho}_{1}, \ldots, \hat{\rho}_{W}\), one clear alternative is to simply use \(\hat{\rho} = \frac{1}{W}\sum_{i = 1}^{W}\hat{\rho}_i\), and using \(\hat{\rho}\) for observable estimation is called the direct mean approach. Our numerical tests find that the direct mean approach has similar performance compared with the median of means, and a similar conclusion was also reported in \cite{struchalin2021experimental}.}

We go through our QST procedure for obtaining \(\rho\).
For \(k \in [n]\), we let \((\sigma^{X}_k, \sigma^{Y}_k, \sigma^{Z}_k)\) denote the Pauli matrices on site \(k\), and we write \((\sigma^{1}_{k}, \sigma^{2}_{k}, \sigma^{3}_{k}, \sigma^{4}_{k}) = (\frac{1}{\sqrt{2}}I_{2}, \frac{1}{\sqrt{2}}\sigma^{X}_{k}, \frac{1}{\sqrt{2}}\sigma^{Y}_{k}, \frac{1}{\sqrt{2}}\sigma^{Z}_{k})\). A density matrix \(\rho\) can be uniquely represented by a tensor \(C \colon [4]^n \to \R\) as follows:
\begin{equation}\label{eqn: def of density matrix}
    \rho = \sum_{i_1, \ldots, i_n =1}^{4}C(i_1, \ldots, i_n)\prod_{l = 1}^{n}\sigma_{l}^{i_l}.
\end{equation}

In particular, the MPS assumption of \(\ket{\psi}\) implies that \(C\) is a tensor train (TT). One represents \(C\) in a tensor diagram as follows:
\begin{equation}\label{eqn: Equation for C in TN diagram}
  \PsiBlock[0.5] \;=\; \MPSChain[0.5]\, .
\end{equation}
Intuitively, our procedure utilizes the TT structure of \(\rho\) and uses the noisy copy \(\hat{\rho}\) from classical shadow to recover \(\rho\). As shown in \Cref{eqn: Equation for C in TN diagram}, \(C\) is defined in terms of tensor components \((G_{k})_{k = 1}^{n}\). 

We illustrate how we obtain each \(G_k\) for a fixed site \(k\). First, we rewrite \Cref{eqn: Equation for C in TN diagram} as a linear equation for \(G_k\). We define two tensors \(C_{<k}, C_{>k}\) by the following diagram
\begin{equation}\label{eqn: def of C_>k}
    \PsiLeftBlock[0.5] \;=\; \MPSLeftChain[0.5] , \quad \PsiRightBlock[0.5] \;=\; \MPSRightChain[0.5]\, ,
\end{equation}
and the equation for \(G_k\) is written as follows:
\begin{equation}\label{eqn: Linear equation for G_k}
  \MPSChainGk[0.5] \;=\; \PsiBlock[0.5]\, .
\end{equation}

The linear system in \Cref{eqn: Linear equation for G_k} is over-determined and is impractical to solve when \(n\) is large. To obtain a practical linear system, we employ sketch tensors \(S_{<k} = \SketchLeftBlock[0.5]\,, \, S_{>k} = \SketchRightBlock[0.5].\) We obtain the desired equation for \(G_k\) by contracting \Cref{eqn: Linear equation for G_k} with \(S_{<k}\) and \(S_{>k}\):
\begin{equation}\label{eqn: sketched equation for G_k v1}
    \SketchBigLeftBlock[0.5]=\SketchBigRightBlock[0.5]\, .
\end{equation}

Simplifying \Cref{eqn: sketched equation for G_k v1}, we get
\begin{equation}\label{eqn: sketched equation for G_k}
  \SketchEquationLHS[0.5] \;=\; \BkBlock[0.5]\,,
\end{equation}
where \(A_{>k}, A_{<k}\) are respectively the contraction of \(C_{>k}, C_{<k}\) by \(S_{>k}, S_{<k}\), and \(B_{k}\) is the contraction of \(C\) by \(S_{>k} \otimes S_{<k}\). 

To obtain \(G_{k}\), we use \(\hat{\rho}\) to approximate all terms in \Cref{eqn: sketched equation for G_k}.
Obtaining \(B_k\) is straightforward. We note from \Cref{eqn: sketched equation for G_k v1} that \(B_{k}\) is a linear measurement of \(C\). By \Cref{eqn: def of density matrix}, one can see that each entry of \(B_k\) is an observable of \(\rho\). Therefore, using \(\hat{\rho}\) allows one to approximately obtain \(B_{k}\). 

To obtain \(A_{>k}, A_{<k}\), we suppose that the sketch tensor \(S_{>j}, S_{<j}\) has been defined for all \(j\). The tensors \(C_{>j}, C_{<j}\) are defined according to \Cref{eqn: def of C_>k}, and \(A_{>j}, A_{<j}\) are defined accordingly. We let \(Z_{k}\) be the contraction of \(C\) by \(S_{<(k+1)} \otimes S_{>k}\). To get \(A_{>k}\), we use the following equation for \(Z_k\):
\begin{equation}\label{eqn: def of Z_k}
    \ZBigLeftBlock[0.5]= Z_k =\ZBigRightBlock[0.5] \,,
\end{equation}
where the first equality is by the definition of \(A_{<(k+1)}, A_{>k}\). The only requirement for determining \((A_{<(k+1)}, A_{>k})\) is that the first equality in \Cref{eqn: def of Z_k} needs to hold. Therefore, to obtain \(A_{>k}\), we use \(\hat{\rho}\) to approximately obtain \(Z_k\). Then, we perform a truncated SVD on \(Z_{k}\) and we set \((A_{<(k+1)}, A_{>k})\) to be the best low-rank approximation of \(Z_k\). One can similarly obtain \(A_{<k}\) by approximating \(Z_{k-1}\) and performing a truncated SVD.

Therefore, approximately obtaining \(A_{<k}, A_{>k}, B_{k}\) for all \(k\) allows one to solve for all \(G_k\). Let \((\hat{G}_{k})_{k=1}^{n}\) be the approximated tensor component obtained from the procedure. The output \(\tilde{\rho}\) is defined as follows:
\begin{equation}\label{eqn: def of approximated density matrix}
    \tilde{\rho} = \sum_{i_{[n]} \in [4]^n}\sum_{\alpha_{1}, \ldots, \alpha_{n-1}}\hat{G}_{1}(i_{1}, \alpha_{1})\hat{G}_{2}(\alpha_{1}, i_{2}, \alpha_{2})\cdots \hat{G}_{n}(\alpha_{n-1}, i_{n})\prod_{l = 1}^{n}\sigma_{l}^{i_l}.
\end{equation}

\subsection{Performance guarantee}
The sketch tomography output \(\tilde{\rho}\) converges to \(\rho\) in the Frobenius norm. We present an informal version below, but we note that the formal version is a non-asymptotic convergence bound: 
\begin{proposition}\label{prop: informal global sample complexity}
(Informal version of the upper bound)
Let \(\rho\) be the density matrix of a matrix product state for \(n\) qubits with \(n\) sufficiently large. Let \(\tilde{\rho}\) denote the output of the sketch tomography procedure, and suppose the observable estimations are done using the classical shadow protocol from \(B\) random Pauli measurements. Let \(\varepsilon, \delta \in (0, 1)\) be accuracy parameters. A sample size of \(B \ge \gO(n^2\log(n/\delta)\varepsilon^{-2})\) ensures that \(\lVert \rho - \tilde{\rho} \rVert_{F} < \varepsilon\) with probability \(1 - \delta\).
\end{proposition}

In particular, the guarantee in \Cref{prop: informal global sample complexity} means that \(\tilde{\rho}\) obtained from sketch tomography enjoys an accurate approximation on global observables regardless of whether the observable \(O\) is a local observable. We further corroborate the accuracy of \(\tilde{\rho}\) in observable estimation tasks by the numerical experiments in \Cref{sec: numerics}. For the formal version of \Cref{prop: informal global sample complexity}, and the proof, we refer the readers to Appendix~\ref{sec: upper}.

\subsection{Information-theoretical lower bound}
\label{subsec: info lower bound}
We further provide a lower bound on the required number of measurements in the proposition below, which serves as a complement to the performance guarantee presented in \Cref{prop: informal global sample complexity} above.

\begin{proposition}[Informal version of the lower bound]
\label{prop:lower bound}
Consider the setting of estimating a quantum state by using the classical shadow protocol based on random Pauli measurements. For a given error tolerance $\epsilon$, there exists some matrix product state with density matrix $\rho$, such that it requires at least (order) $\frac{n}{\epsilon^2}$ measurements to obtain an estimator $\tilde{\rho}$ of $\rho$ satisfying $\left\|\tilde{\rho} - \rho\right\|_{F} \leq \epsilon$. 
\end{proposition}

We note that the lower bound presented above, which is mainly derived from an information-theoretic perspective, depends primarily on the expressive richness of the tensor-train class and the fundamental limitations on the efficiency of the classical shadow protocol. For a formal version of \Cref{prop:lower bound} and a complete proof, we refer the readers to Appendix~\ref{sec: lower proof} below. 

For the sample complexity, we remark that \Cref{prop:lower bound} notably has only a linear dependence on \(n\), the number of qubits, whereas \Cref{prop: informal global sample complexity} has a quadratic dependence on \(n\). We conjecture that using \(\tilde{\rho}\) as a warm start for tensor network training leads to a better dependence on \(n\).

\section{Numerical Experiments}\label{sec: numerics}
We run three challenging MPS tomography experiments to demonstrate the accuracy of sketch tomography in practice. For the first two examples, we include additional benchmarks by training an MPS model with the maximum likelihood estimation (MLE) framework on Pauli measurement data, and we also include the trained MLE model as a benchmark. We show that sketch tomography is as accurate as classical shadow in local observable predictions. For global observables, we show that sketch tomography produces a more accurate prediction than classical shadow and the trained MLE model.

\subsection{1D Heisenberg}\label{sec: 1D Heisenberg}

\begin{figure}
    \centering
    \includegraphics[width=0.8\linewidth]{Figures/Heisenberg_2pt_correlation_plot_extra.pdf}
    \caption{Predictions of two-point functions \(\langle \vec{\sigma}_{1} \cdot \vec{\sigma}_{i} \rangle\) for the ground state of the 1D Heisenberg model with \(n = 20\) lattice sites. One can see that sketch tomography reaches a similar accuracy to classical shadow. The prediction from the MLE model is accurate except at the boundary site \(i = 20\).}
    \label{fig:Heisenberg_2pt_correlation_plot}
    \centering
    \includegraphics[width=0.8\linewidth]{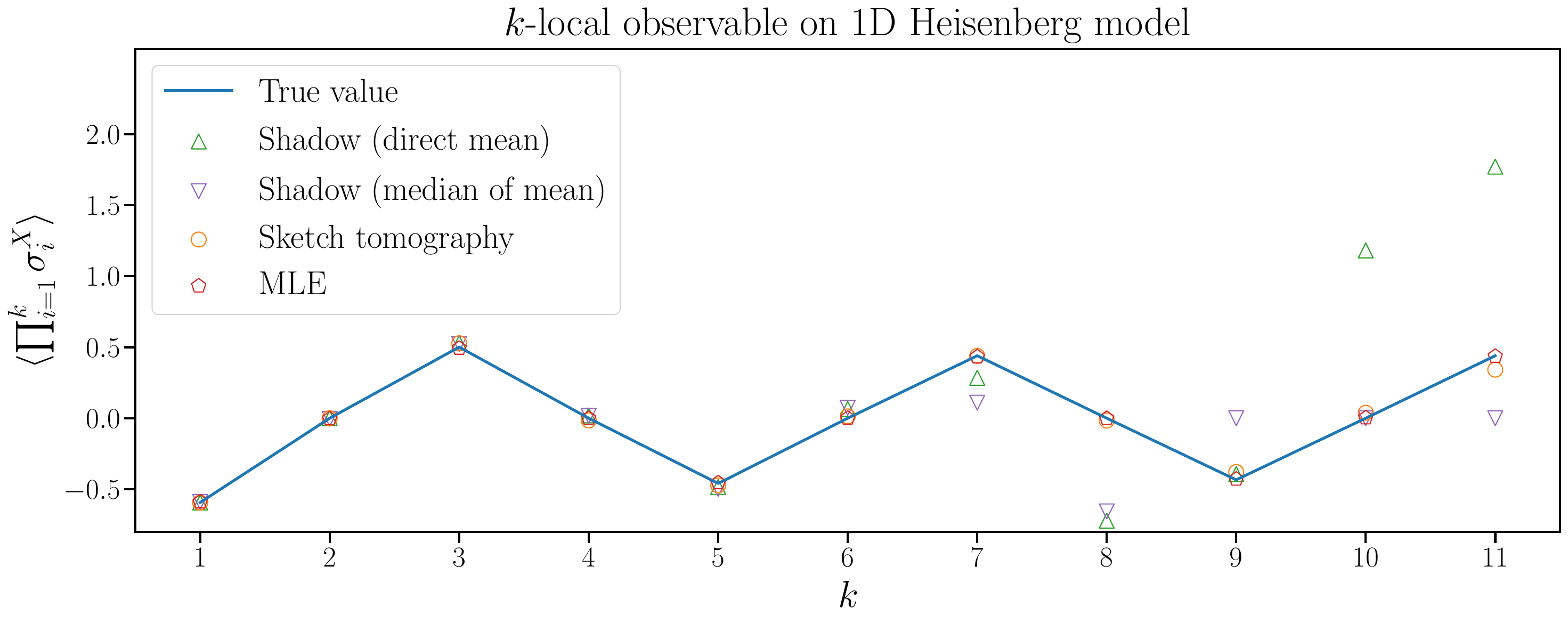}
    \caption{Predictions of observable \(\langle \prod_{i = 1}^{k}\sigma_{i}^{X} \rangle\) for the ground state of the 1D Heisenberg model. The median of means version of the classical shadow splits \(\hat{\rho}\) into \(W = 10\) classical shadow estimators \(\hat{\rho}_{1}, \ldots, \hat{\rho}_{W}\) and reports the median estimation. One can see that using \(\tilde{\rho}\) from QST for observable estimation is more accurate than using \(\hat{\rho}\) from classical shadow.}
    \label{fig:Heisenberg_k_local_plot_plot}
    \centering
    \includegraphics[width=\linewidth]{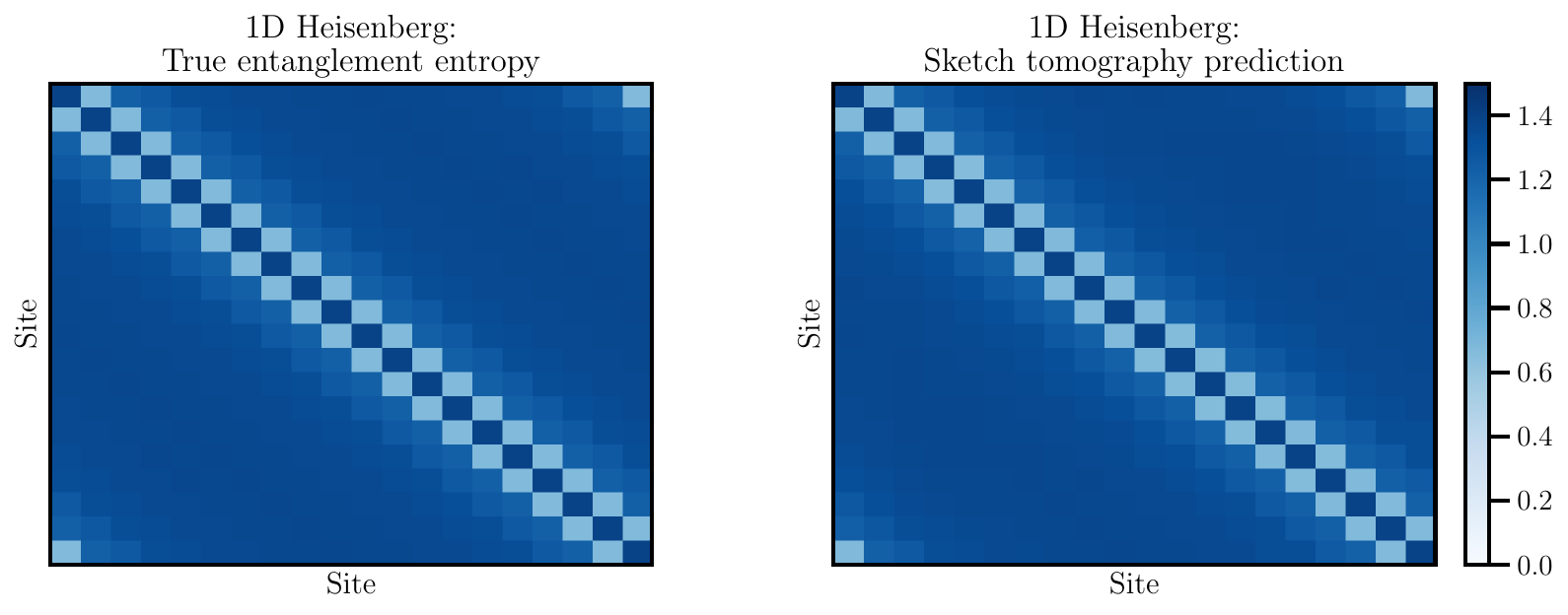}
    \caption{Prediction of second-order Renyi entanglement entropy for all subsystems of size at most two in the 1D Heisenberg model with \(n = 40\) sites. The predicted values using our QST output visually match the true values.}
    \label{fig:Heisenberg_renyi_plot}
\end{figure}

We focus on a 1D antiferromagnetic Heisenberg model with a periodic boundary condition. The Hamiltonian is \(H = \sum_{i-i' \equiv 1\, \mathrm{mod}\, n} \left(\sigma_{i}^{X} \sigma_{i'}^{X}+\sigma_{i}^{Y} \sigma_{i'}^{Y}+\sigma_{i}^{Z} \sigma_{i'}^{Z}\right)\). We consider a system with \(n = 20\) sites. We use the DMRG implementation in the ITensor package \cite{fishman2022itensor} to obtain the ground state \(\ket{\psi}\) as an MPS of maximal internal bond dimension \(a = 40\). We use the obtained \(\ket{\psi}\) as the target state. 

We use the tensor train representation of \(\ket{\psi}\) to simulate \(B = 3 \times 10^5\) random Pauli measurements, and we use the classical shadow protocol to obtain \(\hat{\rho}\) for observable estimation. We use sketch tomography to obtain \(\tilde{\rho}\). We use the MLE framework to train an MPS model \(\ket{\phi}\) on the obtained Pauli measurement data. The training is successful in the sense that \(\ket{\phi}\) has a higher likelihood than \(\ket{\psi}\) in generating the Pauli measurement data. The detailed methodology for training \(\ket{\phi}\) is in \Cref{sec: MLE}.

We test the performance of the considered methods in predicting the two-point correlation function \( \langle \vec{\sigma}_{1} \cdot \vec{\sigma}_{i} \rangle := \frac{1}{3}\left(\langle {\sigma}_{1}^{X}{\sigma}_{i}^{X}\rangle + \langle\sigma_{1}^{Y}{\sigma}_{i}^{Y}\rangle + \langle\sigma_{1}^{Z}{\sigma}_{i}^{Z} \rangle\right)\) for \(i = 2, \ldots, n\). The result is in \Cref{fig:Heisenberg_2pt_correlation_plot}. One can see that \(\tilde{\rho}\) is as successful as \(\hat{\rho}\) at calculating the two-point correlation function of \(\ket{\psi}\).

For global observables, we consider the task of estimating the observable \(O_{k} = \prod_{i = 1}^{k}\sigma_{i}^{X}\). One can see that \(O_k\) is \(k\)-local, and calculating \(\langle O_k \rangle\) is known to be challenging for \(\hat{\rho}\) when \(k\) is large. We plot the result in \Cref{fig:Heisenberg_k_local_plot_plot}, and we see that both \(\tilde{\rho}\) and \(\ket{\phi}\) maintain a good accuracy for large \(k\). Moreover, one can see that using the median-of-means estimator has limited benefit to accuracy when \(k\) is large.

Lastly, we test the performance of sketch tomography in estimating the entanglement entropy \(- \log(\tr{\rho_{A}^2})\), where \(A\) is a subsystem and \(\rho_{A}\) is the partial trace of \(\rho\) over all sites not in \(A\). We use \(- \log(\tr{\tilde{\rho}_{A}^2})\) as the sketch tomography prediction, which can be efficiently calculated with tensor diagrams. We test the performance of all possible subsystems \(A\) of size at most two, and the result is shown in \Cref{fig:Heisenberg_renyi_plot}. The maximal prediction error of \(\tilde{\rho}\) is \(0.010\) and the mean prediction error is \(0.002\), which shows that sketch tomography can accurately predict the entanglement entropy. We also use the MLE output \(\ket{\phi}\) for entanglement entropy prediction. We report that the prediction by \(\ket{\phi}\) has a larger maximal prediction error of \(0.065\) and a mean prediction error of \(0.004\).

\subsection{1D TFIM}\label{sec: 1D TFIM}

Our second example focuses on a 1D ferromagnetic transverse field Ising model (TFIM). The Hamiltonian is \(H = -J \sum_{i=1}^{n-1} \sigma_{i}^{Z} \sigma_{i+1}^{Z} - J \sigma_{1}^{Z} \sigma_{n}^{Z} -  h \sum_{i=1}^{n} \sigma_{i}^{X}\), where we set \(J = h = 1\). We consider a system with \(n = 40\) sites. We use DMRG to obtain the ground state \(\ket{\psi}\) as an MPS of maximal internal bond dimension \(a = 20\).

We simulate \(B = 3 \times 10^5\) random Pauli measurements on \(\ket{\psi}\) and we use the classical shadow protocol to form \(\hat{\rho}\) for observable estimation. We apply our sketching-based procedure and obtain \(\tilde{\rho}\).
We train an MLE model \(\ket{\phi}\) on the generated Pauli measurement data. The training detail for \(\ket{\phi}\) is in Appendix~\ref{sec: MLE}. Importantly, the tensor components of \(\ket{\phi}\) are of the same size as \(\ket{\psi}\), and the training is successful because \(\ket{\phi}\) is better than \(\ket{\psi}\) in the likelihood metric in MLE.

\begin{figure}
    \centering
    \includegraphics[width=0.8\linewidth]{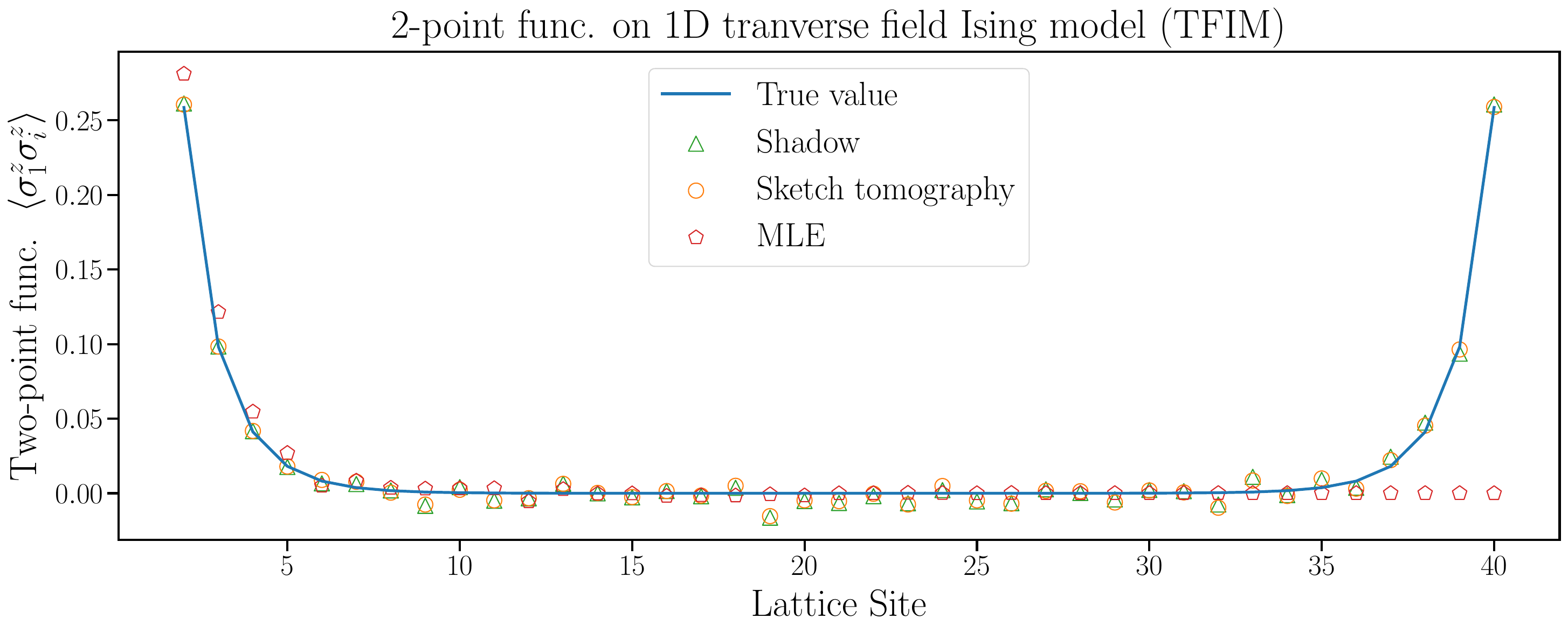}
    \caption{Predictions of two-point functions \(\langle \sigma_{1}^{Z} \sigma_{i}^{Z} \rangle\) for the ground state of the 1D TFIM model with \(n = 40\) lattice sites. One can see that sketch tomography reaches the accuracy of classical shadow.}
    \label{fig:TFIM_2pt_correlation_plot}
    \centering
    \includegraphics[width=0.8\linewidth]{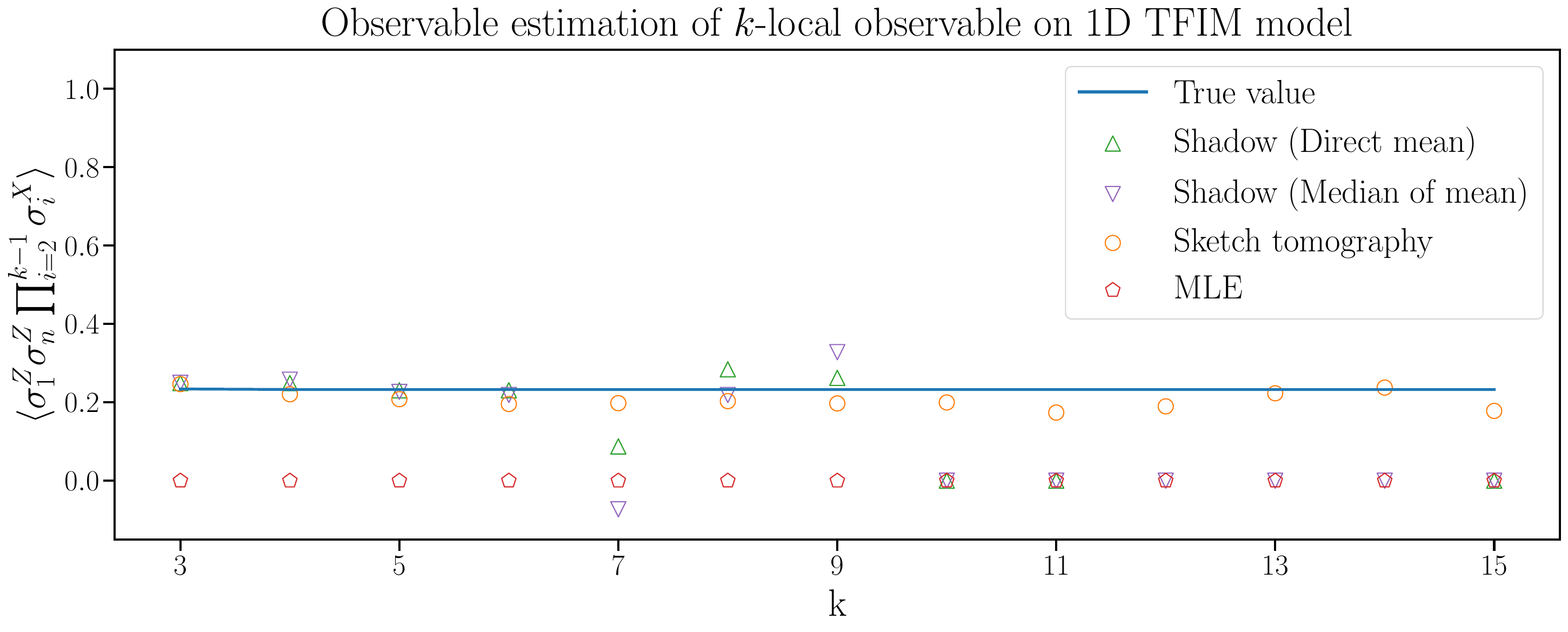}
    \caption{Predictions of \(\langle O_k \rangle = \langle \sigma_{1}^{Z}\sigma_{n}^{Z}\prod_{i = 2}^{k-1}\sigma_{i}^{X} \rangle\) for the ground state of the 1D TFIM model. Details on the median-of-mean of classical shadow are in \Cref{fig:Heisenberg_k_local_plot_plot}. One can see that the prediction from sketch tomography is more accurate than classical shadow and the model trained from the MLE framework.}
    \label{fig:TFIM_k_local_plot_plot}
\end{figure}

We test the performance of the methods in predicting the two-point correlation function \(\langle \sigma_{1}^{Z}\sigma_{j}^{Z}\rangle\) for \(j = 2, \ldots, n\). The result is in \Cref{fig:TFIM_2pt_correlation_plot}. One sees that \(\tilde{\rho}\) is successful at calculating the two-point correlation function. However, the prediction by the MLE model \(\ket{\phi}\) is not accurate. For example, \(\ket{\phi}\) does not capture the strong correlation between the boundary sites \((1, n)\).

We consider the observable estimation task over \(O_{k} = \sigma_{1}^{Z}\sigma_{n}^{Z}\prod_{i = 2}^{k-1}\sigma_{i}^{X}\) for \(k \geq 3\). As in the Heisenberg case, \(O_k\) is a \(k\)-local observable. The result is in \Cref{fig:TFIM_k_local_plot_plot}. One can see that the prediction from sketch tomography is more accurate than both the classical shadow protocol and the MLE model.

Lastly, we repeat the Renyi entanglement entropy calculation considered in the 1D Heisenberg example. For \(\tilde{\rho}\), we report a maximum prediction error of \(0.010\) and a mean prediction error of \(0.003\), which shows that sketch tomography is successful in calculating the entanglement entropy in this case. The MLE output \(\ket{\phi}\) has a larger maximal prediction error of \(0.031\), and its mean prediction error is \(0.011\).

\subsection{2D Heisenberg}\label{sec: 2D Heisenberg}

\begin{figure}
    \centering
    \includegraphics[width=\linewidth]{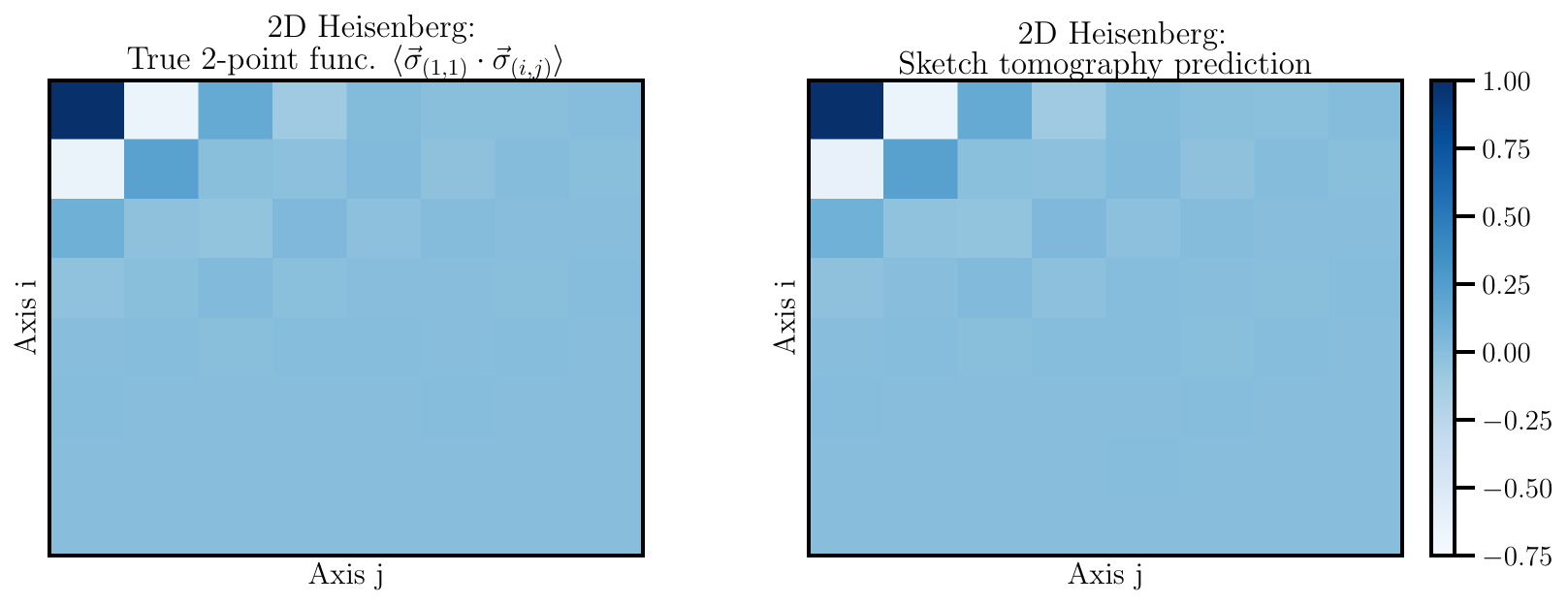}
    \caption{Predictions of two-point functions \(\langle \vec{\sigma_{(1,1)}} \cdot \vec{\sigma_{(i,j)}}\rangle\) for the ground state of the 2D Heisenberg model with \(n = 64\) lattice sites. One can see that our QST procedure successfully approximates the true 2-point correlation function.}
\label{fig:2D_heisenberg_2_pt_correlation_64}
    \centering
    \includegraphics[width=\linewidth]{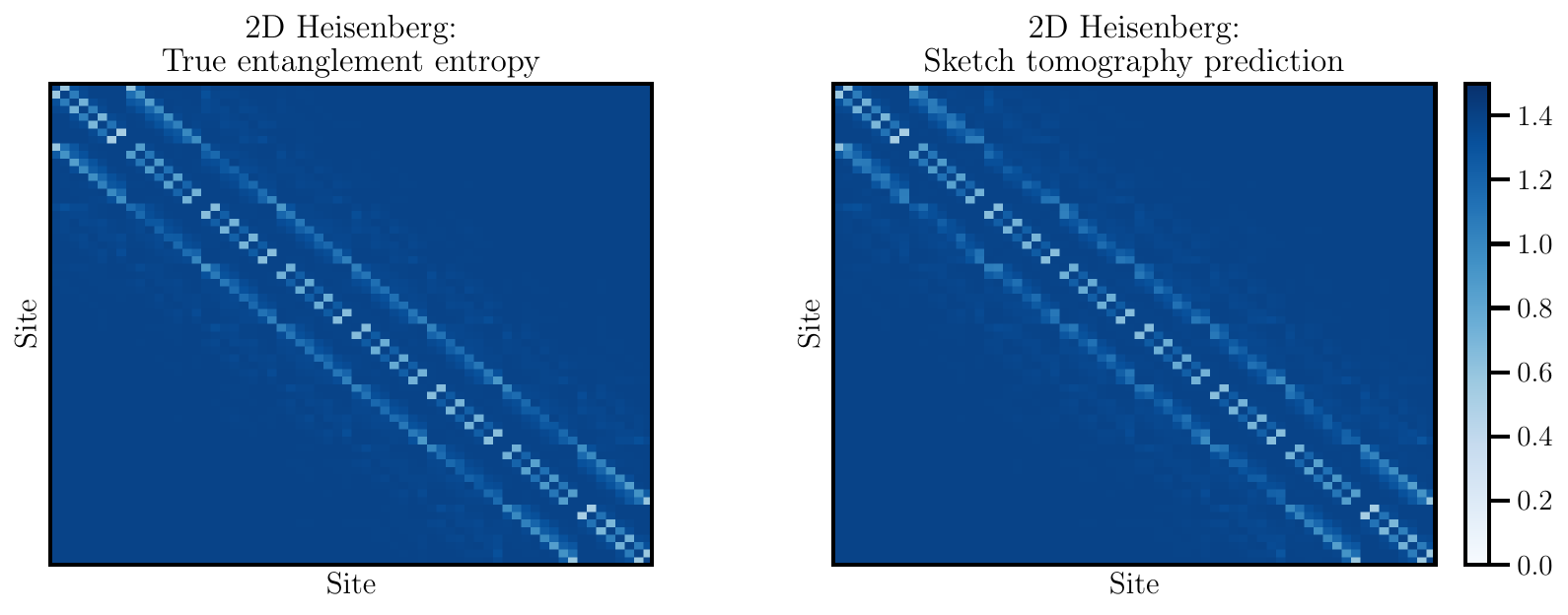}
    \caption{Prediction of second-order Renyi entanglement entropy for all subsystems of size at most two in the 2D Heisenberg model with \(n = 64\) lattice sites. The predicted values using our QST output visually match the true values.}
    \label{fig:2D_Heisenberg_renyi_plot}
\end{figure}

Our third example focuses on a 2D Heisenberg model with an open boundary condition. The Hamiltonian is \(H = \sum_{(i,j) \sim (i',j')} \left(\sigma_{(i,j)}^{X} \sigma_{(i',j')}^{X}+\sigma_{(i,j)}^{Y} \sigma_{(i',j')}^{Y}+\sigma_{(i,j)}^{Z} \sigma_{(i',j')}^{Z}\right)\), where \((i,j) \sim (i',j')\) if \((i,j)\) and \((i',j')\) are adjacent on a \(8 \times 8\) lattice. In this case, the system has \(n = 64\) lattice sites. We use the same model as the 2D Heisenberg experiment considered in \cite{huang2020predicting}, and we likewise use the ground state data for \(\ket{\psi}\) from \cite{carrasquilla2019reconstructing}. The reference ground truth \(\ket{\psi}\) is an MPS of maximal internal bond dimension \(a = 200\). 

The 2D Heisenberg model is significantly more difficult than the other considered cases due to the large parameter size of \(\ket{\psi}\). We simulate \(B = 8 \times 10^5\) random Pauli measurements on \(\ket{\psi}\) and we use the classical shadow protocol to form \(\hat{\rho}\). We apply our sketching-based procedure and obtain \(\tilde{\rho}\).\footnote{We do not provide the training results for MLE because the true MPS model has approximately \( 5\times 10^6\) parameters, which is significantly larger than the sample size. Thus, the comparison with MLE would be inconclusive due to concerns regarding overfitting.}

From \Cref{fig:2D_heisenberg_2_pt_correlation_64}, one can see that our obtained \(\tilde{\rho}\) is successful at calculating the two-point correlation function. The mean prediction error of \(\tilde{\rho}\) is \(0.018\), and the mean prediction error of classical shadow is \(0.013\), which shows that the two methods have comparable performance. The good performance of sketch tomography is noteworthy because representing \(\rho\) in a TT ansatz would require an internal bond of \(r = a^2 = 40000\), but the obtained \(\tilde{\rho}\) only has a maximal internal bond of \(r_{\mathrm{max}} = 50\) for computational efficiency considerations.

Lastly, we use \(\tilde{\rho}\) to calculate the Renyi entanglement entropy. Similar to the 1D Heisenberg case, we iterate over all subsystems of size at most two, and we show the result in \Cref{fig:2D_Heisenberg_renyi_plot}. The maximal prediction error is substantially larger at \(0.210\), but the mean prediction error is comparable to previous cases at only \(0.005\). Therefore, \(\tilde{\rho}\) is successful in predicting the Renyi entanglement entropy for the majority of the considered subsystems.

\section{Conclusion}

We present a quantum state tomography method using the classical shadow protocol under the matrix product state assumption. We demonstrate that our approach is highly accurate and sometimes outperforms the classical shadow protocol in observable estimation. Future work can consider extension to other tensor network structures, such as the hierarchical Tucker \cite{hackbusch2009new}.

\section*{Declarations}

L.Y. is supported by the U.S. Department of Energy, Office of Science, Accelerated Research in Quantum Computing Centers, Quantum Utility through Advanced Computational Quantum Algorithms, Grant No. DESC002557. L.Y., X.T., and H.C. are supported by AFOSR MURI award FA9550-24-1-0254.

\begin{appendices}

\section{Literature review}\label{sec: related work}
This section provides a comprehensive review of related literature, which consists of the following three parts. 

\textbf{Classical shadow in quantum observable estimation}
Classical shadow \cite{huang2020predicting} is a protocol that allows experimenters to predict \(M\) observables of a quantum state \(\rho\) with only \(\log(M)\) copies of \(\rho\). The procedure only requires repeated single-copy measurements of \(\rho\). The terminology is inspired by shadow tomography, which Aaronson coined in \cite{aaronson2018shadow}. One can also view the classical shadow estimator as a projected least-squares predictor \cite{guctua2020fast}. Classical shadow has different procedures to perform random uniform scrambling on the state \(\rho\). In \cite{huang2020predicting}, the authors propose random Pauli measurement and random Clifford scrambling, the former of which can be implemented with a circuit of depth one, whereas the latter is significantly more involved in terms of circuit complexity. One can also consider other random scrambling protocols of intermediate complexity \cite{hu2023classical,akhtar2023scalable,bertoni2024shallow}. While our work only considers the random Pauli measurement setting, future work might also consider other random scrambling protocols.

\textbf{Quantum state tomography}
Quantum state tomography (QST) is a fundamental task related to learning and engineering high-dimensional quantum systems in many-body physics. Specifically, it aims to reconstruct an unknown quantum state from a collection of experimental measurements. An extensive body of work has explored QST from many different aspects. For instance, one prominent line of work formulates QST as a matrix recovery problem~\cite{mirzaee2007finite,gross2010quantum,liu2011universal,alquier2013rank,gonccalves2013quantum,wang2013asymptotic,kalev2015quantum,kueng2016low,kueng2017low,kyrillidis2018provable,guctua2020fast,gupta2021maximal, gupta2022variational,kim2023fast,wang2024efficient,hsu2024quantum,schreiber2025tomography}, where the density matrix is estimated via statistical techniques such as projected least squares, semidefinite programming, compressed sensing~\cite{candes2006stable,donoho2006most,donoho2006compressed,candes2008introduction} and non-convex programming. Moreover, to alleviate the high computational cost of general QST, researchers have adopted simplifying assumptions by considering quantum states that can be well approximated by matrix product states with low bond dimension. This leads to the development of matrix product state tomography~\cite{cramer2010efficient,ohliger2013efficient,hubener2013wick,baumgratz2013scalable,haack2015continuous,ran2020tensor,wang2020scalable,lidiak2022quantum,qin2024quantum,teng2024learning} and variants such as multiscale entangled states tomography~\cite{landon2012practical}. Furthermore, an alternative approach to tackling the high dimensionality of quantum state tomography is to employ machine learning-based methods, such as neural networks~\cite{carleo2017solving,torlai2018neural,choo2018symmetries,park2020geometry,pei2021compact,koutny2022neural,pescia2022neural,chen2023efficient, huang2022provably,chen2023antn,ibarra2025autoregressive} and generative models~\cite{carrasquilla2019reconstructing,ahmed2021quantum}. Building upon the aforementioned methodologies and measurement protocols such as classical shadows~\cite{huang2020predicting}, QST and its variants have inspired numerous related studies and applications, such as theoretical analyses of QST~\cite{haah2016sample,o2016efficient,o2017efficient}, quantum measurement tomography~\cite{fiuravsek2001maximum,d2004quantum,lundeen2009tomography,feito2009measuring,cattaneo2023self,zambrano2025fast} and quantum process tomography~\cite{surawy2022projected,torlai2023quantum,huang2023learning,levy2024classical}. 

We discuss a few related works that our work bears the most resemblance to. Our work considers performing QST procedures based on the classical shadow approximation, and our work uses a sketching algorithm to reconstruct the true state. In \cite{landon2010efficient,lidiak2022quantum}, the authors consider a sketching-based procedure for MPS, but they do not use classical shadow, and the procedure does not have a convergence guarantee. In \cite{qin2024quantum, qin2024sample,kuzmin2024learning}, the authors consider a variational approach to train a matrix product operator (MPO) ansatz based on minimizing measurement error, where the proposed algorithm does not have a formal convergence guarantee, and the system size in the experiments remains relatively small. In \cite{qin2025enhancing}, the authors consider directly fitting an MPO ansatz to fit the classical shadow approximation \(\hat{\rho}\), but the procedure uses classical shadow protocol under Haar-random projective measurements, whereas our work focuses on the Pauli measurement setting.

\textbf{Tensor trains (matrix product states) in quantum physics}
The tensor train (TT) or matrix product state (MPS) ansatz, which originates from the density matrix renormalization group (DMRG) algorithm~\cite{white1992density}, provides an efficient parametrization of entangled quantum states in high-dimensional Hilbert spaces. For a complete discussion of TT/MPS and its variants, we refer the readers to the following review articles~\cite{verstraete2008matrix,orus2014practical,chan2016matrix,orus2019tensor}. As one of the most powerful tools in quantum many-body physics, the TT/MPS ansatz and its variants have been applied to a wide range of related tasks. Examples of such tasks include but are not limited to computing ground states via quantum Monte Carlo methods~\cite{sandvik2007variational, wouters2014projector,yu2024re,jiang2025unbiasing}, studies of electronic structure theory and quantum chemistry~\cite{khoromskaia2011qtt,jolly2025tensorized}, simulating open quantum systems and quantum dynamics in general~\cite{wall2012out,schroder2016simulating,greene2017tensor,leviatan2017quantum,strathearn2018efficient,borrelli2019density,baiardi2019large,luchnikov2019simulation,borrelli2021expanding,ye2021constructing,soley2021functional,dunnett2021efficient,gelss2022solving,lyu2022tensor,ren2022time,mangaud2023survey,shinaoka2023multiscale,lyu2023tensor,liu2024benchmarking,grimm2024accurate,chen2024many,link2024open,zeng2025quantum,gelss2025quantum,wang2025accelerated,park2025simulating,chen2025solving,sroda2025predictor}, solving quantum impurity models~\cite{weichselbaum2009variational,holzner2010matrix,ganahl2014chebyshev,wolf2014chebyshev,wolf2015imaginary,ganahl2015efficient,schwarz2018nonequilibrium,cao2021tree,kohn2022quench,thoenniss2023nonequilibrium,erpenbeck2023tensor,kloss2023equilibrium,wauters2024simulations,ng2023real,chen2024grassmann,eckstein2024solving,yu2025inchworm,kim2025strong,rohshap2025two,matsuura2025tensor}, quantum simulation and quantum computing~\cite{banuls2006simulation,ran2020encoding,yuan2021quantum,pan2022simulation,anselme2024combining,smith2024constant,berezutskii2025tensor}, studies of quantum field theories~\cite{cirac2010infinite,milsted2013matrix,banuls2013matrix,buyens2014matrix,kapustin2017topological,kapustin2018spin}, representing and learning Feynman diagrams~\cite{nunez2022learning,ishida2025low}, etc.

\section{Notations}
\label{app: notations}
In addition to the notations of tensors and tensor networks covered above, we provide a brief review of other mathematical notations in this section. We use $\otimes$ to denote the tensor product. For measuring distances between probability distributions, we use $\KL(\cdot,\cdot)$ to denote the Kullback-Leibler (KL) divergence between any two probability distributions. For two probability distributions $P, Q$, we use $P \ll Q$ to mean that $P$ is absolutely continuous with respect to $Q$. With fixed $d \in \sN$, we use $\mI_d$ to denote the identity matrix of size $d \times d$. For any complex vector $\va \in \mathbb{C}^d$, we use $\va^\intercal$ and $\va^\ast$ to denote the transpose and conjugate transpose of $\va$, respectively. Moreover, the dot product between any two square matrices $A,B \in \mathbb{C}^{d \times d}$ is denoted by $\langle A,B \rangle_F = \tr{A^* B}$, where $A^\ast$ denotes the conjugate transpose of $A$. Regarding the notation of norms, we use $\left\|\cdot\right\|_1$ and $\left\|\cdot\right\|_{F}$ to denote the $l_1$ norm and the Frobenius norm, respectively. In particular, for any vector $v \in \mathbb{C}^n$ and matrix $M \in \mathbb{C}^{m \times n}$, we use $\|v\|$ and $\|M\| = \sup_{\|u\|=1}\|Mu\|$ to denote the $l_2$ norm of $v$ and the induced operator $2$-norm of $M$, respectively. The singular values of $M$ are denoted by $s_1(M) \geq s_2(M) \geq \cdots \geq s_l(M)$, where $l = \min\{m,n\}$. For any $p \in (0,1)$, the Bernoulli distribution with mean $p$ is denoted by $\text{Ber}(p)$. For any two given quantities $f$ and $g$, we write $f \gtrsim g$ when the inequality $f \geq Cg$ holds for some fixed constant $C > 0$. Finally, we use the standard symbols $X,Y,Z$ to denote the Pauli matrices in $\mathbb{C}^{2 \times 2}$, which satisfy
\begin{equation*}
X = \begin{bmatrix}
0 &1\\
1 &0
\end{bmatrix}, \ 
Y = iXZ = \begin{bmatrix}
0 &-i\\
i &0
\end{bmatrix}, \ 
Z = \begin{bmatrix}
1 &0\\
0 &-1
\end{bmatrix}.    
\end{equation*}

\section{Details for the sketch tomography procedure}\label{sec: sketching}
We go through the detailed procedure and derivation for sketch tomography. We assume basic familiarity with the tensor network ansatz and tensor diagrams. 
\subsection{TT representation of \texorpdfstring{\(\rho\)}{rho}}
We first verify the claim in the main text that \(\rho\) has a TT format when the target state \(\ket{\psi}\) is a matrix product state. In terms of a tensor diagram, one can represent \(\ket{\psi}\) as follows:
\begin{equation}\label{eqn: State representation}
    \KetPsiBlock[0.5] \; = \; \KetPsiMPSChain[0.5]\,.
\end{equation}

As is shown in \Cref{eqn: State representation}, there exists a collection of \(n\) tensor components \((F_k)_{k = 1}^{n}\), where \(F_1 \in \sC^{2 \times a_1}\), \(F_k \in \sC^{a_{k - 1} \times 2 \times a_k}\) for \(k = 2, \ldots, n - 1\), and \(F_n \in \sC^{a_{n-1} \times 2}\). The evaluation of any entry in \(\ket{\psi}\) is shown in \Cref{eqn: State representation}, and one can write it equivalently as follows:
\begin{equation}\label{eqn: State representation ver 2}
    \ket{\psi}(j_1, \ldots, j_n) = \sum_{\gamma_{1}, \ldots, \gamma_{n-1}}F_1(j_1, \gamma_1)F_2(\gamma_1, j_2, \gamma_2)\cdots F_n(\gamma_{n-1}, j_n).
\end{equation}

The associated density matrix \(\rho = \ket{\psi}\bra{\psi}\) is the target object that sketch tomography approximates. One can first use \Cref{eqn: State representation} to directly write \(\rho\) in a tensor diagram as follows:
\begin{equation}\label{eqn: density matrix representation}
    \RhoBlock[0.5] \; = \; \RhoMPSChain[0.5]\,.
\end{equation}

We derive the TT format of \(\rho\) by writing \Cref{eqn: density matrix representation} with the Pauli matrices as the basis. For \(k \in [n]\), we let \((\sigma^{X}_k, \sigma^{Y}_k, \sigma^{Z}_k)\) denote the Pauli matrices on site \(k\), and we write \((\sigma^{1}_{k}, \sigma^{2}_{k}, \sigma^{3}_{k}, \sigma^{4}_{k}) = (\frac{1}{\sqrt{2}}I_{2}, \frac{1}{\sqrt{2}}\sigma^{X}_{k}, \frac{1}{\sqrt{2}}\sigma^{Y}_{k}, \frac{1}{\sqrt{2}}\sigma^{Z}_{k})\). The Pauli matrices \((\sigma^{i}_{k})_{i \in [4], k \in [n]}\) allow one to write \Cref{eqn: density matrix representation} in a TT format. We construct tensor components \((G_k)_{k = 1}^{n}\) directly from the MPS ansatz of \(\ket{\psi}\) by the following tensor diagrams:
\begin{equation}\label{eqn: def of Gk ver 2}
    \GkdefLHS[0.5] \;=\; \GkdefRHS[0.5],
\, \GOnedefLHS[0.5] \;=\; \GOnedefRHS[0.5], \, 
\GNdefLHS[0.5] \;=\; \GNdefRHS[0.5]\,.
\end{equation}
In other words, for \(k \not \in \{1, n\}\), we construct \(G_k \in \R^{a_{k-1}^2 \times 4 \times a_{k}^2}\) by
\begin{equation}\label{eqn: construction of G_k}
    G_{k}((\gamma_{k-1}, \gamma_{k-1}'), i, (\gamma_{k}, \gamma_{k}')) = \sum_{j, j'}\sigma^{i}_{k}(j, j')F_{k}(\gamma_{k-1}, j, \gamma_{k})\Bar{F_{k}}(\gamma'_{k-1}, j', \gamma'_{k}),
\end{equation}
and the construction for \(G_1\) and \(G_n\) can be likewise derived from \Cref{eqn: construction of G_k} by respectively omitting the \((\gamma_{k-1},\gamma'_{k-1})\) variable and the \((\gamma_{k},\gamma'_{k})\) variable. 
Lastly, \Cref{eqn: construction of G_k} shows that the entries of \(G_k\) are defined by an inner product between two Hermitian matrices, and so all entries of \(G_k\) are real numbers. 

We now verify that \(\rho\) admits a TT format given by \((G_k)_{k=1}^{n}\). Because multi-linear products of Pauli matrices form a basis in the space of Hermitian matrices in \(\sC^{2^n \times 2^n}\), there exists a tensor \(C \colon [4]^n \to \R\) whereby the term \(\rho\) is given by
\begin{equation*}
    \rho = \sum_{i_1, \ldots, i_n =1}^{4}C(i_1, \ldots, i_n)\prod_{l = 1}^{n}\sigma_{l}^{i_l}.
\end{equation*}
Moreover, the orthonormality of the Pauli matrices implies the following equation \[C(i_1, \ldots, i_n) = 
\sum_{j_1, j_1', \ldots, j_n, j_n'}\rho((j_1, j_1'), \ldots, (j_n, j_n'))\prod_{l = 1}^{n}\sigma_{l}^{i_l}(j_l, j_l').
\]
Therefore, one can plug in the definition of \(\rho\) in \Cref{eqn: density matrix representation} to directly derive \(C\). One has
\begin{equation*}
    \PsiBlockCustom[0.5] \; = \; \RhoPauliMPSChain[0.5] = \MPSChainCustom[0.5]\, ,
\end{equation*}
where the second equality is from \Cref{eqn: def of Gk ver 2}.

\subsection{Sketch tomography algorithm for quantum state tomography}

We cover how to approximate \(\rho\) with the classical shadow protocol. The main idea for the procedure is covered in the main text, and here we go through the full implementation details.

We explain how one applies the classical shadow protocol from Pauli measurements.
Let \(B\) be the number of recorded measurements. The classical shadow protocol uses the measurement outcome to produce a collection of \(2 \times 2\) matrices \(\{M_{l}^{(j)} \in \sC^{2 \times 2}\}_{l \in [n], j \in [B]}\). The resultant classical shadow approximation \(\hat{\rho}\) is written as a sum of separable matrices as follows:
\begin{equation}\label{eqn: formula of rho hat}
    \hat{\rho} = \frac{1}{B}\sum_{j = 1}^{B} \bigotimes_{l  =1}^{n} M_{l}^{(j)}.
\end{equation}
\Cref{eqn: formula of rho hat} allows for an efficient calculation of the partial trace of \(\hat{\rho}\). Therefore, performing observable approximation over \Cref{eqn: formula of rho hat} is efficient for any local observable \(O\). When \(O\) is \(k\)-local, the estimator \(\tr{O\hat{\rho}}\) is unbiased with a variance bounded from above by \(\gO(4^k)\), which is moderate for small \(k\).

In practice, instead of a single classical shadow approximation \(\hat{\rho}\) obtained from \(B\) experiments on \(\rho\), one can repeat the same protocol \(W\) times to form classical shadow approximators \(\hat{\rho}_{1}, \ldots, \hat{\rho}_{W}\). Then, the observable approximation is done by
\[
\langle O \rangle = \tr{O\rho} \approx \mathrm{median}\left\{\tr{O\hat{\rho}_1}, \ldots , \tr{O\hat{\rho}_W}\right\},
\]
where \(\mathrm{median}\{a_1, \ldots, a_W\}\) is the median of the \(W\) scalars \(a_1, \ldots, a_W\). The median-of-mean estimator is unbiased and is more robust with respect to outliers.

We go through our QST procedure for obtaining \(\rho\). For notational compactness, for an index set \(S = \{s_1, \ldots, s_l\} \subset [d]\), we use a multi-index notation by letting \(i_{S} := (i_{s_1}, \ldots, i_{s_l})\), i.e. \(i_{S}\) stands for the subvector of variables with entries from the index set $S$. Using the multi-index notation, any density matrix \(\rho\) can be uniquely represented by a tensor \(C \colon [4]^n \to \R\) as follows:
\begin{equation*}
    \rho = \sum_{i_{[n]} \in [4]^n}C(i_{[n]})\prod_{l = 1}^{n}\sigma_{l}^{i_l}.
\end{equation*}

Given the MPS assumption on \(\ket{\psi}\), the tensor \(C\) can be written in terms of tensor components \((G_{k})_{k = 1}^{n}\), where \(G_{1} \in \R^{ 4 \times r_{1}}\), \(G_{k} \in \R^{r_{k-1} \times 4 \times r_{k}}\) for \(k = 2,\ldots, n-1\), and \(G_{n} \in \R^{r_{n-1} \times 4 }\). The equation for \(C\) is written as follows:
\begin{equation}\label{eq: TT supplemental}
  C(i_{[n]}) =\sum_{\alpha_{[n-1]}}G_{1}(i_{1}, \alpha_{1})G_{2}(\alpha_{1}, i_{2}, \alpha_{2})\cdots G_{n}(\alpha_{n-1}, i_{n}).
\end{equation}

To obtain \(\rho\), we obtain each \(G_{k}\) for \(k = 1,\ldots,n\). We first cover the case where \(k \not \in \{1, n\}\). Due to the structural equation in \Cref{eq: TT supplemental}, there exist tensors \(C_{<k} \colon [4]^{k-1} \times [r_{k-1}] \to \sR\) and \(C_{>k} \colon [r_{k}] \times [4]^{n-k} \to \sR\) so that the following equation holds:
\begin{equation}\label{eq: full equation for Gk}
    \sum_{\alpha_{k-1}, \alpha_{k}}C_{<k}(i_{[k-1]}, \alpha_{k-1})G_{k}(\alpha_{k-1}, i_k, \alpha_{k})C_{>k}(\alpha_{k}, i_{[n] - [k]}) = C(i_{[n]}).
\end{equation}

One then applies sketching to \Cref{eq: full equation for Gk} to form a tractable linear system for \(G_k\). We let \(\tilde{r}_{k-1}, \tilde{r}_{k}\) be two integers such that \(\tilde{r}_{k-1} \geq r_{k-1}\) and \(\tilde{r}_{k} \geq r_{k}\).
We introduce two sketch tensors \(S_{<k} \colon [\tilde{r}_{k-1}] \times [4]^{k-1} \to \R, S_{>k} \colon [4]^{n-k} \times [\tilde{r}_{k}] \to \R\).
By contracting \Cref{eq: full equation for Gk} with \(S_{>k} \) and \(S_{<k}\), one obtains
\begin{equation}\label{eqn: sketched linear system for Gk}
  \sum_{\alpha_{k-1}, \alpha_{k}}A_{<k}(\zeta, \alpha_{k-1})G_{k}(\alpha_{k-1}, i_{k}, \alpha_{k})A_{>k}(\alpha_{k}, \mu) = B_{k}(\zeta, i_k, \mu),
\end{equation}
where \(A_{>k}, A_{<k}\) are respectively the contraction of \(C_{>k}, C_{<k}\) by \(S_{>k}, S_{<k}\) over \(i_{[n] - [k]}, i_{[k-1]}\) and \(B_{k}\) is the contraction of \(C\) by \(S_{>k} \otimes S_{<k}\) over \(i_{[n] - \singleton{k}}\).

In practice, the contractions are done \emph{implicitly} through taking observables. We let \(\{L_{k}^{\zeta}\}_{\zeta \in [\tilde{r}_{k-1}]}\) and \(\{R_{k}^{\mu}\}_{\mu \in [\tilde{r}_{k}]}\) be observables associated with \(S_{<k}, S_{>k}\) through the following equation:
\begin{equation}\label{eqn: implicit construction of sketch tensor}
    L_{k}^{\zeta} = \sum_{i_{[k-1]}}S_{<k}(\zeta, i_{[k-1]})\prod_{l =1}^{k-1}\sigma_{l}^{i_l}, \quad R_{k}^{\mu} = \sum_{i_{[n] - [k]}}S_{>k}(i_{[n] - [k]}, \mu)\prod_{l = k+1}^{n}\sigma_{l}^{i_l}.
\end{equation}
Then, to calculate \(B_k\), we utilize the fact that \(\{\prod_{l = 1}^{n}\sigma_{l}^{i_l}\}_{i_{[n]} \in [4]^n}\) is an orthonormal basis in the space of Hermitian matrices in \(\sC^{2^n \times 2^n}\), and so the following holds for \(B_k\):
\begin{equation*}
        B_{k}(\zeta, i_k, \mu) = \sum_{i_{[n] - \singleton{k}}}C(i_{[n]})S_{<k}(\zeta, i_{[k-1]})S_{>k}(i_{[n] - [k]}, \mu)=\tr{\rho L_{k}^{\zeta}\sigma_{k}^{i_k}R_{k}^{\mu}},
\end{equation*}
where the second equality holds by the definition of \(\{L_{k}^{\zeta}\}_{\zeta \in [\tilde{r}_{k-1}]}\) and \(\{R_{k}^{\mu}\}_{\mu \in [\tilde{r}_{k}]}\). With the classical shadow protocol, we approximate \(B_k\) by
\begin{equation}\label{eqn: definition of B_k}
\begin{aligned}
        B_{k}(\zeta, i_k, \mu) \approx \mathrm{median}\left\{\tr{\hat{\rho}_1 L_{k}^{\zeta}\sigma_{k}^{i_k}R_{k}^{\mu}}, \ldots, \tr{\hat{\rho}_W L_{k}^{\zeta}\sigma_{k}^{i_k}R_{k}^{\mu}}\right\}.
\end{aligned}
\end{equation}

There exists a gauge degree of freedom for the TT representation of \(C\) at each internal bond of the tensor. More explicitly, the pair of tensors \((C_{>k}, C_{<(k+1)})\) determines the value of \((A_{>k}, A_{<(k+1)})\), and so obtaining \(A_{>k}\) is dependent on fixing the gauge on \((C_{>k}, C_{<(k+1)})\). To fix the gauge and obtain \(A_{<k}, A_{>k}\), we assume that all sketch tensors \(\{S_{>1}\} \cup \{S_{>j}, S_{<j}\}_{j=2}^{n-1} \cup \{S_{<n}\}\) have been defined via \Cref{eqn: implicit construction of sketch tensor}. We use sketching of \(C\) to uniquely determine a gauge on \((C_{>k}, C_{<(k+1)})\).

To obtain \(A_{>k}\), we construct a tensor \(Z_{k}\) from contracting \(C\) with \(S_{<(k+1)} \otimes S_{>k}\). By associating \(S_{<(k+1)}\) with observables \(\{L_{k+1}^{\zeta'}\}_{\zeta' \in [\tilde{r}_{k}]}\), we have
\begin{equation}\label{eqn: Z_k formula}
  Z_{k}(\zeta', \mu) = \tr{\rho L_{k+1}^{\zeta'}R_{k}^{\mu}} \approx \mathrm{median}\left\{\tr{\hat{\rho}_1 L_{k+1}^{\zeta'}R_{k}^{\mu}}, \ldots, \tr{\hat{\rho}_W L_{k+1}^{\zeta'}R_{k}^{\mu}}\right\},
\end{equation}
where the equality holds by the definition of \(\{L_{k+1}^{\zeta'}\}_{\zeta' \in [\tilde{r}_{k}]}\) and \(\{R_{k}^{\mu}\}_{\mu \in [\tilde{r}_{k}]}\). 
The SVD of \(Z_{k}\) uniquely determines the best rank-\(r_k\) factorization \(Z_{k} = USV^{\top}\). For example, the choice of setting \(A_{<(k+1)} = U\) and \(A_{>k} = SV^{\top}\) uniquely fixes the gauge of \((C_{>k}, C_{<(k+1)})\). One way to understand the construction is that \(C_{>k}\) is chosen to be the unique tensor for which the contraction of \(C_{>k}\) with \(S_{>k}\) is \(U\).
Finally, obtaining \(A_{<k}\) is done through the best rank-\(r_{k-1}\) factorization of \(Z_{k-1}\).

We now detail the procedure for \(k = 1\) and \(k = n\).
For \(G_1\) and \(G_n\), one can likewise use sketch tensors \(S_{>1}\) and \(S_{<n}\) to form the following linear system
\begin{equation}\label{eqn: sketched linear system boundary case}
\begin{aligned}
&\sum_{\alpha_{1}}G_{1}(i_{1}, \alpha_{1})A_{>1}(\alpha_{1}, \mu) = B_{1}(i_1, \mu),\\
        &\sum_{\alpha_{n-1}}A_{<n}(\zeta, \alpha_{n-1})G_{n}(\alpha_{n-1}, i_{n}) = B_{n}(\zeta, i_n).
\end{aligned}  
\end{equation}

The term \(A_{>1}\) and \(A_{<n}\) are respectively obtained using \(Z_{1}\) and \(Z_{n-1}\) obtained via \Cref{eqn: Z_k formula}. Lastly, the sketch tensors \(S_{>1}, S_{<n}\) are respectively defined implicitly with observables \(\{R_{1}^{\mu}\}_{\mu \in [\tilde{r}_{1}]}\) and \(\{L_{n}^{\zeta}\}_{\zeta \in [\tilde{r}_{n-1}]}\). One obtains \(B_1\) and \(B_n\) by
\begin{equation}\label{eqn: definition of B_k boundary case}
\begin{aligned}
    &B_{1}(i_1, \mu) \approx \mathrm{median}\left\{\tr{\hat{\rho}_1 \sigma_{1}^{i_1}R_{1}^{\mu}}, \ldots, \tr{\hat{\rho}_W \sigma_{1}^{i_1}R_{1}^{\mu}}\right\}, \\ &B_{n}(\zeta, i_n) \approx \mathrm{median}\left\{\tr{\hat{\rho}_1 L_{n}^{\zeta}\sigma_{n}^{i_n}}, \ldots, \tr{\hat{\rho}_W L_{n}^{\zeta}\sigma_{n}^{i_n}}\right\}.
\end{aligned}
\end{equation}

We summarize our approach in \Cref{alg: sketch tomography}. We use the classical shadow to obtain each \(Z_{k}, B_{k}\) and we use SVD on each \(Z_{k}\) to obtain all \(A_{>k}, A_{<k}\). Then, solving the corresponding sketched linear equation for all \(G_k\) allows one to obtain a TT approximation of \(\rho\).

\begin{algorithm}
  \caption{Sketch tomography for quantum state tomography under MPS assumption}
  \label{alg: sketch tomography}
  \begin{algorithmic}[1]
    \REQUIRE Classical shadow approximation \(\{\hat{\rho}_1, \ldots, \hat{\rho}_{W}\}\) from \Cref{eqn: formula of rho hat}.
    \REQUIRE Collection of sketch tensors \(\{S_{>k}, S_{<k}\}\)
    \REQUIRE Target internal ranks $\{r_{k}\}$ for \(k = 1, \ldots, n-1\).
    \FOR{\(k = 1, \ldots, n-1\)}
    \STATE Obtain \(Z_{k}\) by \Cref{eqn: Z_k formula}.
    \STATE Obtain \(A_{<(k+1)}\) and \(A_{>k}\) respectively as the left and right factor of the best rank-\(r_{k}\) factorization of \(Z_{k}\).
    \ENDFOR
    \FOR{\(k = 1, \ldots, n\)}
    \IF{\(k = 1\) or \(k = n\)}
    \STATE Obtain \(B_{k}\) by \Cref{eqn: definition of B_k boundary case}.
    \STATE Obtain \(\hat{G}_k \approx G_k\) by solving the linear system in \Cref{eqn: sketched linear system boundary case}.
    \ENDIF
    \IF{\(k \not \in \{1, n\}\)}
    \STATE Obtain \(B_{k}\) by \Cref{eqn: definition of B_k}.
    \STATE Obtain \(\hat{G}_k \approx G_k\) by solving the linear system in \Cref{eqn: sketched linear system for Gk}.
    \ENDIF
    \ENDFOR
    \STATE Output the approximate density matrix \(\tilde{\rho}\) with tensor component \((\hat{G}_{k})_{k = 1}^{n}\).
  \end{algorithmic}
\end{algorithm}

\section{Proof of the upper bound}\label{sec: upper}
This section goes through the convergence guarantee for the sketch tomography procedure. The ground truth is a density matrix \(\rho\) whose coefficient tensor is a tensor train. Sketch tomography takes in the classical shadow approximation \(\hat{\rho}\) and outputs an approximate density matrix \(\tilde{\rho}\) with tensor component \((\hat{G}_{k})_{k = 1}^{n}\). The organization of this section is as follows. In \Cref{sec: upper assumption}, we state the necessary assumptions for the results of this section. In \Cref{sec: observable estimation error}, we show that each observable estimation task required by \Cref{alg: sketch tomography} is accurate. In \Cref{sec: local error}, we prove that each tensor component of \(\tilde{\rho}\) is close to the tensor component of \(\rho\). In \Cref{sec: global error}, we prove that \(\tilde{\rho}\) is close to \(\rho\) in the Frobenius norm.

\textbf{Notations} 
We summarize the necessary notations for the analysis. For a vector \(v\), the term \(\lVert v \rVert\) is its \(l_2\) norm. For a matrix \(M\), the term \(\lVert M \rVert\) is its operator norm, and the term \(s_j(M)\) is the \(j\)-th singular value of \(M\). For a tensor \(G\), the term \(\lVert G\rVert_{F}\) is the Frobenius norm of \(G\). We also assume the multi-index notation used in Appendix~\ref{sec: sketching}.

\subsection{Assumptions}\label{sec: upper assumption}

We list out the assumptions for the convergence guarantee of sketch tomography. As in Appendix~\ref{sec: sketching}, we use \(\rho\) to represent the target density matrix on \(n\) qubits. For easy reference, we summarize the construction of the Pauli-basis representation of \(\rho\):
\begin{definition}\label{def: target state}
    We let \((\sigma^{X}_k, \sigma^{Y}_k, \sigma^{Z}_k)\) denote the Pauli matrices on site \(k \in [n]\), and we write \((\sigma^{1}_{k}, \sigma^{2}_{k}, \sigma^{3}_{k}, \sigma^{4}_{k}) = (\frac{1}{\sqrt{2}}I_{2}, \frac{1}{\sqrt{2}}\sigma^{X}_{k}, \frac{1}{\sqrt{2}}\sigma^{Y}_{k}, \frac{1}{\sqrt{2}}\sigma^{Z}_{k})\). The density matrix \(\rho\) admits a \emph{coefficient tensor} \(C\) which characterizes \(\rho\) by the following equation:
    \begin{equation}\label{eqn: coefficient tensor defn 1}
        \rho = \sum_{i_{[n]} \in [4]^n}C(i_{[n]})\prod_{l = 1}^{n}\sigma_{l}^{i_l}.
    \end{equation}
    In particular, each entry of \(C\) is constructed from \(\rho\) as follows:
    \begin{equation}
        C(i_1, \ldots, i_n) = \tr{\rho \prod_{l = 1}^{n}\sigma_{l}^{i_l}}.
    \end{equation}
\end{definition}

Following \Cref{def: target state}, we formalize the tensor train (TT) property on \(\rho\) as an assumption. In particular, Appendix~\ref{sec: sketching} shows that the following TT assumption on \(\rho\) holds when \(\rho\) is the density matrix of a matrix product state \(\ket{\psi}\).
\begin{assumption}\label{assumption: TT coeff}
    The coefficient tensor \(C\) defined from \Cref{def: target state} is a tensor train. In particular, there exists a collection of internal rank $\{r_k\}_{k = 1}^{n-1}$ and a collection of tensor components \((G_{k})_{k = 1}^{n}\), where \(G_{1} \in \R^{ 4 \times r_{1}}\), \(G_{k} \in \R^{r_{k-1} \times 4 \times r_{k}}\) for \(k = 2,\ldots, n-1\), and \(G_{n} \in \R^{r_{n-1} \times 4 }\). The tensor \(C\) is defined by \((G_{k})_{k = 1}^{n}\) by the following equation:
    \begin{equation}\label{eq: TT supplemental 2}
      C(i_{[n]}) =\sum_{\alpha_{[n-1]}}G_{1}(i_{1}, \alpha_{1})G_{2}(\alpha_{1}, i_{2}, \alpha_{2})\cdots G_{n}(\alpha_{n-1}, i_{n}).
    \end{equation}
\end{assumption}

Secondly, we place assumptions to constrain the design space for sketch tensors.
As shown in Appendix~\ref{sec: sketching}, the sketch tensors are defined implicitly through observables. The integers $\{\tilde{r}_k\}_{k = 1}^{n-1}$ represent the number of sketch functions used during sketch tomography. The sketching procedure uses sketch tensors \(\{S_{<k}, S_{>k}\}_{k = 1}^{n}\) which are defined implicitly through observables. We summarize the construction here:
\begin{definition}\label{def: sketch tensor association}
    For \(k = 2, \ldots, n\), the sketch tensor \(S_{<k}\) is associated with observables \(\{L_{k}^{\zeta}\}_{\zeta \in [\tilde{r}_{k-1}]}\). Each observable \(L_{k}^{\zeta}\) only involves sites to the left of \(k\), i.e. the set \(\{1, \ldots, k-1\}\). The tensor \(S_{<k}\) is defined from \(\{L_{k}^{\zeta}\}_{\zeta}\) by the following equation:
    \begin{equation}
        S_{<k}(\zeta, i_1, \ldots, i_{k-1}) = \tr{L_{k}^{\zeta}\prod_{l =1}^{k-1}\sigma_{l}^{i_l}}.
    \end{equation}

    Similarly, for \(k = 1, \ldots, n-1\), the sketch tensor \(S_{>k}\) is associated with observables \(\{R_{k}^{\mu}\}_{\mu \in [\tilde{r}_{k}]}\). Each observable \(R_{k}^{\mu}\) only involves sites to the right of site \(k\), i.e. the set \(\{k+1, \ldots, n\}\). The tensor \(S_{>k}\) is defined from \(\{R_{k}^{\mu}\}_{\mu}\) by the following equation:
    \begin{equation}
        S_{>k}(i_{k+1}, \ldots, i_{n}, \mu)= \tr{R_{k}^{\mu}\prod_{l = k+1}^{n}\sigma_{l}^{i_l}}.
    \end{equation}

\end{definition}

The second assumption requires that the associated observables of the sketch tensors should come from local observables. Essentially, this assumption is to ensure that the classical shadow protocol under random Pauli measurement can reliably perform observable estimations required by \Cref{alg: sketch tomography}.
\begin{assumption}\label{assumption: sketch tensor assumption}
    For \(k = 1, \ldots, n - 1\), we let \(\tilde{r}_k = r_k\), where \(r_k\) is the true internal rank in \Cref{def: target state}. Moreover, each of the observables \(\{L_{k}^{\zeta}\}_{\zeta}\) and \(\{R_{k}^{\mu}\}_{\mu}\) in \Cref{def: sketch tensor association} is a linear combination of local observables. Essentially, for \(M = R_{k}^{\mu}\) or \(M = L_{k}^{\zeta}\), we require that one can write
    \[
    M = \sum_{l = 1}^{P}O_{l},
    \]
    where each \(O_{l}\) only involves \(\gO(1)\) sites. Moreover, we require \(\sum_{l =1}^{P} \lVert O_l\rVert = \gO(1)\), where \(\lVert \cdot \rVert\) is the operator norm.
\end{assumption}
\Cref{assumption: sketch tensor assumption} is a mild assumption on the design of sketch tensors. One way to satisfy \Cref{assumption: sketch tensor assumption} is to set \(\{L_{k}^{\zeta}\}_{\zeta}\) and \(\{R_{k}^{\mu}\}_{\mu}\) to be a linear combination of 1-local and 2-local Pauli observables with bounded total weights. When running \Cref{alg: sketch tomography} with sketch tensors satisfying \Cref{assumption: sketch tensor assumption}, one can ensure that all observable estimation done by \(\hat{\rho}\) can achieve a high accuracy with a sufficient number of state copies.

The third assumption chooses how one obtains \(A_{<(k+1)}\) and \(A_{>k}\) from \(Z_{k}\). While this assumption could be precluded, its inclusion significantly simplifies the error analysis. We state it here:
\begin{assumption}\label{assumption: choice of A_k}
    From \Cref{assumption: sketch tensor assumption}, one has \(r_{k} = \tilde{r}_k\) for \(k = 1, \ldots, n-1\). Let \(Z_k \in \R^{r_k \times r_k}\) be from \Cref{eqn: Z_k formula}, we set \(A_{<(k+1)}\) and \(A_{>k}\) to be a rank-\(r_k\) factorization so that
    \[
    A_{<(k+1)} = I_{r_k}, \quad A_{>k} = Z_k.
    \]
    Let \(\hat{Z}_k\) be the finite-sample estimation of \(Z_k\). We set \(\hat{A}_{<(k+1)}\) and \(\hat{A}_{>k}\) to be 
    \[
    \hat{A}_{<(k+1)} = I_{r_k}, \quad \hat{A}_{>k} = \hat{Z}_k.
    \]
\end{assumption}

This particular gauge choice is made only for the sake of analysis. In practice, one may instead take \((A_{<(k+1)}, A_{>k})\) from the SVD of \(Z_k\); this corresponds to an orthogonal change of basis on the internal legs and does not affect the asymptotic scaling of our bounds. However, using SVD complicates the notation for error analysis.

\subsection{Error bound on observable estimation}\label{sec: observable estimation error}
\Cref{assumption: TT coeff} and \Cref{assumption: sketch tensor assumption} allow us to provide an error bound on all of the observable estimation tasks that are performed when running \Cref{alg: sketch tomography}. To obtain \(G_{k}\), we have seen in \Cref{sec: procedure} that one needs to form three sketches \(B_{k} \in \R^{\tilde{r}_{k-1} \times 4 \times \tilde{r}_k}, Z_{k-1} \in \R^{\tilde{r}_{k-1} \times \tilde{r}_{k-1}}, Z_{k} \in \R^{\tilde{r}_{k} \times \tilde{r}_{k}}\) by performing observable estimation on \(\rho\). This section bounds the error rate on the sketch estimation. 

For the error estimation, we assume that the reader is familiar with the definition of the shadow norm \(\lVert \cdot \rVert_{\mathrm{shadow}}\) in \cite{huang2020predicting}. For readers unfamiliar with the concept, one can still follow the derivation by assuming the results on \(\lVert \cdot \rVert_{\mathrm{shadow}}\) from \cite{huang2020predicting}. For a brief explanation, the shadow norm essentially measures the complexity of the observable in a given classical shadow protocol. For random Pauli measurement, estimating \(\tr{\rho O}\) is only accurate when \(\lVert O \rVert_{\mathrm{shadow}}\) is small. Moreover, \(\lVert O \rVert_{\mathrm{shadow}}\) can be reliably bounded from above when \(O\) is a local observable or a sum of local observables. Therefore, \Cref{assumption: sketch tensor assumption} essentially forces the sketch tensors to correspond to observables that have a small shadow norm. 

From the formula in \Cref{def: sketch tensor association}, the sketches \(B_{k}, Z_{k-1}, Z_{k}\) satisfy the following equations:
\begin{equation}\label{eqn: true sketch}
    \begin{aligned}
        &B_{k}(\zeta, i_k, \mu) = \tr{\rho L_{k}^{\zeta}\sigma_{k}^{i_k}R_{k}^{\mu}},\; \\ &Z_{k-1}(\zeta, \mu') = \tr{\rho L_{k}^{\zeta}R_{k-1}^{\mu'}},\;\\ &Z_{k}(\zeta', \mu) = \tr{\rho L_{k+1}^{\zeta'}R_{k}^{\mu}}.
    \end{aligned}
\end{equation}

During the procedure of \Cref{alg: sketch tomography}, we use \(W\) copies of the classical shadow estimator to form the approximated sketches \(\hat{B}_{k}, \hat{Z}_{k-1}, \hat{Z}_{k}\) with the following formula:
\begin{equation}\label{eqn: approximated sketch}
    \begin{aligned}
        &\hat{B}_{k}(\zeta, i_k, \mu) = \mathrm{median}\left\{\tr{\hat{\rho}_1 L_{k}^{\zeta}\sigma_{k}^{i_k}R_{k}^{\mu}},
        \ldots,
        \tr{\hat{\rho}_W L_{k}^{\zeta}\sigma_{k}^{i_k}R_{k}^{\mu}}\right\},
        \\ &\hat{Z}_{k-1}(\zeta, \mu') = 
        \mathrm{median}\left\{\tr{\hat{\rho}_1 L_{k}^{\zeta}R_{k-1}^{\mu'}},
        \ldots,
        \tr{\hat{\rho}_W L_{k}^{\zeta}R_{k-1}^{\mu'}}\right\},
        \\ &\hat{Z}_{k}(\zeta', \mu) = \mathrm{median}\left\{\tr{\hat{\rho}_1 L_{k+1}^{\zeta'}R_{k}^{\mu}},
        \ldots,
        \tr{\hat{\rho}_W L_{k+1}^{\zeta'}R_{k}^{\mu}}
        \right\}.
    \end{aligned}
\end{equation}

\Cref{assumption: sketch tensor assumption} allows us to bound the error in approximating the formed sketches. We show that one has the following result:
\begin{proposition}\label{prop: classical shadow accuracy}
    Let \(\delta, \varepsilon \in (0, 1)\) be accuracy parameters. Let \(\tilde{r}_{\mathrm{max}} = \max_{k \in [n-1]}\tilde{r}_k\). In \Cref{alg: sketch tomography}, set \(W = 2\log(12n\tilde{r}_{\mathrm{max}}^2/\delta)\) and suppose each \(\hat{\rho}_{w}\) with \(w \in [W]\) is a classical shadow approximator formed from \(B\) measurements. Moreover, suppose \(\{S_{<k}, S_{>k}\}_{k = 1}^{n}\) are sketch tensors satisfying \Cref{assumption: sketch tensor assumption}. Then, there exists a problem-independent constant \(c_S\) such that \(B \geq \frac{c_S}{\varepsilon^2}\) ensures that the following events jointly hold with probability \(1 - \delta\):
    \begin{enumerate}
        \item For any \(k = 1, \ldots, n - 1\) and \(\zeta, \mu \in [\tilde{r_k}]\), \[\lvert Z_{k}(\zeta, \mu) - \hat{Z}_{k}(\zeta, \mu)\rvert \leq \varepsilon.\]
        \item For any \(k = 1, \ldots, n\) and \(\zeta \in [\tilde{r}_{k-1}], i_k \in [4], \mu \in [\tilde{r}_{k}]\),
        \[
        \lvert B_{k}(\zeta, i_k, \mu) - \hat{B}_{k}(\zeta, i_k, \mu) \rvert \leq \varepsilon.
        \]
    \end{enumerate}
\end{proposition}
\begin{proof}
    The proof requires the reader to be familiar with the shadow norm \(\lVert \cdot \rVert_{\mathrm{shadow}}\) in \cite{huang2020predicting}.
    The result is essentially a direct corollary of \Cref{assumption: sketch tensor assumption} and Theorem 1 in \cite{huang2020predicting}. Moreover, while the statement sets \(c_S\) to be a problem-independent constant, we provide the value of \(c_S\) by specifying the \(\gO(1)\) constant in \Cref{assumption: sketch tensor assumption}. 

    We consider an observable \(L_{k+1}^{\zeta}R_{k}^{\mu}\) corresponding to the term \(Z_k(\zeta, \mu)\). By \Cref{assumption: sketch tensor assumption}, one can write
    \[
    L_{k+1}^{\zeta} = \sum_{l = 1}^{P}O_l, \quad R_{k}^{\mu}=\sum_{l' = 1}^{P'}U_{l'},
    \]
    where each \(O_l, U_{l'}\) only involves at most \(D_1\) sites. Moreover, we suppose \(D_2\) is a constant so that \[\sum_{l =1}^{P} \lVert O_l\rVert \leq D_2, \quad \sum_{l' =1}^{P'} \lVert U_{l'}\rVert \leq D_2.\]

    By \Cref{def: sketch tensor association}, one can see that each \(O_l\) only involves site \(\{1, \ldots, k\}\), and each \(U_l\) only involves site \(\{k+1, \ldots, n\}\). Thus, the observable \(L_{k+1}^{\zeta}R_{k}^{\mu}\) can be written as 
    \[
    L_{k+1}^{\zeta}R_{k}^{\mu} = \sum_{l= 1 }^{P}\sum_{l' = 1}^{P'}O_lU_{l'}.
    \]
    For each index \((l,l')\), one can see that the term \(O_lU_{l'}\) only involves \(2D_1\) sites. Moreover, as \(O_l,U_{l'}\) acts on non-overlapping sites, one has 
    \(\lVert O_l U_{l'} \rVert  = \lVert O_l \rVert\lVert U_{l'} \rVert.\) Therefore, one has 
    \[
    \sum_{l= 1 }^{P}\sum_{l' = 1}^{P'} \lVert O_lU_{l'}\rVert = \sum_{l= 1 }^{P}\sum_{l' = 1}^{P'}\lVert O_l \rVert\lVert U_{l'} \rVert = \left(\sum_{l =1}^{P} \lVert O_l\rVert\right)\left(\sum_{l' =1}^{P'} \lVert U_{l'}\rVert\right) \leq D_2^2.
    \]

    By Proposition 3 in \cite{huang2020predicting}, as each term \(O_lU_{l'}\) involves only \(2D_1\) sites, one has 
    \[\lVert O_l U_{l'}\rVert_{\mathrm{shadow}} \leq 2^{2D_1} \lVert O_l U_{l'}\rVert.\]
    As \cite{huang2020predicting} has shown that the shadow norm satisfies the triangle inequality, one has 
    \begin{equation}
        \lVert L_{k+1}^{\zeta}R_{k}^{\mu}\rVert_{\mathrm{shadow}} \leq \sum_{l= 1 }^{P}\sum_{l' = 1}^{P'}\lVert O_l U_{l'}\rVert_{\mathrm{shadow}} \leq \sum_{l= 1 }^{P}\sum_{l' = 1}^{P'} 2^{2D_1} \lVert O_l U_{l'}\rVert \leq 4^{D_1} D_2^2.
    \end{equation}

    We similarly bound the shadow norm for the observable \(L_{k}^{\zeta}\sigma_{k}^{i_k}R_{k}^{\mu}\) corresponding to the term \(B_k(\zeta, i_k, \mu)\). Again, using \Cref{assumption: sketch tensor assumption}, one can write
    \[
    L_{k}^{\zeta} = \sum_{l = 1}^{\tilde{P}}\tilde{O}_l,
    \]
    so that \(\tilde{O}_l\) only acts on \(D_1\) sites and \(\sum_{l =1}^{\tilde{P}} \lVert \tilde{O}_l\rVert \leq D_2\).
    Then, one has 
    \[
    L_{k}^{\zeta}\sigma_{k}^{i_k}R_{k}^{\mu} = \sum_{l= 1 }^{\tilde{P}}\sum_{l' = 1}^{P'}\tilde{O}_l\sigma_{k}^{i_k}U_{l'}.
    \]
    We can see that \(O_l\sigma_{k}^{i_k}U_{l'}\) acts on \(2D_1 + 1\) sites, and so Proposition 3 in \cite{huang2020predicting} similarly shows
    \[
    \lVert \tilde{O}_l \sigma_{k}^{i_k} U_{l'}\rVert_{\mathrm{shadow}} \leq 2^{2D_1+1} \lVert \tilde{O}_l \sigma_{k}^{i_k} U_{l'}\rVert = 2^{2D_1+1/2} \lVert \tilde{O}_l\rVert \lVert U_{l'}\rVert,
    \]
    where the last equality uses the separability of the observable and the fact that \(\lVert \sigma_k^i\rVert = \frac{1}{\sqrt{2}}\) for \( i \in [4]\). Then, by the triangle inequality on \(\lVert \cdot \rVert_{\mathrm{shadow}}\), one has
    \begin{equation}
        \lVert L_{k}^{\zeta}\sigma_{k}^{i_k}R_{k}^{\mu} \rVert_{\mathrm{shadow}} \leq  \sum_{l= 1 }^{\tilde{P}}\sum_{l' = 1}^{P'}\lVert \tilde{O}_l \sigma_{k}^{i_k} U_{l'}\rVert_{\mathrm{shadow}} \leq \sum_{l= 1 }^{\tilde{P}}\sum_{l' = 1}^{P'}2^{2D_1+1/2} \lVert \tilde{O}_l\rVert \lVert U_{l'}\rVert \leq 4^{D_1 + 1/4} D_2^2.
    \end{equation}

    Therefore, when running \Cref{alg: sketch tomography}, we require the classical shadow protocol to perform \(M\) observable estimation tasks on observable \(Q_1, \ldots, Q_M\). In this case, one has \(M \leq 6n\tilde{r}_{\mathrm{max}}^2\) and \(\lVert Q_i \rVert_{\mathrm{shadow}} \leq 4^{D_1 + 1/4}D_2^2\). We can now use Theorem 1 in \cite{huang2020predicting}, and one can see that taking \(c_S = 68 \times 16^{D_1}D_2^4\) is sufficient for \Cref{prop: classical shadow accuracy} to hold.
\end{proof}

For the constant \(c_S\), one typically takes \(D_1 = 2\) and \(D_2 = 1\) to ensure the classical shadow procedure is successful. In practice, instead of enforcing \Cref{assumption: sketch tensor assumption}, one can also directly design sketch tensors so that they correspond to observables with a small shadow norm.

\subsection{Error bound on tensor component \texorpdfstring{$G_k$}{Gk}}\label{sec: local error}
This section proves the component-wise convergence result. \Cref{alg: sketch tomography} outputs a series of tensor components \((\hat{G}_{k})_{k = 1}^{n}\) to form an approximation \(\tilde{\rho}\). We prove that the output \(\hat{G}_{k}\) accurately approximates \(G_{k}\).

\begin{proposition}
\label{prop:local-perturbation}(Local error bound on sketch tomography)
Let $\rho \in \mathbb{C}^{2^n \times 2^n}$ be a density matrix satisfying \Cref{assumption: TT coeff}. Moreover, let \(\{S_{<k}, S_{>k}\}_{k = 1}^{n}\) be sketch tensors satisfying \Cref{assumption: sketch tensor assumption}, and let \Cref{alg: sketch tomography} satisfy \Cref{assumption: choice of A_k}. Let the true sketch $\{B_{k}\}_{k = 1}^{n}$ and $\{Z_{k}\}_{k = 1}^{n -1}$ be as in \Cref{eqn: true sketch}. Moreover, let the approximated sketch $\{\hat{B}_{k}\}_{k = 1}^{n}$ and $\{\hat{Z}_{k}\}_{k = 1}^{n -1}$ be as in \Cref{eqn: approximated sketch}, and let \(\{\hat{G}_k\}_{k = 1}^{n}\) be the obtained tensor component from \Cref{alg: sketch tomography}.

Let \(\varepsilon_1, \varepsilon_2\) be two accuracy parameters so that the classical shadow estimate satisfies \[\max_{k \in [n-1]}\lVert Z_{k} - \hat{Z}_{k}\rVert \leq \varepsilon_1, \quad \max_{k \in [n]}\lVert B_{k} - \hat{B}_{k}\rVert_F \leq \varepsilon_2.\] Moreover, suppose that \(\varepsilon_1 \leq \frac{1}{2}\min_{k\in [n-1]}s_{r_k}(Z_k)\). Then, for \(k = 1, \ldots, n\), one has
\begin{equation}
\label{eqn: component bound}
    \lVert \hat{G}_k - G_k \rVert_{F}^2 \le 2c_Z^2(c_G^2\varepsilon_1^2 + \varepsilon_2^2),
\end{equation}
where the constant \(c_Z\) is given by \(c_Z = \max\left(1,  \max_{k \in [n-1]}\frac{2}{s_{r_k}(Z_k)}\right)\), and \(c_G\) is given by \(c_G =\max(1, \max_{k \in [n]}\lVert G_k \rVert_F)\). 

In particular, let \(\varepsilon \in (0, 1)\) be an accuracy parameter and let \(c_S, r_{\mathrm{max}}\) be the same term as in \Cref{prop: classical shadow accuracy}. In \Cref{alg: sketch tomography}, set \(W = 2\log(12n\tilde{r}_{\mathrm{max}}^2/\delta)\) and set \(B \geq 2 c_Sr_{\mathrm{max}}^2 c_Z^2(c_G^2 + 4)\varepsilon^{-2}\). Then, with probability \(1 - \delta\), the following holds for all \(k  = 1,\ldots, n\):
\begin{equation}
\label{eqn: component bound 2}
    \lVert \hat{G}_k - G_k \rVert_{F} \le \varepsilon.
\end{equation}
\end{proposition}

For the error analysis, we need to invoke a result on the stability of the least-squares solution under perturbations of the linear system. 

\begin{lemma}(Corollary to Theorem 3.48 of \cite{wendland_2017})
\label{lemma: Least-Squares stability}
    Let \(A \in \R^{m \times r}\) be of rank \(r\). Let \(\Delta A\) be a perturbation with \(\lVert A^{\dagger} \rVert\lVert \Delta A \rVert \leq \frac{1}{2}\). Let \(b, \Delta b \in \R^{m}\). Let \(v\) satisfy \(Av = b\), and let \(\hat{v}\) be the least-squares solution to \((A+\Delta A)x = b + \Delta b\). Then one has
    \[
    \lVert \hat{v} - v \rVert \leq 2 \lVert A^{\dagger} \rVert\left(\lVert \Delta A \rVert\lVert v \rVert + \lVert \Delta b \rVert \right).
    \]
\end{lemma}

We now present the proof of \Cref{prop:local-perturbation}.
\begin{proof}
We first prove \Cref{eqn: component bound}, and we then show that \Cref{eqn: component bound 2} holds as a consequence. The main idea for the proof lies in the error analysis on the sketched linear equation for \(G_k\) and \(\hat{G}_k\).
We first consider the case where \(k\) is not a boundary node. By \Cref{assumption: choice of A_k}, the linear equation for \(G_k\) reads as follows:
\begin{equation}\label{eqn: sketched equation for G_k supplement}
  \sum_{\alpha_{k}}G_{k}(\alpha_{k-1}, i_{k}, \alpha_{k})Z_{k}(\alpha_{k}, \mu) = B_{k}(\alpha_{k-1}, i_k, \mu).
\end{equation}

Likewise, obtaining \(\hat{G}_k\) is done by solving a perturbed linear system via least-squares. From \(\hat{Z}_{k}\) as in \Cref{eqn: approximated sketch}, the linear equation for \(\hat{G}_k\) reads:
\begin{equation}\label{eqn: sketched equation for hat G_k supplement}
  \sum_{\alpha_{k}}\hat{G}_{k}(\alpha_{k-1}, i_{k}, \alpha_{k})\hat{Z}_{k}(\alpha_{k}, \mu) = \hat{B}_{k}(\alpha_{k-1}, i_k, \mu).
\end{equation}

For any \(\alpha_{k-1} \in [r_{k-1}], i_k \in [4]\), we take \(b, \hat{b}\) to be the slice of \(B_k\) and \(\hat{B}_{k}\) where we take the first index to be \(\alpha_{k-1}\) and second index to be \(i_k\). In MATLAB notation, one might write \(b = B_k(\alpha_{k-1}, i_k, :), \hat{b} = \hat{B}_k(\alpha_{k-1}, i_k, :)\). Likewise, we take \(v, \hat{v}\) to be the slice of \(G_k\) and \(\hat{G}_{k}\) where we take the first index to be \(\alpha_{k-1}\) and second index to be \(i_k\). Then, one can see that \(v\) is the exact solution to \(Z_k^{\top}x = b\), whereas \(\hat{v}\) is the least-squares solution to \(\hat{Z}_k^{\top}x = \hat{b}\). Thus, by invoking \Cref{lemma: Least-Squares stability}, one has
\begin{equation*}
    \lVert \hat{v} - v \rVert \leq 2 \lVert A^{\dagger} \rVert\left(\lVert \Delta A \rVert\lVert v \rVert + \lVert \Delta b \rVert \right) \leq \frac{2}{s_{r_k}(Z_k)} \left(\varepsilon_1\lVert v \rVert + \lVert \Delta b \rVert \right).
\end{equation*}

By the Cauchy-Schwarz inequality, one has 
\begin{equation}\label{eqn: perturbation bound on a slice}
    \lVert \hat{v} - v \rVert^2 \leq 2\left(\frac{2}{s_{r_k}(Z_k)}\right)^2 \left(\varepsilon_1^2\lVert v \rVert^2 + \lVert \Delta b \rVert^2 \right).
\end{equation}
Thus, summing over all \((\alpha_{k-1}, i_k)\) indices, one has
\begin{equation*}
    \lVert G_k - \hat{G}_k \rVert^2_F \leq 2\left(\frac{2}{s_{r_k}(Z_k)}\right)^2 \left(\varepsilon_1^2\lVert G_k \rVert^2_F + \lVert \Delta B_k \rVert^2_F \right) \leq 2\left(\frac{2}{s_{r_k}(Z_k)}\right)^2 \left(\varepsilon_1^2\lVert G_k \rVert^2_F + \varepsilon_2^2 \right).
\end{equation*}

For \(k = n\), one can see that \(G_k = B_k\) and \(\hat{G}_k = \hat{B}_k\), and so \(\lVert G_k - \hat{G}_k \rVert^2_F \leq \varepsilon_2^2\). For \(k = 1\), one can repeat the same calculation as in the \(1 < k < n\) case. One has 
\begin{equation}
  \sum_{\alpha_{1}}G_{1}(i_{1}, \alpha_{1})Z_{1}(\alpha_{1}, \mu) = B_{1}(i_1, \mu), \quad \sum_{\alpha_{1}}\hat{G}_{1}(i_{1}, \alpha_{1})\hat{Z}_{1}(\alpha_{1}, \mu) = \hat{B}_{1}(i_1, \mu).
\end{equation}
For any \(i_1 \in [4]\), one can take we takes \(b, \hat{b}\) to be the slice of \(B_1\) and \(\hat{B}_{1}\) where we take the first index to be \(i_1\), and one takes \(v, \hat{v}\) to be the slice of \(G_1\) and \(\hat{G}_{1}\) where we take the first index to be \(i_1\). By invoking \Cref{lemma: Least-Squares stability} and repeating the exact same calculation for \( 1 < k < n\), one obtains \(\lVert G_1 - \hat{G}_1 \rVert^2_F \leq 2\left(\frac{2}{s_{r_1}(Z_1)}\right)^2 \left(\varepsilon_1^2\lVert G_1 \rVert^2_F + \varepsilon_2^2 \right)\). We have thus shown that \Cref{eqn: component bound} holds for all \(k \in [n].\)

We now show that \Cref{eqn: component bound 2} holds with high probability. First, we let \(\tilde{\varepsilon} >0\) be a parameter to be specified. We employ \Cref{prop: classical shadow accuracy}. For \(B \geq c_S\tilde{\varepsilon}^{-2}\), simple norm equivalence shows that the following two bounds jointly hold with probability \(1 - \delta\):
\[\max_{k \in [n-1]}\lVert Z_{k} - \hat{Z}_{k}\rVert \leq r_{\mathrm{max}}\tilde{\varepsilon}, \quad \max_{k \in [n]}\lVert B_{k} - \hat{B}_{k}\rVert_F \leq 2r_{\mathrm{max}}\tilde{\varepsilon}.\]

Suppose that \(r_{\mathrm{max}}\tilde{\varepsilon} \leq \frac{1}{2}\min_{k\in [n-1]}s_{r_k}(Z_k)\). Then, plugging in \Cref{eqn: component bound}, one sees that one has
\begin{equation*}
    \lVert \hat{G}_k - G_k \rVert_{F}^2 \le 2c_Z^2(c_G^2r_{\mathrm{max}}^2\tilde{\varepsilon}^2 + 4r_{\mathrm{max}}^2\tilde{\varepsilon}^2) = 2c_Z^2r_{\mathrm{max}}^2(c_G^2 + 4) \tilde{\varepsilon}^2.
\end{equation*}
Thus, for \Cref{eqn: component bound 2} to hold with probability \(1 - \delta\), one takes \(\tilde{\varepsilon}^{-2} = 2c_Z^2r_{\mathrm{max}}^2(c_G^2 + 4) \varepsilon^{-2}\), which shows that our choice of \(B\) leads to \(r_{\mathrm{max}}\tilde{\varepsilon} \leq \frac{1}{2}\min_{k\in [n-1]}s_{r_k}(Z_k)\) with probability \(1 - \delta\). Therefore, the choice of \(B\) allows \Cref{eqn: component bound 2} to hold with probability \(1 - \delta\).
\end{proof}

\subsection{Global error bound on estimating \texorpdfstring{$\rho$}{rho}}\label{sec: global error}
We have established that the tensor components \((\hat{G}_k)_{k = 1}^{n}\) converge to \((G_k)_{k = 1}^{n}\) at an \(\gO(B^{-1/2})\) rate when given \(\gO(B)\) copies of \(\rho\). This subsection shows how the local component-wise convergence ensures that the final output \(\tilde{\rho}\) converges to \(\rho\). The analysis in this section uses a novel perturbational result for tensor train.
This part of the error analysis heavily uses tensor reshaping, and we frequently reshape tensors into matrices. Therefore, we introduce an additional matricization notation that simplifies the analysis. 

\begin{definition}
    Let \(n\) be the dimension of an \(n\)-tensor \(C \colon \prod_{k = 1}^{n}[m_k] \to \R\), and let \(I, J\) be two non-overlapping index sets with \(I \cup J = [n]\). We use \(C(i_I; i_J)\) to denote the \emph{unfolding matrix} with bipartition \(I \cup J = [n]\). Namely, we group the indices in \(I\) and \(J\) respectively as the row index and column index. The matrix \(C(i_I; i_J)\) is of size \((\prod_{k \in I}m_k) \times (\prod_{l \in J}m_l)\).
\end{definition}

We present a result on how local component-wise error propagates to the global tensor train error in the Frobenius norm. To the best of our knowledge, this result is a novel contribution first derived in this work. We anticipate that this bound is of independent interest for the tensor network community. Minimal modifications would generalize the result to tree-based tensor networks as defined in \cite{tang2025wavelet}. 

\begin{lemma}
\label{lemma: perturbational bound for TT}
    Let \(d\) be the dimension, and let \((F_k)_{k = 1}^{d}\) be a series of tensor components such that \(F_1 \in \R^{m_1 \times r_1}\), \(F_k \in \R^{r_{k-1} \times m_k \times r_k}\) for \(k = 2, \ldots, d-1\), and \(F_d \in \R^{r_{d - 1} \times m_d}\). Let \(D\) be a tensor train formed by \((F_k)_{k = 1}^{d}\) via the following equation
    \begin{equation}
      \PsiBlockGeneral[0.5]{$D$}{d} \;=\; \MPSChainGeneral[0.5]{F}{d}.
    \end{equation}
    
    For \(k = 1, \ldots, d\), let \(\Delta F_k\) be a perturbation to \(F_k\) with \(\hat{F}_k = F_k + \Delta F_k\). Let \(\tilde{D}\) be a tensor train formed by \((\hat{F}_k)_{k = 1}^{d}\) via the following equation
    \begin{equation}
      \PsiBlockGeneral[0.5]{$\tilde{D}$}{d} \;=\; \MPSChainGeneral[0.5]{\hat{F}}{d}.
    \end{equation}

    We suppose that the following conditions hold:
    \begin{enumerate}
        \item For \(k = 1\), one has \(m_1 \geq r_{1}\). Moreover, the unfolding matrix \(F_1(i_1; \alpha_{1}) \in \R^{m_1 \times r_{1}}\) is of rank \(r_{1}\).
        \label{item: condition case k =1 in perturbation bound}
        \item For \(k = 2, \ldots, d - 1\), one has \(m_k \geq r_{k-1}r_{k}\). Moreover, the unfolding matrix \(F_k(i_k; (\alpha_{k-1}, \alpha_k))\) is of rank \(r_{k-1}r_k\).
        \label{item: condition case 1<k<d in perturbation bound}
        \item For \(k = d\), one has \(m_d \geq r_{d-1}\). Moreover, the unfolding matrix \(F_d(i_d; \alpha_{d-1}) \in \R^{m_d \times r_{d-1}}\) is of rank \(r_{d-1}\).
        \item There exists an \(\varepsilon > 0\) so that for any \(k \in [d]\), one has
        \begin{equation}
            \lVert F_k - \hat{F}_k\rVert_{F} \leq \varepsilon.
        \end{equation}
    \end{enumerate}

    Under the given conditions, define a constant \(c_F\) by 
    \[c_F = \max\left(\lVert F_1(i_1; \alpha_{1})^{\dagger}\rVert, \max_{k = 2, \ldots, d-1} \lVert F_k(i_k; (\alpha_{k-1}, \alpha_k))^{\dagger} \rVert,\lVert F_d(i_d; \alpha_{d-1})^{\dagger} \rVert \right).\]
    Then, assuming that \(\varepsilon \leq c_{F}^{-1}d^{-1}\), one has
    \[
    \frac{\lVert D - \tilde{D}\rVert_{F} }{\lVert D\rVert_{F}} \leq 2 d c_F\varepsilon.
    \]
\end{lemma}
\begin{proof}
    We first perform the analysis via a telescoping sum construction. Let \(D_0 = D\), \(D_d = \tilde{D}\). For \(k = 1, \ldots, d-1\), we let \(D_k\) be a tensor train with tensor component given by \((\hat{F}_l)_{l = 1}^{k}\cup (F_l)_{l = k+1}^{d}\). In terms of the diagram, one writes
    \begin{equation}
      \PsiBlockGeneral[0.5]{$D_k$}{d} \;=\; \MPSChainPerturbed[0.5].
    \end{equation}
    By the triangle inequality, one has 
    \[
    \lVert D - \tilde{D}\rVert_{F} = \lVert D_0 - D_d\rVert_{F} \leq \sum_{k = 0}^{d-1}\lVert D_k - D_{k+1}\rVert_{F}.
    \]

    For each of the difference term \(\lVert D_k - D_{k+1}\rVert_{F}\), we define tensors \(D_{<k}\), \(\tilde{D}_{<k}\) by the following equations:
    \[
    \begin{aligned}
        &\PsiLeftBlockGeneral[0.5]{D} \;=\;
        \MPSLeftChainGeneral[0.5]{F}, \\&\PsiLeftBlockGeneral[0.5]{\tilde{D}} \;=\; \MPSLeftChainGeneral[0.5]{\hat{F}}.
    \end{aligned}
    \]
    Similarly, we define \(D_{>k}\) by
    \[
    \begin{aligned}
        \PsiRightBlockGeneral[0.5]{D}{d} \;=\;
        \MPSRightChainGeneral[0.5]{F}{d}.
    \end{aligned}
    \]
    As a result, for \(k = 2, \ldots, d - 1\), one has
    \begin{equation}\label{eqn: telescoping term diagram}
        \PsiBlockGeneral[0.5]{$D_{k}$}{d} \;-\; \PsiBlockGeneral[0.5]{$D_{k-1}$}{d} \;=\; \MPSChainPerturbation[0.5].
    \end{equation}

    We first consider the boundary term \(\lVert D_0 - D_{1}\rVert_{F}\). Similar to \Cref{eqn: telescoping term diagram}, one has
    \begin{equation}\label{eqn: telescoping term diagram case k  = 0}
        \PsiBlockGeneral[0.5]{$D_{1}$}{d} \;-\; \PsiBlockGeneral[0.5]{$D_{0}$}{d} \;=\; \MPSChainPerturbationLeftCase[0.5].
    \end{equation}

    Deriving the bound for the \(\lVert D_0 - D_{1}\rVert_{F}\) term in \Cref{eqn: telescoping term diagram case k  = 0} is instructive, and our subsequent bound uses essentially the same technique. We let \(J_1 = \{2, \ldots, n\}\) and we consider the unfolding matrices \(D_0(i_1; i_{J_1})\), \(D_1(i_1; i_{J_1})\) and \(D_{>1}(\alpha_1; i_{J_1})\). The equation in \Cref{eqn: telescoping term diagram case k  = 0} reads
    \[
    D_1(i_1; i_{J_1}) - D_0(i_1; i_{J_1}) = \Delta F_1(i_1;\alpha_1)D_{>1}(\alpha_1; i_{J_1}).
    \]

    Let \(A \in \R^{m \times r}\) be a matrix of rank \(r\), and let \(B \in \R^{r \times L}\) be a matrix where \(L\) could potentially be very large. For \(v\) being an arbitrary column of \(B\), one has
    \[
    \frac{\lVert Av \rVert^2}{\lVert v \rVert^2} \geq \lVert A^{\dagger} \rVert^{-2}.
    \]
    By summing \(\lVert Av \rVert^2 \geq \lVert A^{\dagger} \rVert^{-2}\lVert v \rVert^2\) over all choices of \(v\), one has 
    \[
    \frac{\lVert AB \rVert_F}{\lVert B \rVert_F} \geq \lVert A^{\dagger} \rVert^{-1}.
    \]
    
    Then, let \(\Delta A\) be a perturbation to \(A\) with \(\hat{A} = A + \Delta A\). One has 
    \[
    \lVert (A - \hat{A})B\rVert_{F} \leq  \lVert (A - \hat{A})\rVert \lVert B\rVert_{F} \leq \lVert A^{\dagger}\rVert \lVert \Delta A \rVert\lVert AB \rVert_F.
    \]

    In other words, for a well-conditioned matrix \(A\), one has a relative error bound
    \begin{equation}\label{eqn: relative error matrix}
        \frac{\lVert (A - \hat{A})B\rVert_{F}}{\lVert AB\rVert_{F}} \leq \lVert A^{\dagger}\rVert\lVert \Delta A \rVert.
    \end{equation}
    Then, by condition (\ref{item: condition case k =1 in perturbation bound}) in the statement of \Cref{lemma: perturbational bound for TT}, we see that we can apply \Cref{eqn: relative error matrix} with \(F_1, \hat{F_1}\) in place of \(A, \hat{A}\) and with \(D_{>1}(\alpha; i_{J_1})\) in place of \(B\). One has 
    \begin{equation}
    \label{eqn: TT perturbation bound k=1 case}
        \lVert D_1 -D_0 \rVert_F \leq \lVert \Delta F_1 \rVert\lVert F_1^{\dagger}\rVert\lVert F_1(i_1;\alpha_1)D_{>1}(\alpha_1;i_{J_1}) \rVert_F = \lVert \Delta F_1 \rVert\lVert F_1^{\dagger}\rVert\lVert D \rVert_{F},
    \end{equation}
    where the last equality is true since the contraction between \(F_{1}\) and \(D_{<1}\) on the \(\alpha_1\) index forms \(D\).

    For \(1 < k < n\), we first rewrite \Cref{eqn: telescoping term diagram} in terms of a matrix equation. We let \(J_{k} = \{k+1, \ldots, d\}\), and then one can write  
    \[
    \begin{aligned}
        &D_{k}(i_k;(i_{[k-1]}, i_{J_k})) - D_{k-1}(i_k;(i_{[k-1]}, i_{J_k}))\\ = 
        &\Delta F_k(i_k;(\alpha_{k-1}, \alpha_k))\tilde{D}_{<k}\otimes D_{>k}((\alpha_{k-1}, \alpha_{k}); (i_{[k-1]}, i_{J_k}),
    \end{aligned}
    \]
    In particular, the unfolding matrix term \(\tilde{D}_{<k}\otimes D_{>k}((\alpha_{k-1}, \alpha_{k}); (i_{[k-1]}, i_{J_k})\) is the Kronecker product between the two unfolding matrices \(\tilde{D}_{<k}(\alpha_{k-1}; i_{[k-1]})\) and \(D_{>k}(\alpha_{k}; i_{J_k})\). One can check that the matrix equation is equivalent to \Cref{eqn: telescoping term diagram}. 
    
    To simplify the notation, in what follows, we assume that \(D_{k-1}, D_k\) respectively represent the unfolding matrix \(D_{k-1}(i_k;(i_{[k-1]}, i_{J_k})), D_{k}(i_k;(i_{[k-1]}, i_{J_k}))\). We assume that \(F_{k}, \Delta F_k\) respectively represent the unfolding matrix \( F_k(i_k;(\alpha_{k-1}, \alpha_k)), \Delta F_k(i_k;(\alpha_{k-1}, \alpha_k))\). We assume that \(D_{<k}, \tilde{D}_{<k}\) respectively represent the unfolding matrix \(D_{<k}(\alpha_{k-1}; i_{[k-1]}), \tilde{D}_{<k}(\alpha_{k-1}; i_{[k-1]})\). Likewise, we assume that \(D_{>k}\) represent the unfolding matrix \(D_{>k}(\alpha_{k}; i_{J_k})\). Then, one can write \Cref{eqn: telescoping term diagram} compactly as the matrix equation
    \[
    D_{k} - D_{k-1} = \Delta F_{k}\left(\tilde{D}_{<k} \otimes D_{>k}\right).
    \]
    By the triangle inequality, we have
    \begin{equation}
    \label{eqn: TT perturbation bound middle case step 1}
        \lVert \Delta F_{k}\left(\tilde{D}_{<k} \otimes D_{>k}\right) \rVert_F \leq \lVert \Delta F_{k}\left((\tilde{D}_{<k} - D_{<k})  \otimes D_{>k}\right) \rVert_F + \lVert \Delta F_{k}\left(D_{<k}  \otimes D_{>k}\right) \rVert_F.
    \end{equation}
    We see that condition (\ref{item: condition case 1<k<d in perturbation bound}) in the statement of \Cref{lemma: perturbational bound for TT} allows us to use \Cref{eqn: relative error matrix}. In particular, we take \(F_k, \hat{F_k}, \Delta F_k\) to be in place of \(A, \hat{A}, \Delta A\). For the place of \(B\), we consider both \(D_{<k}\otimes D_{>k}\) and \(\tilde{D}_{<k}\otimes D_{>k}\). 
    
    In the first case, we take \(B\) to be \(D_{<k}\otimes D_{>k}\), and we have
    \begin{align*}
        &\lVert \Delta F_{k}\left((\tilde{D}_{<k} - D_{<k})  \otimes D_{>k}\right)  \rVert_F \\\leq &\lVert \Delta F_k \rVert\lVert F_k^{\dagger}\rVert\lVert F_k \left(\tilde{D}_{<k} - D_{<k}\right)\otimes D_{>k}
        \rVert_F,
        \\= &\lVert \Delta F_k \rVert\lVert F_k^{\dagger}\rVert\lVert D_{k-1} - D
        \rVert_F,
    \end{align*}
    where the last equality is because the contraction between \(F_k\) and \(\tilde{D}_{<k}\otimes D_{>k}\) on the \((\alpha_{k-1}, \alpha_k)\) index forms \(D_{k-1}\) (see \Cref{eqn: Equation for C in TN diagram} for a diagram illustration), whereas the contraction between \(F_k\) and \(D_{<k} \otimes D_{>k}\) on the \((\alpha_{k-1}, \alpha_k)\) index forms \(D\).

    Similarly, taking \(B\) to be \(D_{<k}\otimes D_{>k}\) in \Cref{eqn: relative error matrix}, we have
    \begin{align*}
        &\lVert \Delta F_{k}\left(D_{<k}  \otimes D_{>k}\right)  \rVert_F \\\leq &\lVert \Delta F_k \rVert\lVert F_k^{\dagger}\rVert\lVert F_k\left( D_{<k}  \otimes D_{>k}\right)
        \rVert_F,
        \\= &\lVert \Delta F_k \rVert\lVert F_k^{\dagger}\rVert\lVert D
        \rVert_F,
    \end{align*}
    where the last equality is because the contraction between \(F_k\) and \(D_{<k} \otimes D_{>k}\) on the \((\alpha_{k-1}, \alpha_k)\) index forms \(D\).

    Thus, combining the two results, we have 
    \begin{equation}
    \label{eqn: TT perturbation bound middle case step 2}
        \lVert D_{k-1} - D_k  \rVert_F \leq \lVert \Delta F_k \rVert\lVert F_k^{\dagger}\rVert\left(\lVert D_{k-1} - D
        \rVert_F+ \lVert D
        \rVert_F\right).
    \end{equation}

    For the case of \(k = d\), by repeating the calculation in the \(1 < k < d\) case, one has 
    \begin{equation}
    \label{eqn: TT perturbation bound k=d case}
        \lVert D_{d-1} - D_d  \rVert_F \leq \lVert \Delta F_d(i_d; \alpha_{d-1}) \rVert \lVert F_d(i_d; \alpha_{d-1})^{\dagger}\rVert\left(\lVert D_{d-1} - D
        \rVert_F+ \lVert D
        \rVert_F\right).
    \end{equation}

    The rest of the proof is by a simple induction-type argument. We write \(b_{k} = \lVert D_{k-1} - D_{k} \rVert.\) One can see that \Cref{eqn: TT perturbation bound k=1 case} can be written as 
    \[
    b_1 \leq  \lVert \Delta F_1 \rVert\lVert F_1^{\dagger}\rVert\lVert D \rVert_{F} \leq c_F \varepsilon\lVert D \rVert_{F},
    \]
    where the last inequality uses the definition of \(c_F\) and \(\varepsilon\) in the statement of \Cref{lemma: perturbational bound for TT}.
    Likewise, \Cref{eqn: TT perturbation bound middle case step 2} shows that, for \(k = 2, \ldots, d-1\), one has
    \[
    b_k \leq  \lVert \Delta F_k \rVert\lVert F_k^{\dagger}\rVert\left(\lVert D_{k-1} - D \rVert_{F}+\lVert D \rVert_{F}\right) \leq c_F \varepsilon \left(\sum_{l = 1}^{k-1}b_{l} + \lVert D \rVert_{F}\right).
    \]
    Lastly, for \(k = d\), we have
    \[
    \begin{aligned}
    b_k \leq  &\lVert \Delta F_d(i_d; \alpha_{d-1}) \rVert \lVert F_d(i_d; \alpha_{d-1})^{\dagger}\rVert \left(\lVert D_{d-1} - D \rVert_{F}+\lVert D \rVert_{F}\right) \\\leq &c_F \varepsilon \left(\sum_{l = 1}^{d-1}b_{l} + \lVert D \rVert_{F}\right).
    \end{aligned}
    \]

    Our induction hypothesis is that \(b_k \leq c_F \varepsilon(1 + c_F \varepsilon)^{k-1}\lVert D \rVert_{F}\). One sees that this inequality holds when \(k = 1\). Then, suppose that the statement is true for all indices up to \(k\). Then, for \(k+1\), one has
    \begin{align*}
        &b_{k+1} \leq c_F \varepsilon \left(\sum_{l = 1}^{k}b_{l} + \lVert D \rVert_{F}\right)\leq c_F \varepsilon \left(\sum_{l = 1}^{k}\varepsilon c_F(1+c_F \varepsilon)^{l-1} + 1\right)\lVert D \rVert_{F}
        = c_F \varepsilon \left(1+c_F\varepsilon\right)^{k}\lVert D \rVert_{F},
    \end{align*}
    where the last equality uses the formula for geometric series summation. Thus, the claim is true for \(k \in [d]\). 

    Lastly, we plug in the telescoping sum and obtain
    \[
    \lVert D - \tilde{D}\rVert_{F} \leq \sum_{k = 1}^{d}b_k \leq \sum_{k = 1}^{d}c_F \varepsilon(1 + c_F \varepsilon)^{k-1}\lVert D \rVert_{F} \leq \left((1+c_F\varepsilon)^{d} - 1\right)\lVert D \rVert_{F}.
    \]
    Under the assumption that \(c_F\varepsilon < 1/d\), one further has 
    \[
    \frac{\lVert D - \tilde{D}\rVert_{F}}{\lVert D \rVert_{F}} \leq \left((1+c_F\varepsilon)^{d} - 1\right) \leq  e^{c_Fd\varepsilon} -1  \leq 2c_Fd\varepsilon,
    \]
    where the second inequality is due to \(1 + t \leq e^{t}\), and the third inequality is because \(e^x \leq 1+2x\) for \(x \in [0, 1]\).
\end{proof}

One can see that \Cref{lemma: perturbational bound for TT} can not be directly applied to our setting unless the coefficient tensor \(C\) has maximal rank \(r \leq 2\). However, the result in \Cref{lemma: perturbational bound for TT} becomes applicable after one merges several tensor components into a bigger tensor component. To use \Cref{lemma: perturbational bound for TT}, we shall merge connected tensor components. For example, if one has \(L\) tensor components \(G_{k}, \ldots, G_{k+L-1}\), then merging these components leads to
\[
\MPSChainIllustrationTwo[0.5] \;=\;\MPSChainIllustration[0.5],
\]
which one views as a 3-tensor of size \(r_{k-1} \times 4^L \times r_{k+L-1}\). One sees that it suffices to take merge \(L = \ceil{\log_2{(r_{\mathrm{max}}})}\) blocks to ensure that the condition in \Cref{lemma: perturbational bound for TT} can hold.

We are now ready to prove our main result.

\begin{proposition}
\label{prop:global-perturbation}(Global error bound on sketch tomography)
Assume that one is in the setting of \Cref{prop:local-perturbation}. Let \(\varepsilon_3\) be an accuracy parameter so that \Cref{alg: sketch tomography} satisfies \[\max_{k \in [n]}\lVert G_{k} - \hat{G}_{k}\rVert_{F} \leq \varepsilon_3.\]

Suppose that \(r_{\mathrm{max}}\) is as in \Cref{prop:local-perturbation}. We let \(L = \ceil{\log_2(r_{\mathrm{max}})}\) and we suppose that \(n > L^2\). Then, there exists \(a, b \in \mathbb{Z}_{\geq 0}\) so that \(n = aL + b(L+1)\). For \(j = 1, \ldots, a,\) we construct \(F_{j}\) to be a tensor component constructed by merging tensor components \(G_{(j-1)L+1}, \ldots, G_{(j-1)L+L}\). For \(j = a+1, \ldots, a+b\), we \(F_{j}\) to be a tensor component constructed by merging tensor components \(G_{aL + (j-1-a)(L+1)+1}, \ldots, G_{aL +(j-a-1)(L+1)+L+1}\). Similarly, we define \(\hat{F}_{j}\) to be tensor components obtained by merging \((\hat{G}_{k})_{k = 1}^{n}\). Then, if \(\varepsilon_3 \leq \frac{c_G}{L+1}\), one has
\begin{equation}
\label{eqn: component bound 3}
    \max_{j \in [a+b]}\lVert F_j - \hat{F}_j\rVert_{F} \leq 2 (L+1)c_G^{L}\varepsilon_3.
\end{equation}

In particular, let \(\varepsilon \in (0, 1)\) be an accuracy parameter and let \(c_S, r_{\mathrm{max}}, c_Z, c_G\) be the same terms as in \Cref{prop:local-perturbation}. Moreover, let \(c_F\) be as in \Cref{lemma: perturbational bound for TT} with the definition of \((F_j)_{j = 1}^{a+b}\) given as above. In \Cref{alg: sketch tomography}, set \(W = 2\log(12n\tilde{r}_{\mathrm{max}}^2/\delta)\) and set \[B \geq 128n^2c_G^{2L} c_F^2 c_Sr_{\mathrm{max}}^2 c_Z^2(c_G^2 + 4)\varepsilon^{-2},\] 
Then, with probability \(1 - \delta\), one has 
\begin{equation}
\label{eqn: total bound}
    \frac{\lVert \tilde{\rho} - \rho \rVert_{F}}{\lVert\rho\rVert_{F}}  \le \varepsilon.
\end{equation}

\end{proposition}

\begin{proof}    
    We first prove \Cref{eqn: component bound 3}. For two matrices \(A, B\) one has the simple bound 
    \[
    \lVert AB \rVert_F \leq \lVert A \rVert\lVert B \rVert_{F} \leq \lVert A \rVert_F \lVert B \lVert_{F}.
    \]

    Therefore, for an arbitrary tensor network \(E\) consisting of tensor components \((M_k)_{k  =1}^{K}\), one has the simple bound
    \begin{equation}
    \label{eqn: global bound step 1}
    \lVert E \rVert_{F} \leq \prod_{k = 1}^{K}\lVert M_k \rVert_{F},
    \end{equation}
    and the inequality is quite loose in general. Moreover, let \(E'\) be a tensor network of the same tensor network structure as \(E\), and let \(E'\) consist of tensor components \((M'_k)_{k  =1}^{K}\). In particular, each of the tensor components \(M_k'\) is of the same shape as \(M_k\). We claim that
    \begin{equation}
    \label{eqn: global bound step 2}
    \lVert E' - E \rVert_{F} \leq \sum_{S \subseteq [K], |S| > 0} \prod_{k \in S}\lVert M_k' - M_k \rVert_{F} \prod_{j \not \in S}\lVert M_j \rVert_{F}.
    \end{equation}
    We prove \Cref{eqn: global bound step 2}. Let \(S \subseteq [K]\) be a nonempty subset and let \(E_{S}\) be the tensor network of the same tensor network structure as \(E\). Let \(E_S\) be of tensor component \((M_k' - M_k)_{k  \in S} \cup (M_j)_{j  \not \in S}\). By multi-linearity, one has \(E' - E = \sum_{S \subseteq [K], |S|>0}E_S\), and hence one has
    \[
    \lVert E' - E \rVert_{F} \leq \sum_{S \subseteq [K], |S| > 0} \lVert E_S \rVert_{F} \leq \sum_{S \subseteq [K], |S| > 0} \prod_{k \in S}\lVert M_k - M_k' \rVert_{F} \prod_{j \not \in S}\lVert M_j \rVert_{F},
    \]
    where the first inequality is by the triangle inequality, and the second inequality is by \Cref{eqn: global bound step 1}. In particular, if one writes \(\zeta = \max_{k \in [K]}\lVert M_k \rVert_F\) and \(\gamma = \max_{k \in [K]}\lVert M_k - M_k' \rVert_F\), then \Cref{eqn: global bound step 2} implies 
    \[
    \lVert E' - E \rVert_{F} \leq \sum_{S \subseteq [K], |S| > 0} \gamma^{|S|}\zeta^{K - |S|} = (\gamma + \zeta)^{K} - \zeta^{K},
    \]
    where the last equality is by the binomial theorem. In particular, we consider the case where \(\gamma \leq \frac{\zeta}{K}\), where one has 
    \[
    \lVert E' - E \rVert_{F} \leq (\gamma + \zeta)^{K} - \zeta^{K} = \zeta^{K}\left(( 1+ \gamma/\zeta)^K - 1\right) \leq \zeta^{K}\left(\exp(\gamma K/\zeta ) - 1\right) \leq 2\gamma K\zeta^{K-1},
    \]
    where the second inequality is because \((1+x) \leq \exp(x)\), and the second inequality is because \(\exp(x) \leq 1+2x\) for \(x \in [0, 1]\).

    We can now bound the magnitude of \(\lVert F_j - \hat{F}_j\rVert_{F}\). We see that \(F_j, \hat{F}_j\) are two tensor networks of the same structure, and the tensor components share the same shape. There are at most \(L+1\) components. Each component of \(F_j\) is bounded in the Frobenius norm by \(C_G\), and the difference in each tensor component is likewise bounded in the Frobenius norm by \(\varepsilon_3\). Thus, if \(\varepsilon_3 \leq \frac{c_G}{L+1}\), one has 
    \[
    \lVert F_j - \hat{F}_j\rVert_{F} \leq 2 (L+1)c_G^{L}\varepsilon_3.
    \]
    
    We now show that \Cref{eqn: total bound} holds with high probability. First, we let \(\tilde{\varepsilon}\) be a parameter to be specified. We employ \Cref{prop:local-perturbation}. For \(B \geq 2c_Sc_Z^2r_{\mathrm{max}}^2(c_G^2 + 4) \tilde{\varepsilon}^{-2}\), the following bound holds with probability \(1 - \delta\):
    \[\max_{k \in [n]}\lVert G_{k} - \hat{G}_{k}\rVert_{F} \leq \tilde{\varepsilon}.\]   
    
    Plugging in \Cref{eqn: component bound 3}, for \(\tilde{\varepsilon} \leq c_G/(L+1)\), one has
    \begin{equation*}
        \max_{j \in [a+b]}\lVert \hat{F}_j - F_j \rVert_{F} \le 2 (L+1)c_G^{L}\tilde{\varepsilon}.
    \end{equation*}
    Let \(\tilde{C}\) denote the tensor train consisting of tensor components \((\hat{G}_k)_{k = 1}^{n}\). One can see that \(C, \tilde{C}\) can be respectively viewed as a tensor train with tensor components \((F_j)_{j = 1}^{a+b}\) and \((\hat{F}_j)_{j = 1}^{a+b}\). Using \Cref{lemma: perturbational bound for TT}, if \(\tilde{\varepsilon} \leq \frac{1}{2}(L+1)^{-1}c_G^{-L}c_F^{-1}(a+b)^{-1}\), we see that one has
    \[
    \lVert \rho - \tilde{\rho} \rVert = \lVert C - \tilde{C} \rVert \leq  4(a+b) c_F(L+1)c_G^{L}\tilde{\varepsilon} \lVert C \lVert_{F} \leq 8nc_Fc_G^{L}\tilde{\varepsilon}\lVert \rho \lVert_{F},
    \]
    where the last inequality uses \(\lVert C \lVert_{F} = \lVert \rho \lVert_{F}\) and \((a+b)(L+1) \leq 2n\).
    Thus, for \Cref{eqn: total bound} to hold with probability \(1 - \delta\), one takes \(\tilde{\varepsilon}^{-2} = 64n^2c_F^2c_G^{2L}\varepsilon^{-2}\). Plugging in \Cref{prop:local-perturbation}, one sees that our choice of \(B\) allows \Cref{eqn: total bound} to hold with probability \(1 - \delta\).

    Lastly, the informal version in \Cref{prop: informal global sample complexity} holds as a consequence of \Cref{eqn: total bound}. In particular, when \(\rho\) is the density matrix of a pure state \(\ket{\psi}\), one has \(\lVert \rho \rVert_{F} = 1\), which is why the statement in \Cref{prop: informal global sample complexity} does not contain a \(\lVert \rho \rVert_{F}\) factor.

\end{proof}

We give some remarks on the bound that we have obtained in \Cref{prop:global-perturbation}. The \(c_G^{L}\) dependency is mild because one has \(L = \ceil{\log_2(r_{\mathrm{max}})}\). However, the \(c_G^{L}\) dependency is due to the loose bound in \Cref{eqn: global bound step 1}. We conjecture that a more refined analysis can improve the \(c_G^{L}\) dependence to an \(\gO(1)\) dependence that is independent of \(L\). We leave a more refined analysis for future work. 

Moreover, the assumption of \(n \geq L^2\) is not necessary. For the regime of \(L \leq n < L^2\), one is not guaranteed to write an integer less than \(L^2 - L - 1\) by a weighted sum of \(L\) and \(L+1\). Therefore, one would require a more complicated construction for the merging operation to construct \((F_j)_{j = 1}^{d}\). That said, for \(n \leq L^2\), one can obtain a similar result as obtained in \Cref{prop:global-perturbation}, but the worst case would involve an \(\gO(c_G^{4L})\) dependence.

\section{Detail on maximum likelihood estimation training using MPS}\label{sec: MLE}
\subsection{Methodology}
We follow the procedure to utilize the Pauli measurement data for MLE training.
One has access to \(B\) unitary transformations \(\{U^{(j)} \in U(2^n)\}_{j = 1}^{B}\) and the corresponding binary measurement outcomes \(b^{(j)} = (b^{(j)}_1, \ldots, b^{(j)}_n) \in \{0, 1\}^{n}\). The \(j\)-th measurement outcome is also a state \(\ket{b^{(j)}} \in \sC^{2^n}\) by the computational basis encoding. In the MLE framework, one minimizes a negative-log-likelihood (NLL) loss defined as follows
\begin{equation}\label{eqn: NLL}
    \NLL(\ket{\phi}) = -\frac{1}{B}\sum_{j = 1}^{B}{\log\left(\lvert\bra{b^{(j)}} U^{(j)}\ket{\phi}\rvert^2\right)} + \log(\braket{\phi | \phi}),
\end{equation}
and \(\ket{\phi}\) is optimized over an MPS class. We illustrate the key concepts in a tensor diagram. The ansatz \(\ket{\phi}\) is parameterized by a collection of tensor component \((F_k)_{k = 1}^{n}\) and the representation of \(\ket{\phi}\) is as follows:
\[
\KetPhiBlock[0.5] \; = \; \KetPhiMPSChain[0.5]\,.
\]

The normalization constant \(Z = \braket{\phi | \phi}\) is evaluated in a tensor diagram as follows:
\begin{equation}\label{eqn: def of Z in phi}
    Z \; =\; \KetPhiZDef[0.5]\, .
\end{equation}

Similarly, the evaluation of \(\bra{b^{(j)}} U^{(j)}\ket{\phi}\) is done in a tensor diagram. Let \(U^{(j)} = \otimes_{k = 1}^{n}U_{k}^{(j)}\) be the unitary transformation for the \(j\)-th measurement. One can write the diagram for the evaluation as follows:
\begin{equation}\label{eqn: def of eval in phi}
    \bra{b^{(j)}} U^{(j)}\ket{\phi} = \KetPhiEvalDef[0.5]\,.
\end{equation}

The diagrams in \Cref{eqn: def of Z in phi} and \Cref{eqn: def of eval in phi} show that \(\NLL\) in \Cref{eqn: NLL} is efficient to evaluate and differentiate against in an MPS class. For the optimization algorithm, we use a one-site DMRG approach summarized in \Cref{alg:DMRG}, whereby DMRG refers to the fact that the mixed canonical form has been used and that the parameter update is done sequentially.

\begin{algorithm}[H]
\caption{One-site DMRG algorithm for optimizing \(\NLL\)}
\label{alg:DMRG}
\begin{algorithmic}[1]
\REQUIRE Loss function \(\mathcal{L}\).
\REQUIRE Current tensor component \((F_k)_{k = 1}^{n}\).
\REQUIRE Learning rate \(\eta\)
\WHILE{Not converged}
\FOR{\(i = 1,\ldots, n\) \textbf{and then} \(i = n,\ldots,1\)}
\STATE Transform \(\ket{\phi}\) to the mixed canonical form centered at site \(i\).
\STATE Perform gradient descent update: \(F_{i} \gets F_{i} - \eta \frac{\partial \NLL}{\partial F_{i}}\).
\ENDFOR
\ENDWHILE

\RETURN Final tensor component \((F_k)_{k = 1}^{n}\).
\end{algorithmic}
\end{algorithm}

\subsection{Implementation detail}

We report the implementation details for our MLE benchmark. For the 1D Heisenberg experiment and the 1D TFIM experiment in the main text, we choose a learning rate of \(\eta = 0.1\) and we obtain \(\ket{\phi}\) by running many iterations of \Cref{alg:DMRG} until \(\mathcal{L}(\ket{\phi}) \leq \mathcal{L}(\ket{\psi})\). By the definition of the loss function in the MLE procedure, this means that the obtained state \(\ket{\phi}\) is better at explaining the generated data than the true state \(\ket{\psi}\). The ansatz for \(\ket{\phi}\) is taken to be in the same class as \(\ket{\psi}\) in the sense that \(\ket{\phi}\) is optimized over the complex MPS and all tensor components of \(\ket{\phi}\) are of the same shape as \(\ket{\psi}\). 

The main text shows that the obtained MLE solution is not accurate in certain observable estimation tasks. One can conclude from the numerical performance that, for practical settings, an MPS model that is competitive in the log likelihood metric can have poor observable estimation accuracy. MLE is trained on the log likelihood metric, and it does not include a penalization term for how well it predicts individual observables. Therefore, when the sample size is not sufficiently large, the fact that \(\ket{\psi}\) is successful in the log likelihood metric is insufficient to guarantee an accurate observable prediction. On the other hand, MLE training is more computationally challenging when the sample size is large. We remark that the conducted experiment is already more ideal than practical settings by using the log likelihood level of the true model \(\ket{\psi}\) as a termination criterion. In practice, one might even prematurely terminate training when in a local minimum, which leads to a worse quality for the MLE solution. Therefore, while MLE is a good practical proposal for quantum state tomography, our proposed sketch tomography procedure is, in some cases, more advantageous due to its performance guarantee in \Cref{prop: informal global sample complexity} and due to practical performances shown in the main text.

We explain why we do not include a benchmark in the 2D Heisenberg experiment. To calculate the evaluation diagram used in \Cref{eqn: def of eval in phi}, one needs to at least store a complex tensor of size \(a \times 2 \times a\). Therefore, the memory complexity of calculating \(B\) samples in parallel is on the order of \(4a^2B\) floating points. For \(B = 10^4\) samples, the memory requirement is 6.4 GB, which is already quite substantial. Therefore, training the MLE model has a substantial memory complexity even for a small sample size. Moreover, as mentioned in the main text, the Pauli measurement samples for \(\ket{\psi}\) are too small and will likely lead to overfitting when one chooses \(\ket{\phi}\) to be in the same MPS class as \(\ket{\psi}\). Therefore, to have a fair assessment of MLE, one would also need a much larger sample size, which would drastically increase the computational complexity to perform the training. Therefore, we choose to keep the sample size at \(B = 8 \times 10^5\) and only present the result for sketch tomography. For researchers who wish to train an MLE benchmark in this 2D experiment, we suggest that training the MPS model with a mini-batch stochastic gradient descent strategy might allow one to compare MLE with shadow tomography.

\section{Proof of the lower bound}\label{sec: lower proof}

In this section, we provide both a formal version and a complete proof of the information-theoretic lower bound presented in \Cref{prop:lower bound} of \Cref{subsec: info lower bound} above. We begin by presenting a mathematical definition of tensor trains as follows, which serves as a complement to the definition based on tensor diagrams presented in~\eqref{eqn: Equation for C in TN diagram} of \Cref{sec: procedure} above. 
\begin{definition}[Mathematical definition of Tensor Trains]
\label{def: class of TTs}
In general, any TT $\rho \in \sC^{2^n \times 2^n}$ associated with a quantum system of $n$ qubits has the following form: 
\begin{equation}
\label{eq: defn of MPO}
\begin{aligned}
&\rho(x_1,x_2,\cdots,x_n;y_1,y_2,\cdots,y_n)\\
:= &\sum_{\alpha_1 =1}^{r_1}\cdots\sum_{\alpha_{n-1}=1}^{r_{n-1}}G_1(x_1,y_1,\alpha_1)G_{2}(\alpha_{1}, x_{2},y_{2},\alpha_{2}) \cdots G_n(\alpha_{n-1},x_n,y_n),
\end{aligned}
\end{equation}
where $G_k \in \mathbb{R}^{r_{k-1} \times 2 \times 2 \times r_k}$ for $1 \leq k \leq n$ with $r_0 = r_n =1$ are called the tensor components associated with the TT $\rho$. The vector $\vr := (r_0,r_1,\cdots,r_n)$ is said to be the ranks of the TT. Throughout this section, we use \(\mathcal{M}_{n}(R,C)\) to denote the class of all density matrices on $n$ qubits that admit a TT representation as in~\eqref{eq: defn of MPO} with bounded rank and Frobenius norm. Specifically, 
\begin{equation}
\label{eqn: defn of MPO class}
\mathcal M_n(R,C):=
\Biggl\{\, \rho\in\mathbb C^{2^n\times 2^n}\ \Big|\ 
\begin{aligned}
&\rho \text{ is a density matrix of form}~\eqref{eq: defn of MPO},\\
&\|\rho\|_F \leq C \text{ and } r_k \leq R\ (1\le k\le n-1)
\end{aligned}
\Biggr\}.    
\end{equation}
\end{definition}
With the formal definition of TTs given above, we then present a mathematically rigorous version of the information-theoretic lower bound as below.
\begin{proposition}
[Formal version of the lower bound]
\label{prop:lower bound formal}
Let $\gU$ denote the random Pauli basis measurement primitive. For any matrix product state with density matrix $\rho$, we take $\left\{\widehat{\rho}_{(i)}\right\}_{i=1}^{B}$ to be the $B$ estimations of $\rho$ produced by applying the classical shadow protocol associated with $\gU$ over $B$ measurements. Then for any constant $r \geq 2$ and estimator of $\rho$ denoted by $\widehat{\Theta}:\{\widehat{\rho}_{(i)}\}_{i=1}^{B} \rightarrow \sC^{2^n \times 2^n}$, we have the following minimax lower bound: 
\begin{equation}
\label{eqn: minimax lower bound in main thm}
\inf_{\widehat{\Theta}}\sup_{\rho \in \gM_n(r,1)}\mathbb{E}_{\{\widehat{\rho}_{(i)}\}_{i=1}^{B}}\left[\left\|\widehat{\Theta} - \rho\right\|_{F}\right] \gtrsim \sqrt{\frac{n}{B}},
\end{equation}
under the assumption that $B \geq n$, where the notation $\mathbb{E}_{\{\widehat{\rho}_{(i)}\}_{i=1}^{B}}[\cdot]$ above means that the expectation is taken with respect to the $B$ estimations $\{\widehat{\rho}_{(i)}\}_{i=1}^{B}$.
\end{proposition}

Before discussing the proof of \Cref{prop:lower bound formal} above, we need to present a few lemmas. In particular, \Cref{lem: tensor product of MPS} below discusses properties of the TT class specified in \Cref{def: class of TTs} above under tensor products, while \Cref{lem: rank bound on general TT} provides a universal upper bound on the rank of any TT in $\mathbb{C}^{2^n \times 2^n}$.

\begin{lemma}
\label{lem: tensor product of MPS}
Fix integers $n_1,n_2$ and $R_1, R_2$ in $\sN^+$. Then for any two constants $C_1,C_2 \in (0,1)$ and TTs $\rho^{(1)}, \rho^{(2)}$ satisfying $\rho^{(1)} \in \gM_{n_1}\left(R_1,C_1\right)$ and $\rho^{(2)} \in \gM_{n_2}(R_2,C_2)$, we have that the tensor product $\rho^{(1)} \otimes \rho^{(2)} \in \gM_{n_1+n_2}\left(\max\{R_1,R_2\},C_1C_2\right)$.
\end{lemma}
\begin{proof}
Note that the bound on Frobenius norm directly follows from the fact that $\left\|\rho^{(1)} \otimes \rho^{(2)}\right\|_F = \|\rho^{(1)}\|_F\|\rho^{(2)}\|_F$. Hence, it suffices to check that the tensor product $\rho^{(1)} \otimes \rho^{(2)}$ can be expressed by some TT whose ranks are all bounded by $\max\{R_1,R_2\}$. Following the notations used in~\eqref{eq: defn of MPO}, we assume that the tensor components of $\rho^{(1)}$ and $\rho^{(2)}$ are given by $\{C_{j}\}_{j=1}^{n_1}$ and $\{D_j\}_{j=1}^{n_2}$ respectively. Specifically, the two TTs can be expressed in the form of~\eqref{eq: defn of MPO} as follows:  
\begin{equation}
\begin{aligned}
&\rho^{(1)}(x_1,x_2,\cdots,x_{n_1};y_1,y_2,\cdots,y_{n_1})\\
:= &\sum_{\alpha_1 =1}^{r_1}\cdots\sum_{\alpha_{n_{1}-1}=1}^{r_{n_{1} - 1}}C_1(x_1,y_1,\alpha_1)C_{2}(\alpha_{1}, x_{2},y_{2},\alpha_{2}) \cdots C_{n_1}(\alpha_{n_{1}-1},x_{n_1},y_{n_1}),\\
&\rho^{(2)}(x_1,x_2,\cdots,x_{n_2};y_1,y_2,\cdots,y_{n_2})\\
:= &\sum_{\beta_1 =1}^{r'_1}\cdots\sum_{\beta_{n_2-1}=1}^{r'_{n_2 - 1}}D_1(x_1,y_1,\beta_1)D_{2}(\beta_{1}, x_{2},y_{2},\beta_{2}) \cdots D_{n_2}(\beta_{n_2-1},x_{n_2},y_{n_2}),
\end{aligned}
\end{equation}
Now let's consider tensor components $\left\{E_{j}\right\}_{j=1}^{n_1+n_2}$ defined as follows:
\begin{equation}
\begin{aligned}
E_{n_1}(\gamma,x,y,1) &= C_{n_1}(\gamma,x,y), \ \left(1 \leq \gamma \leq r_{n_1-1}, x \in \{0,1\}, y \in \{0,1\}\right)\\
E_{n_1+1}(1,x,y,\gamma) &= D_1(x,y,\gamma), \ \left(1 \leq \gamma \leq r'_{1}, x \in \{0,1\}, y \in \{0,1\}\right)\\
E_{k} &= C_k \ (1 \leq k \leq n_1-1), \ E_{k'} = D_{k'-n_1} \ (n_1 + 2 \leq k' \leq n_1+n_2).  
\end{aligned}
\end{equation}
Then we have that the tensor product $\rho^{(1)} \otimes \rho^{(2)}$ satisfies 
\begin{equation}
\begin{aligned}
&\left(\rho^{(1)} \otimes \rho^{(2)}\right)(x_1,\cdots,x_{n_1}, x_{n_1+1}, \cdots, x_{n_1+n_2};y_1,\cdots,y_{n_1}, y_{n_1+1}, \cdots, y_{n_1+n_2})\\
= &\rho^{(1)}(x_1,\cdots,x_{n_1};y_1,\cdots,y_{n_1})\rho^{(2)}(x_{n_1+1},\cdots,x_{n_1+n_2};y_{n_1+1},\cdots,y_{n_1+n_2})\\
= &\left(\sum_{\alpha_1 =1}^{r_1}\cdots\sum_{\alpha_{n_{1}-1}=1}^{r_{n_{1} - 1}}C_1(x_1,y_1,\alpha_1) \cdots C_{n_1}(\alpha_{n_{1}-1},x_{n_1},y_{n_1})\right)\\
&\left(\sum_{\beta_1 =1}^{r'_1}\cdots\sum_{\beta_{n_2-1}=1}^{r'_{n_2 - 1}}D_1(x_{n_1+1},y_{n_1+1},\beta_1) \cdots D_{n_2}(\beta_{n_2-1},x_{n_1+n_2},y_{n_1+n_2})\right)\\
= &\sum_{\gamma_1=1}^{r_1}\cdots\sum_{\gamma_{n_1-1} = 1}^{r_{n_1-1}}\sum_{\gamma_{n_1}=1}^{1}\sum_{\gamma_{n_1+1}=1}^{r_1'}\cdots\sum_{\gamma_{n_1+n_2-1}=1}^{r'_{n_2-1}}E_1(x_1,y_1,\gamma_1)\\
&E_{2}(\gamma_{1}, x_{2},y_{2},\gamma_{2}) \cdots E_{n_1}(\gamma_{n_{1}+n_2-1},x_{n_1+n_2},y_{n_1+n_2}).
\end{aligned}    
\end{equation}
Given that $\rho^{(1)} \in \gM_{n_1}(R_1,C_1)$ and $\rho^{(2)} \in \gM_{n_2}(R_2,C_2)$, we have $r_j \leq R_1 \ (1 \leq j \leq n_1-1)$ and $r'_j \leq R_2 \ (1 \leq j \leq n_2-1)$. Then we can directly deduce that $\rho^{(1)} \otimes \rho^{(2)}$ can be expressed by some TT with ranks all bounded by $\max\{R_1,R_2\}$. This concludes our proof. 
\end{proof}

\begin{lemma}
\label{lem: rank bound on general TT}
For any fixed integer $m \in \mathbb{N}^+$ and $\rho \in \sC^{2^m \times 2^m}$, we have $\rho \in \gM_m(2^m,\|\rho\|_F)$     
\end{lemma}
\begin{proof}
From the definition of the class $\gM_m(2^m,\|\rho\|_F)$ specified in~\eqref{eqn: defn of MPO class} above, we know it suffices to show that any $\rho \in \sC^{2^m \times 2^m}$ can be expressed as a TT whose ranks are all bounded by $2^m$. In fact, if we use $A_k$ to denote the $k$-th unfolding matrix of $\rho$, where its row and column are formed by the first $k$ sites and the last $m-k$ sites respectively, then we have that $A_k \in \sC^{4^k \times 4^{m-k}}$ and satisfies the following bound
\begin{equation}
\text{rank}(A_k) \leq \min\{4^k, 4^{m-k}\} \leq \sqrt{4^m} = 2^m    
\end{equation}
Then we apply Theorem 2.1 in~\cite{oseledets2011tensor} directly to deduce that $\rho$ can be expressed as a TT with ranks bounded by $2^m$, as desired. 
\end{proof}

In addition to the results on TTs presented above, we also need a few results from nonparametric statistics in our proof. Specifically, we need Theorem 20 from~\cite{fischer2020sobolev}, which is presented as below:
\begin{theorem}[Fano's Method]
\label{lem: Fano's method}
Fix an integer $M \geq 2$ and let $(\Omega,\gA)$ be some measurable space. Given $(M+1)$ probability measures $\{P_j\}_{j=0}^{M}$ satisfying $P_j \ll P_0$ for any $1 \leq j \leq M$ and 
\begin{equation}
\frac{1}{M}\sum_{j=1}^{M}\KL\left(P_j, P_0\right) \leq \alpha_\ast    
\end{equation}
for some $\alpha^\ast \in (0,\infty)$. Then for all classifiers represented by some measurable function $\Psi: \Omega \rightarrow \{0,1,2,\cdots,M\}$, the following bound holds:
\begin{equation}
\max_{0 \leq j \leq M}P_j\left(\omega \in \Omega: \Psi(\omega) \neq j\right) \geq \frac{\sqrt{M}}{1+\sqrt{M}}\left(1- \frac{3\alpha_\ast}{\log(M)}-\frac{1}{2\log(M)}\right).    
\end{equation}
\end{theorem}

One other result we need here is the Varshamov-Gilbert (VG) Lemma, which is presented as Theorem 2.9 in~\cite{tsybakov2004introduction}. 
\begin{theorem}[VG Lemma]
\label{thm: VG lemma}
Fix some $D \in \mathbb{N}^+$ that is divisible by $8$. Then there exists a subset $\mathcal{V}=\left\{\tau^{(0)},\cdots,\tau^{(2^{\frac{D}{8}})}\right\}$ of binary strings from the $D$-dimensional hypercube $\{0,1\}^D$, such that $\tau^{(0)}=(0,0,\cdots,0)$ and
\begin{equation}
\label{eqn: VG lemma l1 norm difference}
\left\|\tau^{(j)}-\tau^{(k)}\right\|_{1} = \sum_{l=1}^D\left|\tau^{(j)}_l-\tau^{(k)}_l\right| \geq \frac{D}{8}
\end{equation}
for any $0 \leq j \neq k \leq 2^{\frac{D}{8}}$.  
\end{theorem}
\begin{remark}
Without loss of generality, above we assume that $D$ is divisible by $8$, as this only changes the previously stated lower bound up to a constant. Also, we note that the $l_1$ norm $\|\cdot\|_{1}$ above essentially measures the difference between any two binary strings, so the claim above essentially provides a way of constructing a collection of binary strings that are pairwisely separated from each other. 
\end{remark}

We note that the VG lemma is also a key component in the proof of the information-theoretic lower bound established in~\cite{cai2016optimal}, where it is employed to construct a family of pairwise-separated density matrices. Such a family, which is often referred to as a covering, plays a central role in characterizing fundamental information-theoretic limits of estimation problems from not only statistics~\cite{tsybakov2004introduction} but also quantum physics~\cite{barthel2018fundamental,lowe2025lower}. With all essential tools listed above, we now present a complete proof of the lower bound provided in \Cref{prop:lower bound formal} as follows.

\textit{Proof of \Cref{prop:lower bound formal}}
Without loss of generality, here we assume that $n$ is some integer divisible by $8$, as this will only impact the information-theoretic lower bound up to some constant. We begin by taking $\gamma = \frac{1}{1600B}$, which satisfies $\epsilon \in \left(0,\frac{1}{1600n}\right)$ from the given assumption $B \geq n$. Then we consider the following two density matrices $\sigma,\omega$ in $\mathbb{C}^{2 \times 2}$ defined as follows:
\begin{equation}
\label{eqn: defn of basis density matrices}
\sigma = \frac{1}{2}\mI_2 -\frac{1}{2}\sqrt{\gamma}X + \frac{1}{2}\sqrt{1-\gamma}Y, \ \omega := \frac{1}{2}\mI_2 + \frac{1}{2}\sqrt{\gamma}X + \frac{1}{2}\sqrt{1-\gamma}Y  
\end{equation}
Moreover, applying the VG Lemma listed in \Cref{thm: VG lemma} above yields a collection $\mathcal{V}_{n} := \left\{\tau^{(0)},\cdots,\tau^{(2^{\frac{n}{8}})}\right\}$ of $\left(2^{\frac{n}{8}} + 1\right)$ binary strings in $\{0,1\}^{n}$ such that
\begin{equation}
\label{eqn: VG lemma l1 norm lower bound}
\left\|\tau^{(j)}-\tau^{(k)}\right\|_{1} = \sum_{l=1}^{n}\left|\tau^{(j)}_l-\tau^{(k)}_l\right| \geq \frac{n}{8},
\end{equation}
holds for any $ 0 \leq j \neq k \leq 2^{\frac{n}{8}}$. Based on the two density matrices $\sigma, \omega$ and the collection $\mathcal{V}_{n}$ of binary strings constructed above, we further define a density matrix $\rho^{(j)}$ in $\sC^{n \times n}$ associated with the string $\tau^{(j)}$ for any $j \in \left\{0,1,\cdots,2^{\frac{n}{8}}\right\}$, which takes the form of a tensor product as follows: 
\begin{equation}
\label{eqn: defn of density matrices via tensor product}
\begin{aligned}
\rho^{(j)} &:= \left(\tau^{(j)}_1 \sigma + \left(1-\tau^{(j)}_1\right)\omega\right) \otimes \cdots \otimes \left(\tau^{(j)}_n \sigma + \left(1-\tau^{(j)}_n\right)\omega\right).
\end{aligned}    
\end{equation}
Let $\mathcal{C}_{n} := \left\{\rho^{(0)}, \rho^{(1)}, \cdots,\rho^{(2^{\frac{n}{8}})}\right\}$ be the corresponding collection of density matrices defined in~\eqref{eqn: defn of density matrices via tensor product} above. We begin by first verifying that $\gC_n \subseteq \gM_n(r,1)$ for constant $r \geq 2$. On the one hand, since $\left\|\sigma\right\|_{F} = \left\|\omega\right\|_{F} =  \frac{1}{2}\left(1+\epsilon +1-\epsilon\right) = 1$, a direct computation of the Frobenius norm $\left\|\rho^{(j)}\right\|_{F}$ yields that
\begin{equation}
\label{eqn: bound on frobenius norm of each rho_j}
\begin{aligned}
\left\|\rho^{(j)}\right\|_{F} = \prod_{i=1}^{n}\left\|\tau^{(j)}_i \sigma + \left(1-\tau^{(j)}_i\right)\omega\right\|_{F} = 1,
\end{aligned}
\end{equation}
for any $0 \leq j \leq 2^{\frac{L}{8}}$. On the other hand, we apply \Cref{lem: rank bound on general TT} to the $i$-th component $\tau^{(j)}_i \sigma + \left(1-\tau^{(j)}_i\right)\omega$ in the tensor product $\rho^{(j)}$ to deduce that 
\begin{equation}
\tau^{(j)}_i \sigma + \left(1-\tau^{(j)}_i\right)\omega \in \gM_{1}(2,1) \subset \gM_{1}(r,1),    
\end{equation} 
for any $0 \leq j \leq 2^{\frac{L}{8}}$ and $1 \leq i \leq L$. Then we may further apply \Cref{lem: tensor product of MPS} to the tensor product $\rho^{(j)} = \left(\tau^{(j)}_1 \sigma + \left(1-\tau^{(j)}_1\right)\omega\right) \otimes \cdots \otimes \left(\tau^{(j)}_L \sigma + \left(1-\tau^{(j)}_L\right)\omega\right)$ to deduce that $\rho^{(j)} \in \gM_{1 \times n}(r,1) = \gM_{n}(r,1)$ for any $0 \leq j \leq 2^{\frac{L}{8}}$, as desired. 

Now we proceed to verify that density matrices in the class $\gC_n$ are pairwisely separated before applying Fano's method. Specifically, for any $0 \leq j \neq k \leq 2^{\frac{L}{8}}$, we may compute the Frobenius norm of the difference $\rho^{(j)} - \rho^{(k)}$ as below:
\begin{equation}
\label{eqn: lower bound between difference of density matrices}
\begin{aligned}
\left\|\rho^{(j)} - \rho^{(k)}\right\|^2_{F} &= \left\langle \rho^{(j)} - \rho^{(k)}, \rho^{(j)} - \rho^{(k)} \right\rangle_F \\
&= \left\|\rho^{(j)}\right\|^2_{F} + \left\|\rho^{(k)}\right\|^2_{F} - 2\left\langle \rho^{(j)}, \rho^{(k)}\right\rangle_F = 2- 2\left\langle \rho^{(j)}, \rho^{(k)} \right\rangle_F\\
&= 2-2\prod_{i=1}^{n}\left\langle \tau^{(j)}_i \sigma + \left(1-\tau^{(j)}_i\right)\omega, \tau^{(k)}_i \sigma + \left(1-\tau^{(k)}_i\right)\omega \right\rangle_F\\
&= 2-2\left\langle\sigma, \omega\right\rangle_{F}^{\left\|\tau^{(j)}-\tau^{(k)}\right\|_{1}}.
\end{aligned}    
\end{equation}
Plugging in the expressions of $\sigma$ and $\omega$ in~\eqref{eqn: defn of basis density matrices} above gives us that
\begin{equation}
\label{eqn: computation of basis dot product}
\left\langle\sigma, \omega\right\rangle_{F} = \frac{1}{2} -\frac{1}{2}\gamma + \frac{1}{2}(1-\gamma) = 1-\gamma
\end{equation}
Then we may substitute~\eqref{eqn: computation of basis dot product} and the lower bound in~\eqref{eqn: VG lemma l1 norm lower bound} into~\eqref{eqn: lower bound between difference of density matrices}, which yields the following lower bound
\begin{equation}
\label{eqn: lower bound on separation distance}
\begin{aligned}
\left\|\rho^{(j)} - \rho^{(k)}\right\|^2_{F} &\geq 2-2\left(1-\gamma\right)^{\frac{n}{8}} \geq 2 \times \frac{1}{2}\times \gamma \times \frac{n}{8} \geq \frac{\gamma n}{8},
\end{aligned}
\end{equation}
where the second last inequality follows from the assumption $B \geq n$ and the fact that 
$$1-(1-t)^n \geq nt - \frac{n(n-1)}{2}t^2 \geq \frac{1}{2}nt$$
holds for any $n \in \sN$ and $t \in (0, \frac{1}{n}]$. Then we can define $\delta:= \sqrt{\frac{\gamma n}{8}}$ to be the separation distance between any two density matrices. 

Furthermore, we now consider the probability distributions associated with the samples obtained by measuring the collection $\gC_n$ of pairwisely separated density matrices via the classical shadow protocol based on $\mathcal{U}$. Specifically, we use the binary string 
$b^{(j)}_k = \left(b^{(j)}_k(l)\right)_{l=1}^{n} \in \{0,1\}^{n}$ 
to denote the outcome obtained from the $k$-th measurement of the density matrix $\rho^{(j)}$ under the classical shadow protocol based on $\mathcal{U}$ for any $1 \leq k \leq B$. From the tensor product structure of each $\rho^{(j)}$ in $\mathcal{C}_n$, we can deduce that the joint distribution formed by $\left\{b^{(j)}_k\right\}_{k=1}^{B}$ satisfies
\begin{equation}
P_{j} := \prod_{k=1}^{B}P_{b^{(j)}_k} = \prod_{k=1}^{B}\prod_{l=1}^{n}P_{b^{(j)}_k(l)}   
\end{equation}
for any $0 \leq j \leq 2^{\frac{L}{8}}$. Then a direct computation further implies
\begin{equation}
\label{eqn: upper bound on KL between product of dists}
\begin{aligned}
\KL(P_j, P_0) &= \KL\left(\prod_{k=1}^{B}\prod_{l=1}^{n}P_{b^{(j)}_k(l)}, \prod_{k=1}^{B}\prod_{l=1}^{n}P_{b^{(0)}_k(l)}\right)\\
&= \sum_{k=1}^{B}\sum_{l=1}^{n}\KL\left(P_{b^{(j)}_k(l)}, P_{b^{(0)}_k(l)}\right).
\end{aligned}    
\end{equation}
From our construction of the density matrices $\rho^{(j)} \ (1 \leq j \leq 2^{\frac{L}{8}})$ given above, we have that each $P_{b^{(j)}_k(l)}$ must be a Bernoulli distribution. Moreover, we note that the difference between $\rho^{(j)}$ and $\rho^{(0)}$ is determined by the difference between $\sigma$ and $\omega$ for any $1 \leq j \leq 2^{\frac{n}{8}}$. By using $\text{Ber}(p)$ to denote the Bernoulli distribution with mean $p$ for any $p \in (0,1)$, we can use the expressions of $\sigma$ and $\omega$ given in~\eqref{eqn: defn of basis density matrices} above to deduce that  $\KL\left(P_{b^{(j)}_k(l)}, P_{b^{(0)}_k(l)}\right)$ can be upper bounded as follows
\begin{equation}
\begin{aligned}
\KL\left(P_{b^{(j)}_k(l)}, P_{b^{(0)}_k(l)}\right) &\leq \KL\left(\text{Ber}\left(\frac{1+\sqrt{\gamma}}{2}\right), \text{Ber}\left(\frac{1-\sqrt{\gamma}}{2}\right)\right)\\
&= \frac{1+\sqrt{\gamma}}{2}\log\left(\frac{\frac{1+\sqrt{\gamma}}{2}}{\frac{1-\sqrt{\gamma}}{2}}\right) + \frac{1-\sqrt{\gamma}}{2}\log\left(\frac{\frac{1-\sqrt{\gamma}}{2}}{\frac{1+\sqrt{\gamma}}{2}}\right)\\
& = \sqrt{\gamma}\log\left(\frac{1+\sqrt{\gamma}}{1-\sqrt{\gamma}}\right) \leq \frac{2\gamma}{1-\gamma} \leq 4\gamma,
\end{aligned}    
\end{equation}
where the second last inequality above follows from the fact that $\log\left(\frac{1+t}{1-t}\right) \leq \frac{2t}{1-t^2}$ for $t \in (-1,1)$ and the last inequality above follows from our choice of $\gamma$. Substituting the upper bound derived above into~\eqref{eqn: upper bound on KL between product of dists} then yields that 
\begin{equation}
\begin{aligned}
\frac{1}{2^{\frac{n}{8}}}\sum_{j=1}^{2^{\frac{n}{8}}}\KL(P_j, P_0) \leq 4nB\gamma = \frac{n}{400}. 
\end{aligned}    
\end{equation}
By letting $\mathcal{B}:= \{\widehat{\rho}_{(i)}\}_{i=1}^{B} $ be the collection of $B$ outcomes obtained via the $B$ measurements under the classical shadow protocol and taking $\alpha^\ast:= \frac{n}{400}$, $M:= 2^{\frac{n}{8}}$ here, we can further apply Fano's method listed in \Cref{lem: Fano's method} above to deduce the following lower bound on the probability of misclassification for any classifier $\Theta: \mathcal{B} \rightarrow \{0,1,\cdots,M\}$:
\begin{equation}
\label{eqn: lower bound on misclassification probability}
\begin{aligned}
\max_{0 \leq j \leq M}P_j\left(\Theta(\mathcal{B}) \neq j\right) 
&\geq \frac{2^{\frac{n}{16}}}{1+2^{\frac{n}{16}}}\left(1 - \frac{3n}{400 \times \frac{n}{8}\log(2)} -\frac{1}{\frac{n}{2}\log(2)}\right) \\
&\geq \frac{1}{2}\left(1- \frac{3}{50\log(2)} - \frac{1}{2\log(2)}\right) \geq \frac{1}{20}
\end{aligned}    
\end{equation}
where the second last inequality above follows from the assumption $n \geq 8$. 

Finally, we will prove the desired minimax lower bound by combining the lower bound on the misclassification probability proved in~\eqref{eqn: lower bound on misclassification probability} above with Markov's inequality and inequality~\eqref{eqn: lower bound on separation distance} proved earlier. Specifically, for any estimator $\widehat{\Theta}:\{\widehat{\rho}_{(i)}\}_{i=1}^{B} \rightarrow \sC^{2^n \times 2^n}$ of the target density matrix, we define a corresponding classifier $\widetilde{\Theta}:\mathcal{B} \rightarrow \{0,1,\cdots,M\}$ as follows:
\begin{equation}
\widetilde{\Theta} \in \argmin_{0 \leq j \leq M} \left\|\widehat{\Theta} -\rho^{(j)}\right\|_{F}    
\end{equation}
Given that the collection $\gC_n = \{\rho^{(j)}\}_{j=0}^{M}$ has been proved to be $\delta$-seperated in~\eqref{eqn: lower bound between difference of density matrices}, we then have the following inequality for any $0 \leq j \leq M$:
\begin{equation}
P_j\left(\widetilde{\Theta}(\mathcal{B}) \neq j\right) \leq P_j\left(\left\|\widehat{\Theta} -\rho^{(j)}\right\|_{F} \geq \frac{\delta}{2}\right)     
\end{equation}
from the triangle inequality. Applying Markov's inequality to the RHS above and the lower bound in~\eqref{eqn: lower bound on misclassification probability} to the LHS above further implies 
\begin{equation}
\label{eqn: application of Markov's ineq}
\begin{aligned}
\max_{0 \leq j \leq M}\frac{\mathbb{E}_{P_j}\left[\left\|\widehat{\Theta} -\rho^{(j)}\right\|_{F}\right]}{\frac{\delta}{2}} &\geq \max_{0 \leq j \leq M}P_j\left(\left\|\widehat{\Theta} -\rho^{(j)}\right\|_{F} \geq \frac{\delta}{2}\right)\\
&\geq \max_{0 \leq j \leq M}P_j\left(\widetilde{\Theta}(\mathcal{B}) \neq j\right) \geq \frac{1}{20}. 
\end{aligned}
\end{equation}
Finally, plugging in~\eqref{eqn: application of Markov's ineq} indicates that
\begin{equation}
\label{eqn: final minimax lower bound}
\begin{aligned}
\inf_{\widehat{\Theta}}\sup_{\rho \in \gM_n(r,1)}\mathbb{E}_{\{\widehat{\rho}_{(i)}\}_{i=1}^{B}}\left[\left\|\widehat{\Theta} - \rho\right\|_{F}\right] &\geq \inf_{\widehat{\Theta}}\max_{0 \leq j \leq M}\mathbb{E}_{P_j}\left[\left\|\widehat{\Theta} - \rho^{(j)}\right\|_{F}\right]\\
&\geq \frac{\delta}{40} = \frac{1}{40}\sqrt{\frac{\gamma n}{8}} \gtrsim \sqrt{n\gamma} \gtrsim \sqrt{\frac{n}{B}},
\end{aligned}
\end{equation}
as desired. This concludes our proof of the information-theoretic lower bound.

\begin{remark}
In order to obtain the lower bound on the number of measurements presented in \Cref{prop:lower bound} above, we can set $\epsilon = \sqrt{\frac{n}{B}}$ to be the lower bound proved in \Cref{prop:lower bound formal} here. Solving for the number of measurements then yields $B = \frac{n}{\epsilon^2}$. Moreover, we note that the proof strategy adopted here is similar to that of~\cite{cai2016optimal}, which mainly relies on standard tools such as Fano's method and the VG lemma from nonparametric statistics. An alternative way to establish lower bounds of similar form is to follow ideas used in prior studies like~\cite{huang2020predicting,soleimanifar2022testing,chen2025stabilizer,bakshi2025learning,lowe2025lower} by leveraging tools from quantum information theory, which we leave as future work. 
\end{remark}

\end{appendices}

\bibliography{sn-bibliography}%

\end{document}